\documentclass[a4paper,12pt]{article}
\pdfoutput=1
\usepackage[utf8]{inputenc}
\usepackage{jheppub}
\usepackage{graphicx}
\usepackage{caption}
\usepackage{subcaption}
\usepackage{hyperref}

\newcommand{\be}{\begin{equation}}
\newcommand{\ee}{\end{equation}}
\newcommand{\bea}{\begin{eqnarray}}
\newcommand{\eea}{\end{eqnarray}}

\def\la{\langle}
\def\ra{\rangle}
\def\Tr{{\mathrm{Tr}}}

\def\CD{\mathcal{D}}

\def\CF{\mathcal{F}}
\def\CG{\mathcal{G}}

\def\CL{\mathcal{L}}

\def\CN{\mathcal{N}}

\def\CO{\mathcal{O}}
\def\CP{\mathcal{P}}

\def\CS{\mathcal{S}}

\title{Quasinormal Modes and Correlators in the Shear Channel of Spacetime-Filling Branes}

\author{Nikola I. Gushterov}

\affiliation{Rudolf Peierls Centre for Theoretical Physics, University of Oxford \\
Clarendon Laboratory,
Parks Road,
Oxford OX1 3PU\\
United Kingdom}

\emailAdd{nikola.gushterov@physics.ox.ac.uk}

\abstract{
We study the shear momentum diffusion and related modes of a strongly coupled $(2+1)$-dimensional 
conformal field theory at finite temperature and chemical potential, using a dual holographic description. 
We consider a space-time filling charged black brane solution  of Einstein's gravity in $(3+1)$-dimensional asymptotically Anti-de Sitter space coupled to a $U(1)$ gauge field via a Dirac-Born-Infeld action. In addition to temperature and  chemical potential, the holographic model has two other parameters: the tension of the brane, and the non-linearity parameter controlling the higher-derivative terms of the $U(1)$ field. By varying the parameters, one can, in particular,  interpolate between 
the Reissner-Nordstrom-AdS background and the background of probe branes embedded into AdS space. We find analytically the retarded two-point functions of the shear (transverse to the direction of spatial momentum) components of the energy-momentum tensor and the global $U(1)$ current of the $(2+1)$-dimensional field theory in the hydrodynamic approximation. We also find numerically the location of the poles of the correlators (quasinormal modes) for a wide range of the parameters, focusing on the effects of the back-reaction and non-linearities. We show, in particular, that the shear diffusion constant agrees with the hydrodynamic form for a wide range of parameters, including temperature and backreaction.

}

\preprint{OUTP-18-05P}

\begin{document}
\maketitle
\flushbottom
\section{Introduction}
\label{sec:intro}

Gauge-gravity duality \citep{Maldacena:1997re,Gubser:1998bc,Witten:1998qj} has been extensively used to get insights into the nature of strongly interacting quantum liquids at finite temperature $T$, and chemical potential $\mu$. In the strongly coupled regimes where the usual perturbation techniques fail and numerical methods  are often not reliable due to the ``sign problem", holographic techniques provide a framework where toy models can be built to understand general principles governing these systems, using gauge theories in the 't Hooft's large $N_c$ limit \citep{Aharony:1999ti}.

To model a strongly interacting  $d-$dimensional  Conformal Field Theory ($CFT_d$) in flat Minkowski spacetime at non-zero $T$ and $\mu$, one usually considers a dual description in terms of a charged black brane in asymptotically Anti-de Sitter spacetime in one extra dimension,  $(AdS_{d+1})$. Using the holographic dictionary, the temperature of the system is dual to the Hawking temperature of the black brane and the chemical potential is related to the flux of the electric field through the asymptotic $AdS$ boundary \citep{Aharony:1999ti}. The dictionary does not necessarily  specify the string-theoretic origin of this electric flux which gives rise to a variety of phenomenological ``bottom-up" models, usually referred to as the $AdS/CMT$ correspondence \citep{Hartnoll:2009sz, McGreevy:2009xe, Benini:2012iq, Hartnoll:2016apf}. Such models were  used extensively to study transport properties at strong coupling.

We will concentrate on the $d=3$ case, and study $CFT_3$ with adjoint matter ($N_c$) and charged fundamental matter  ($N_f$), holographically dual to a spacetime-filling black brane, with a gauge field governed by a non-linear Dirac-Born-Infeld ($DBI$) electrodynamics  \citep{Born:1934gh, Dirac:1962iy, Cederwall:1996uu, Ketov:2001dq} in AdS ($AdS_{4}DBI$), described by the following action \citep{Pal:2012zn, Tarrio:2013tta, Kundu:2017cfj, Pal:2017hai, Fernando:2003tz, Dey:2004yt}
\begin{equation}
S_{DBI} =\frac{1}{2\kappa^{2}}\int d^{4}x\sqrt{-g}\left(R+\frac{6}{L_{_0}^{2}}\right)-T_D\int d^{4}x\sqrt{-\det\left(g_{\mu\nu}+\alpha F_{\mu\nu}\right)}.
\label{main-action}
\end{equation}
Here, $L_{_0}$ is the bare $AdS$ radius, $R$ is the Ricci scalar of the spacetime metric $g_{\mu\nu}$  with determinant $g$, $T_D$ is the tension of the brane,  $F_{\mu\nu} = \partial_{\mu} A_{\nu}-\partial_{\nu} A_{\mu}$  is the field strength of the U(1) gauge field $A_{\mu}$ and $\alpha$ is the ``non-linearity" parameter which  controls the higher derivative corrections to 
  the standard $U(1)$ kinetic term. In holographic interpretation,
 $\alpha$ can be thought of as being  proportional to an inverse power of $\lambda$, 
 where  $\lambda\propto  g_{YM}^2 N_c \gg1 $ is the 't Hooft coupling of the dual gauge theory.
Despite the non-linearities in the electromagnetic sector, in $d=4$ space-time dimensions this system shares several remarkable properties with the linear Maxwell theory: the electric-magnetic self-duality, causal propagation without shock waves, and the existence of soliton-like solutions
 \citep{Gaillard:1997rt,Ketov:2001dq,Gibbons:1995cv}.
To make the self-duality manifest and to express the action in a more familiar form, we can use the identity for $4\times4$ antisymmetric matrices\footnote{For $N\times N$ antisymmetric matrix $F=F^{\mu}_{~\nu}$, there will be in general extra terms. We can use the well-known identity $\det\left(I + \alpha F\right)= \exp\left(\Tr[\log(I + \alpha F)]\right)=\exp\left(-\sum_{m=1}^{\infty}\frac{\alpha^{2m}}{2m}\Tr\left[F{}^{2m}\right]\right)$, and then expand the exponential to order $\alpha^N$, so that the last coefficient will be equal to $\det(\alpha F)$. In the case of $N=4$, we recover (\ref{eqdet}) after realizing that $\det(F)=\frac{1}{8} \left(\text{Tr}[F^2]^2-2 \text{Tr}[F^4]\right)$ and $\Tr[F^2]=\Tr[F^{\mu}_{~\nu}F^{\nu}_{~\alpha}]=-F_{\mu\nu}F^{\mu\nu}$.}
\begin{equation}\label{eqdet} 
\det(\delta_{\,\,\nu}^{\beta}+\alpha F_{~\nu}^{\beta})=1+\frac{\alpha^{2}}{2}F_{\mu\nu}F^{\mu\nu}+\alpha^4 \det\left(F_{~\nu}^{\mu}\right)
\end{equation}
to rewrite the action as
\begin{equation}
\label{action-x}
S_{DBI} = \frac{1}{2\kappa^{2}}\int d^{4}x\sqrt{-g}\left[R+\frac{6}{L_{_0}^{2}}-2\kappa^{2}T_D\CL_{DBI}\right],
\end{equation}
where we have defined
\begin{equation}\label{LDBI}
\CL_{DBI} \equiv \sqrt{1+\frac{\alpha^{2}}{2}F_{\mu\nu}F^{\mu\nu}+\CP\alpha^4\det (F_{~\nu}^{\mu})}
, \qquad \text{with}\qquad \det (F_{~\nu}^{\mu})\propto(\vec{B}\cdot\vec{E} )^{2}.
\end{equation}

Since we are only interested in translationally invariant systems at finite chemical potential $\mu$, we consider a homogeneous electric field in the bulk $F_{r t}=E(r)$. One might be tempted to set $ \det (F_{~\nu}^{\mu})=0$ in the previous expression, given that it won't change the geometry or the thermodynamics of the system. However, as we will show in Section~\ref{fluctuations}, the transverse or shear perturbations of $\delta A_{\mu}$ can source the $\det (F_{~\nu}^{\mu})$ term which will modify the equations of motion (E.o.M.) for the fluctuations of the gauge field and therefore the transport properties in the shear channel. For this reason we have introduced an extra parameter $\CP$ in Eq.~\eqref{LDBI}, in order to explore the effect that this term will have on our system\footnote{In contrast, the longitudinal channel studied in \citep{Tarrio:2013tta,andy} does not get contributions from this term and therefore it is safe to set $\CP = 0$, before computing the E.o.M. for the fluctuations.}.
In the linear limit $\alpha^2 F_{\mu\nu}F^{\mu\nu}\ll1$ with $\alpha^2\kappa ^2 T_D\sim1$, we recover the Einstein-Maxwell action,
\begin{equation}
 S_{RN}=\frac{1}{2\kappa^{2}}\int d^{4}x\sqrt{-g}\left(R+\frac{6}{L^{2}}-{\frac{\alpha^{2}\kappa^{2}T_D }{2}}F_{\mu\nu}F^{\mu\nu}\right)\,.
\end{equation}
The transverse  fluctuations of the AdS Reissner-Nordstrom ($AdS_{4}RN$) solution to the corresponding equations of motion were 
  previously studied in refs.~\citep{Davison:2013bxa,Edalati:2010hk}.

Note that the first term in the small $\alpha$ expansion of the action (\ref{action-x}) changes 
the effective cosmological constant and the $AdS$ radius to
\begin{equation}
 L^{2}=\frac{L_{_0}^{2}}{\left(1-{\frac{\kappa^{2}T_D L_{_0}^{2}}{3}}\right)}.
\end{equation}
By requiring the space-time to be asymptotically $AdS$ and to have a conformal boundary at infinity, we see that the tension (in units of $L_{_0}=1$) must be bounded:
\begin{equation}
{\kappa^{2} T_D} < 3.
\label{dof-bound}
\end{equation}
To understand the meaning of this bound in the dual field theory, we can reason by analogy with the case of strongly coupled 
$\CN$=4 SYM theory in $d=3+1$ dimensions, where the number of adjoint degrees of freedom is encoded in the entropy density $s\propto N_c^2$ and is related through the holographic dictionary to the (inverse of) Newton's constant $1/\kappa_{5}^2$. Adding fundamental degrees of freedom in the limit $N_c/N_f\ll 1$ corresponds to considering probe brane geometries \citep{Karch:2002sh}, with the source of charged fundamental degrees of freedom being proportional to the charge density, $\rho\propto N_f N_c$ and related to the tension of the brane $ T_D$. In units of $AdS$ radius, and after restoring powers of $T$ using dimensional analysis, this can be summarized as
\begin{equation}
s \propto\frac{T^3}{\kappa_{5}^2} \propto {N_c ^2 T^3}, \qquad \text{and} \qquad  \rho\propto  T_D T^3 \propto N_f N_c T^3.
\end{equation}
Comparing these two expressions, we see that the quantity
\begin{equation}
 \frac{\rho}{s} \propto \kappa_{5}^2  T_D \propto \frac{N_f}{N_c}
\end{equation}
controls the ratio of fundamental to adjoint degrees of freedom and makes it manifestly clear that in the probe limit, $\kappa_{5}^2  T_D\ll1$, we have $N_f \ll N_c$, or equivalently $\rho \ll s$. We can then recast the bound \eqref{dof-bound} on the brane tension
in the form of  a bound on the number of fundamental degrees of freedom $N_f$ in the (hypothetical) dual theory
\begin{equation}
 \frac{N_f}{N_c} \lesssim 3.
 \label{nfncb}
\end{equation}
Further insight can be obtained when we recall that gauge-gravity duality can be viewed as a geometrization of  Renormalization Group (RG) flow of the dual theory \citep{deBoer:1999tgo,Fukuma:2002sb}. The extra holographic coordinate, $r$, plays the role of the energy scale $\Lambda$ at which we probe the system, and the possible conformal fixed points along the RG flow are then mapped to locally $AdS$ spacetimes. In this view, the fact that we have an asymptotically $AdS$ boundary at $r\rightarrow\infty$, implies that we are dealing with a theory well-defined in the $UV$, whose beta function must be negative \citep{tHooft:1998qmr,Politzer:1973fx,Gross:1973id}
\begin{equation}
\beta\equiv\frac{\partial  g_{YM}^{}}{\partial \log{\Lambda}} \propto  g_{YM}^{3}(A_f N_f- A_c N_c)<0\,,
\end{equation}
where $A_{c}$ and $A_f$ are non-negative constants related to Casimir invariants.
By requiring $\beta<0$, we obtain a bound on $N_f/ N_c$,
\begin{equation}
 \frac{N_f}{N_c}\lesssim\frac{A_c}{A_f},
\end{equation}
similar to Eq.~\eqref{nfncb}. These arguments can be viewed as a heuristic explanation of the bound \eqref{dof-bound} on the brane tension.
%
%

As we will show in the following, in the $(2+1)$-dimensional case, taking into account the backreaction of  probe branes leads  to a finite contribution to the entropy density, whose form at low temperatures (i.e. at $T^2\ll\rho$) is given by 
$$
s \sim \rho/\alpha+ T^2/\kappa^2\,.
$$
The probe brane limit condition $\kappa^2  T_D\ll1$ ($s\gg\rho$) then implies
\begin{equation}\label{T_res}
\frac{\kappa^2}{\alpha}\ll \frac{T^2}{\rho} \ll 1\,.
\end{equation}
This condition breaks down explicitly at $T^2/\rho=0$ (for $\kappa^2  T_D\neq0$), reflecting the fact that the low temperature limit and the probe limit do not commute. At zero temperature, the near horizon geometry  in the exact probe limit $\kappa^2 T_D =0$  is given by $AdS_4$, and differs from the probe limit case $\kappa^2 T_D\ll1$, where we have instead an $AdS_2 \times \mathbb{R}^2$ geometry, similar to the one observed in the $AdS_{4}RN$ case \citep{Faulkner:2009wj}. 

To summarize, we have two interesting limiting cases  in our system: the  linear, or $AdS_4RN$, regime, and the  space-time filling probe brane regime. They are characterized by the following conditions
$$
AdS_4 RN~Limit: \qquad \alpha^2 F_{\mu\nu}F^{\mu\nu}\ll 1\,, \qquad \frac{\alpha^2 \kappa^2  T_D}{L^2}\sim1\,,
$$
$$
 Probe~Brane~Limit: \qquad \alpha^2 F_{\mu\nu}F^{\mu\nu}\sim 1\,, \qquad {\kappa^{2}T_D L_{_0}^{2}}\ll 1\,.
$$
%
%
%

In this paper, we explore the effects of non-linearities and finite back-reaction on correlation functions 
of the shear components of the energy-momentum tensor $T^{a b}$ and of a global U(1) current $J^{a}$ in a $CFT_3$ at finite temperature and density dual to the action \eqref{main-action}.
For momenta $q$ and frequency $\omega$ small compared to the temperature, $\omega,q\ll T$, we can use the hydrodynamic approximation \citep{LLfluids,Kovtun:2012rj}, which predicts \cite{Policastro:2002se,Son:2006em} that in a CFT at finite $T$ and $\mu$, the two-point functions of the appropriate\footnote{Transverse with respect to the direction of the spatial momentum.} components of the energy-momentum tensor and  the U(1) current exhibit a pole corresponding to a diffusion mode with the dispersion relation
\begin{equation}\label{Dhydro}
\omega=-i\frac{\eta}{\left(\epsilon+P\right)}q^{2}+ \CO(q^4),
\end{equation}
where $\epsilon$ is the energy density, $P$ is the pressure and $\eta$ is the shear viscosity of the dual field theory. This pole is common for the shear channel correlators $G_{T^{ab}T^{cd}}^{R}$,  as well as the correlators $G_{T^{ab}J^{c}}^{R}$ and $G_{J^aJ^{b}}^{R}$, where $J^a$ is the transverse component of the current, and should not be confused with the charge diffusion pole in correlators of the longitudinal components of the current (see ref.~\cite{Son:2006em} for details).

The shear viscosity controls the attenuation and, with the spatial momentum in $(2+1)$-dimensional field theory directed along the x-axis,  can be determined by the low frequency limit of the zero momentum retarded Green's function of $T^{xy}$ using the Kubo formula
\begin{equation}
\eta \equiv - \lim_{\omega \rightarrow0} \frac{1}{\omega} \text{Im}\left[G_{T^{xy}T^{xy}}^{R}(\omega, q=0)\right].
\end{equation}

One of the most celebrated achievements of holography, is the realization that for many strongly interacting holographic fluids, there is an universal value for $\eta / s$ \citep{Policastro:2001yc, Kovtun:2003wp, Kovtun:2005ev,Buchel:2003tz,Starinets:2008fb} given by 
\begin{equation}\label{eta/s}
\frac{\eta}{s} = \frac{\hbar}{4 \pi k_B}\approx 6.08 \times 10^{-13}~K \cdot s.
\end{equation}
The shear viscosity to entropy density ratio in strongly correlated quantum liquids such as ultracold quantum gases and the quark-gluon plasma generated at LHC and RHIC, has been determined to have a value remarkably close to the one predicted by holography \citep{Adams:2012th}, which can be viewed as an indication that the systems are strongly coupled. An interesting  feature of  holographic theories is that momentum and charge transport properties, which in weakly interacting systems are normally determined by the mean-free path in kinetic theory \citep{PhysRevB.56.8714}, in the infinite coupling limit are essentially determined by thermodynamics and equilibrium data (for example, the shear viscosity is determined in this limit by the entropy density, up to a coupling-independent constant) \citep{Kovtun:2008kx}. The universality of this statement can be understood
 in the framework of the membrane paradigm \citep{Iqbal:2008by}, once we realize that in the low frequency/long wavelength limit, the response of the boundary fluid is precisely determined by the horizon geometry of its dual black hole, and therefore by the entropy density of the dual field theory.

In section (\ref{RGFs}), we confirm this result for the $(2+1)$-dimensional  system under consideration by computing the retarded Green's functions of both $T^{a b}$ and $J^a$ in the hydrodynamic limit and observe that the structure of their singularities only depends on thermodynamic data and agrees with the universal result (\ref{eta/s}).
In fact, as was already observed in \citep{Davison:2013bxa}, in the large chemical potential regime, $\mu \gg \omega$, the quadratic part of the diffusive mode $\mu\CD$ is described by the hydrodynamic prediction even at $T\lesssim \omega\ll\mu$, outside of the hydrodynamic regime in which it was first derived, $\omega\ll T$. We  also confirm this result by numerically computing the charge diffusion mode and fitting it to a quartic polynomial of the form $\omega= -i\CD q^2 -i\Delta q^4$, in order to  extract $\CD$ and compare it to the hydrodynamic prediction.

The presence of the $AdS_2$ piece in the near horizon geometry at $T/\mu=0$, indicates the existence of light modes described by an effective $CFT_1$ or ``semi-local quantum liquid" \citep{Iqbal:2011in}, and therefore we expect the retarded correlators of both $T^{a b}$ and $J^a$ to have a continuous spectrum and exhibit branch cuts along the negative imaginary axis \citep{Edalati:2010hk,Brattan:2010pq,Denef:2009yy}. To avoid this regime and be consistent with (\ref{T_res}), we will always consider finite temperatures $T/\mu>0$ throughout this paper. Despite this restriction, we will still be able to observe the emergence of a 
branch cut in the form of an infinite set of purely imaginary poles (located below the diffusive mode) approaching the origin and becoming denser as we lower the temperature.

\vspace{0.25cm}
This paper is organized as follows. In section \ref{solution}, we review the charged spacetime-filling black brane solution and its thermodynamics. After computing the probe limit of the entropy density and showing the thermodynamic stability of the system, in section \ref{fluctuations}  we present the linearised equations of motion for the gauge-invariant fluctuations of the background, which we  use to compute the correlators in the hydrodynamic limit.  In section \ref{numerics}, we explain the details of the numerical method and explore numerically  the effects of the non-linearities on  quasi-normal modes in the mixed correlator for various values of momenta $q/\mu$ and temperature $T/\mu$. We conclude with summary, discussion and outlook in section  \ref{summary}.
\section{Charged Spacetime-Filling Black Brane Solution $(AdS_4 DBI)$}\label{solution}
\subsection{The equilibrium solution and stability}
The equations of motion derived from the action \eqref{main-action}  can be solved analytically \citep{Tarrio:2013tta,Pal:2012zn}. The equation of motion for the gauge field reads
\begin{equation}
\partial_{\mu}\left(\sqrt{-g}\left[\frac{F{}^{\mu\nu}-{\cal{P}}\alpha^2\sqrt{|\det{
 F_{~\nu}^{\mu}|}}\tilde{F}^{\mu\nu}}{\CL_{DBI}}\right]\right)=0\,,
\end{equation}
where the dual tensor is defined as
\begin{equation}
\tilde{F}^{\mu\nu}\equiv\frac{1}{2\sqrt{-g}}\epsilon^{\mu\nu\alpha\beta}F_{\alpha\beta}, \qquad \text{with} \qquad |\det F_{~\nu}^{\mu}| = \frac{1}{16} \left(\tilde{F}^{\mu\nu} F_{\mu\nu}\right)^2.
\end{equation}
The equation for the metric tensor is 
\begin{equation}\label{einst}
R_{\mu\nu}-\frac{1}{2}Rg_{\mu\nu}-\frac{3}{L_{_0}^2} g_{\mu\nu}=\kappa^2  T_D \left[\frac{\alpha^{2}F_{\mu\beta}F_{\nu}^{~\beta}-g_{\mu\nu}{\cal{P}}\alpha^4|\det F_{~\nu}^{\mu}|}{\CL_{DBI}}-g_{\mu\nu}\CL_{DBI}\right]\,.
\end{equation}
To solve this system, we assume an Ansatz consisting of a radially symmetric homogeneous  electric field, $F_{rt} = E(r)$, 
and  the metric 
\begin{equation}
ds^{2}=\frac{L^{2}}{r^{2}f(r)}dr^{2}-\frac{r^{2}}{L^{2}}f(r)dt^{2}+\frac{r^{2}}{L^{2}}d\vec{x}^{2}\,.
\end{equation}
The solution to the equations of motion is  then given by \citep{Pal:2012zn, Tarrio:2013tta, Kundu:2017cfj, Pal:2017hai, Fernando:2003tz, Dey:2004yt}
\begin{align}
E(r)&=\frac{r_H^{2}}{r^{2}}\frac{Q }{L^{2}}\frac{1}{\sqrt{1+\frac{r_H^{4}}{r^{4}}{\CS^2}}}
\equiv \frac{r_H^{2}}{r^{2}}\frac{Q }{L^{2}}\frac{1}{\CG[r]}\,,\\
f(r)&=1-M\frac{r_H^{3}}{r^{3}}+\frac{\kappa^{2} T_D L^{2}}{3}\left(1-{}_2F_1\left[-\frac{1}{2},-\frac{3}{4};\frac{1}{4};-\frac{r_H^{4}}{r^{4}} \CS ^2 \right]\right)\,,
\end{align}
where the parameters $M$ and  $r_H$ are uniquely defined by the largest root of $f(r_H)=0$. They are related to the entropy density $s$ and temperature $T$ of the system via
\begin{equation}\label{thermo1}
s=\frac{2\pi}{\kappa^{2}}\left(\frac{r_H}{L}\right)^{2},\qquad
T=\frac{3 r_H}{4\pi L^{2}}\left(1-\frac{\kappa^{2} T_D L^{2}}{3}\left[\sqrt{1+{\cal S}^{2}}-1\right]\right).
\end{equation}
The parameter $Q$ is related to the chemical potential $\mu$ and the charge density $\rho$:
\begin{align}
\mu&=\int_{r_H}^{\infty}E(r)dr=\frac{Q r_H}{L^{2}} 
{}_2F_1\left[\frac{1}{2},\frac{1}{4};
\frac{5}{4};-\CS ^2\right]
\equiv \frac{Q r_H}{L^{2}}\CF,\\
\rho&=\frac{1}{V}\frac{\delta S_{DBI}}{\delta E(r)}
=\alpha T_D\left(\frac{r_H}{L}\right)^{2}\CS
= N_D\frac{Q s}{\pi}
\end{align}
where in all the expressions we have used the dimensionless quantities
\begin{equation}
{\cal S}=\frac{\alpha Q}{L^{2}},\qquad N_D=\frac{\alpha^2 \kappa^2  T_D}{2 L^2}
\end{equation}
along with
\begin{equation}
\CF={}_2F_1\left[\frac{1}{2},\frac{1}{4};\frac{5}{4};-{\cal S}^{2}\right],\qquad \CG[r]=\sqrt{1+\frac{r_H^{4}}{r^{4}}{\cal S}^{2}}.
\end{equation}
The reason to define $\CS$ and $N_D$  is that in the limit $\CS\ll1$  with  $N_D\sim1 $,
we recover\footnote{It is useful to remember that ${}_2F_1[a,b;c;x]\sim 1+ \frac{ab}{c}x+ \CO(x^2)$ for small $x$.}
the $AdS_4 RN$ results studied in \citep{Edalati:2010hk, Davison:2013bxa}.
Using the renormalized on-shell action  \citep{Tarrio:2013tta}, $M$ is related to the boundary energy density $\epsilon = \la T^{tt}\ra$ and pressure $P = \la T^{xx}\ra =\la T^{yy}\ra$ of the dual $CFT_3$, with 
\begin{equation}
\epsilon=2P=\frac{r_H^{3}}{\kappa^{2}L^{4}}M,
\end{equation}
which is consistent with $\la T^a_a\ra=0$ as required by conformal invariance. Combining all the expressions above, we can furthermore show that the thermodynamic identity
\begin{equation}
\epsilon+P=Ts+\mu \rho
\end{equation}
is satisfied, which is a consistency check of our results and will also be  useful in later sections, when computing the temperature dependence of the charge diffusion constant.

Another important problem already studied in \citep{Pal:2017hai, Tarrio:2013tta, Pal:2012zn}, is the thermodynamic stability of the solution. For this purpose, we will need to compute the specific heat at fixed charge, $c_{\rho}$, and fixed chemical 
potential, $c_{\mu}$, together with the charge susceptibility $\chi$ of the system. They are given by

\begin{align}
c_{\rho}&=T\left(\frac{\partial s}{\partial T}\right)_{\rho}= 2s\left(1+\frac{\kappa^{2} T_D
r_H}{2\pi T}\frac{{\cal S}^{2}}{\sqrt{1+{\cal S}^{2}}}\right)^{-1}\geq0\,,\\
c_{\mu}&=T\left(\frac{\partial s}{\partial T}\right)_{\mu}= 2s\left(1+\frac{\kappa^{2} T_D
r_H}{2\pi T}\left(\frac{\CF{\cal S}^{2}}{1+\CF\sqrt{1+{\cal S}^{2}}}\right)\right)^{-1}\geq0\,,\\
\chi&=\left(\frac{\partial\rho}{\partial\mu}\right)_{T}=
\frac{2 \rho }{\mu} { \left(1+ \frac{1}{\CF\sqrt{1+\CS^2}} \left(1+\frac{\kappa^{2} T_D
r_H}{2\pi T}\frac{{\cal S}^{2}}{\sqrt{1+{\cal S}^{2}}}\right)^{-1}\right)}^{-1}\ge0.
\end{align}
These expressions are positive and free of singularities. The two specific heats being non-negative, together with the positiveness of the susceptibility, indicate that the system is compressible and thermodynamically stable, in agreement with \citep{Tarrio:2013tta}, as long as we consider $ T_D\geq0$ \footnote{The thermodynamics with $ T_D<0$ has been studied in ref.~\citep{Pal:2017hai}.}.

\subsection{Thermodynamics in the low temperature probe limit}
\label{probe}
For the probe limit analysis, it is convenient to use a different metric where we rescale $f_{_0}(r)/L_{_0}^2=f(r)/L^2$ and the volume $V_{_0}/L_{_0}^2=V/L^2$ such that the new line element becomes
\begin{equation}
ds^{2}=\frac{L_{_0}^{2}}{r^{2}f_{_0}(r)}dr^{2}-\frac{r^{2}}{L_{_0}^{2}}f_{_0}(r)dt^{2}+\frac{r^{2}}{L_{_0}^{2}}d\vec{x}_{_0}^{2}.
\end{equation}
In analogy with the previous case, we have
\begin{equation}
T=\frac{3 r_H}{4\pi L_{_0}^{2}}\left(1-{\frac{\kappa^{2} T_D L_{_0}^{2}}{3}}\sqrt{1+{\cal S}^{2}}\right),
\end{equation}
and the densities such as $\rho$ and $s$ have an extra factor of $L^{2}/L_{_0}^{2}$ coming from the volume
\begin{equation}
\rho=\alpha  T_D \left(\frac{r_H}{L_{_0}}\right)^{2}{\cal S},\qquad s=\frac{2\pi}{\kappa^{2}}\left(\frac{r_H}{L_{_0}}\right)^{2}
\end{equation}
We begin with the case where we first take the probe limit ${\kappa^{2} T_D}\ll1$ with $\rho/s\sim\kappa^2 T_D$ ($\CS\sim1$), and then we will expand at low temperatures $\rho/s\gg\kappa^2 T_D$ ($\CS\gg1)$. Finally, we will substitute the exact probe results for $\CS_{_0}$ at ${\kappa^{2} T_D=0}$. This discontinuous limit is necessary to account for the discontinuous transition in the the near horizon geometries of the $\kappa^2 T_D =0$ case, which is given by $AdS_4$, and the probe limit $\kappa^2 T_D\ll1$ case, where we have instead an $AdS_2 \times \mathbb{R}^2$ geometry similar to $AdS_{4}RN$ \citep{Faulkner:2009wj}. The correct way of considering the probe limit is to set $\kappa^{2} T_D=0$ in Einstein's equations \eqref{einst} while considering the on-shell action to leading order in $\kappa^{2} T_D$.
By computing the first correction to $r_H$ for small back-reaction we obtain

\begin{equation}
r_H=r_{_0}\left(1+{\frac{\kappa^{2} T_D L_{_0}^{2}}{3}}\sqrt{1+{\cal S}^{2}}+{\cal O}\left({\kappa^{4} T_D^2}\right)\right)\,.
\end{equation}
Substituting this result  into the entropy density formula gives
\begin{equation}
s=s_{_0}\left(1+\frac{2 \kappa^{2} T_D L_{_0}^{2}}{3}\sqrt{1+{\cal S}^{2}}+{\cal O}\left({\kappa^{4} T_D^2}\right)\right)\,.
\end{equation}
In the exact probe limit, we know that ${\cal S}_{_0}$ and the entropy $s_{_0}$ are related to the temperature  via the black brane horizon radius $r_{_0}$ defined as
\begin{equation}
r_{_0}=\frac{4\pi L_{_0}^{2}T}{3},\qquad s_{_0}=\frac{2\pi}{\kappa^{2}}\left(\frac{r_{_0}}{L_{_0}}\right)^{2}, \qquad {\cal S}_{_0}=\frac{2\pi \rho_{_0}}{\alpha{\kappa^{2} T_D}s_{_0}}.
\end{equation}
%
%
Substituting this into the expression for $s$ with $\CS\to\CS_{_0}\gg1$, we obtain
\begin{equation}\label{s0eq}
s=\frac{2\pi L_{_0}^{2}}{\kappa^{2}}\left(\frac{4\pi}{3}\right)^{2} T^2+\frac{4\pi }{3}\frac{\rho_{_0}L_{_0}^{2}}{\alpha}+\frac{\alpha  T_D ^2 L_{_0}^{6}}{2\rho_{_0}}\left(\frac{4\pi }{3}\right)^{5}{T^{4}}+ \CO \left(T^8\right) 
\end{equation}
This expression agrees exactly with the one found in \citep{Karch:2008fa,Karch:2009zz}, after setting $L_{_0}=1$ and  making the substitution $\rho_{_0}\rightarrow { \CN }\alpha d$ and $ \alpha \rightarrow 2 \pi \alpha'$.

As already mentioned in the Introduction, the first term comes from the adjoint matter and is proportional to $1/\kappa^2 \propto N_c^2$ while the constant term is related to the charged matter and is proportional to $\rho_{_{0}} \propto N_c N_f $.  Notice also that we could have done the same calculation with $s$ and $\rho$, without taking care of the extra volume normalization in the probe limit. In that case (\ref{s0eq}) would have an extra charge-independent ``contact term" coming from the volume normalization, responsible for the term $\Omega_{c.t}$ neglected in \citep{Karch:2008fa,Karch:2009zz}.
%

On the other hand, if we had first considered the low temperature limit and then the probe limit $\kappa^2 T_D\ll1$, the result would have been different, and we would have had a similar situation to the one in the $AdS_4{RN}$ case, where we have finite entropy density at zero temperature
\begin{equation}
s= \frac{2\pi }{3}\frac{\rho_{_0}L_{_0}^{2}}{\alpha} +\frac{8 \pi ^2}{3}\sqrt{\frac{\rho_{_0}}{3 \alpha}}\frac{L_{_0}^2}{\kappa}T+ \CO (T^2).
\end{equation} 

Another important observaton is that the charge-dependent low-temperature contribution to the heat capacity, $c\equiv T\partial_{_T}(s)$, changes when we consider different order of limits. In the exact probe limit, we recover the result from probe branes \citep{Karch:2008fa,Karch:2009zz}, given by $c\sim T^4/\rho_{_0}$, while including backreaction we obtain instead $c\sim T\sqrt{\rho_{_0}}$. This linear behaviour in  temperature is consistent with having charged fermionic matter in the degenerate ground state, despite having no signal of a holographic Fermi surface in any of the correlators \citep{Hartnoll:2016apf, Davison:2013uha}, although the system does exhibit spectral weight at $\omega = 0$ up to a finite momenta, similar to a smeared Fermi-Dirac distribution \citep{Anantua:2012nj}.

We do not observe any dependence on the parameter $\CP$ in any of the thermodynamics relations derived in this section. This is expected since the thermodynamics is related to the on-shell renormalized action, and therefore can only depend on equilibrium data. However, as we will explore in the next section, once we allow fluctuations in the system, we can effectively ``source" the $\CP\det (F_{~\nu}^{\mu})$ term and therefore change the transport properties of the dual theory.

\section{Linearized Fluctuations and Retarded Green's Functions}\label{fluctuations}
\subsection{Gauge-invariant fluctuations}
Through the holographic dictionary, the U(1) conserved current $J^a$ and the conserved energy-momentum tensor $T^{ab}$ are  sourced by the boundary values of the fluctuations of the bulk gauge field $\delta A_\mu|_{r=\infty}$ and the background metric $\delta g_{\mu\nu}|_{r=\infty}$, respectively. Since we are interested in computing the correlators, we will start by solving the linearised E.o.M for the coupled fluctuations of the bulk metric and gauge field, given by a general perturbation of the form
\begin{equation}
g_{\mu\nu}\rightarrow g_{\mu\nu}+\delta g_{\mu\nu}=\bar{g}_{\mu\nu}(r)+h_{\mu\nu}(t,x,r)\,,
\end{equation}
\begin{equation}
A_{\mu}\rightarrow A_{\mu}+\delta A_{\mu}=\bar{A}_{\mu}(r)+a_{\mu}(t,x,r),
\end{equation}
where the spacetime indices of the fluctuations are raised and lowered using the unperturbed metric $\bar{g}_{\mu\nu}$.
After Fourier transforming the fluctuations along the filed theory coordinates $x^a=(t,x,y)$ and using the isotropy in the $x-y$ plane to fix the momentum propagation along the $x$-axis, $q^a=(\omega,q,0)$, we can assume the linear fluctuations to be of the following form  
\begin{equation}
h_{\mu\nu}(t,x,r)\sim e^{-i\omega t}e^{iqx}h_{\mu\nu}(r),\qquad a_{\mu}(t,x,r)\sim e^{-i\omega t}e^{iqx}a_{\mu}(r).
\end{equation}
Now that we have fixed the direction of spatial momentum, we observe that the fluctuations naturally decouple and split into two groups, according to their parity under $y\rightarrow-y$ transformation \citep{Herzog:2002fn,Edalati:2010hk}. 
In this paper, we will only consider  the group of the odd parity fluctuations, 
$$
h^{y}{}_{t}, \qquad h^{x}{}_{y}, \qquad a_{y}\,,
$$
responsible for the momentum
 diffusion mode in the dual $CFT_3$ (\ref{Dhydro}) \citep{Herzog:2002fn}. The even parity or sound modes, and their relation to the holographic zero sound \citep{Karch:2008fa,Karch:2009zz}, have been studied in detail in \cite{andy} and for the case of $AdS_4RN$ in \citep{Edalati:2010pn}.

It will be convenient to  introduce a dimensionless variable $u\equiv r/r_H$, so that the horizon is now at $u=1$, and work in the radial gauge 
\begin{equation}
 a_{u}=0,\qquad h_{u\mu}=0\,.
 \label{rad-gauge}
\end{equation}
Furthermore, it will be useful to define the dimensionless variables 
\begin{equation}
\bar{\omega}\equiv\frac{\omega}{\mu},\qquad\bar{q}\equiv\frac{q}{\mu},\qquad\bar{T}\equiv \frac{T}{\mu}.
\end{equation}
%
This choice of normalization is useful in the large chemical potential, low temperature regime, $\bar{q},\bar{\omega}\ll1$, $\bar{T}\ll1$, where the standard hydrodynamic normalization diverges:  $\omega/T=\bar{\omega}/\bar{T}\rightarrow\infty$. 

Using these notations, we obtain three second-order differential equations as well as one first-order constraint equation coming from the radial gauge fixing \footnote{The gauge fixing conditions \eqref{rad-gauge} must be implemented after the E.o.M have been computed for general fluctuations of the fields, otherwise we will miss the constraint equation (\ref{D}). }
\begin{equation}\label{A}
f\left[u^{4}h^{y}{}_{t}'\right]'+\frac{4Q^{2} f { {N_D} \CF}}{\mu} a_{y}'-Q^{2}\bar{q}{ \CF^{2}}\left[\bar{q}h^{y}{}_{t}+\bar{\omega}h^{x}{}_{y}\right]=0\,,
\end{equation}
\begin{equation}\label{B}
f\left[u^{4}fh^{x}{}_{y}'\right]'+Q^{2}\bar{\omega}{ \CF^{2}}\left[\bar{q}h^{y}{}_{t}+\bar{\omega}h^{x}{}_{y}\right]=0\,,
\end{equation}
\begin{equation}\label{C}
u^{2}f\left[u^{2}f{ \CG}a_{y}'\right]'+\frac{\mu u^{2}f}{{ \CF}}h^{y}{}_{t}'+Q^{2}{ \CF^{2} \CG}\left[\bar{\omega}^{2}-\bar{q}^{2}f\right]a_{y}+{\CP}\frac{\bar{q}^{2}Q^{2}{ \CF^{2}{ \CG'}}uf}{2}a_{y}=0\,,
\end{equation}
\begin{equation}\label{D}
\frac{4Q^{2}\bar{\omega}{ {N_D} \CF}}{\mu} a_{y}+u^{4}\bar{q}f h^{x}{}_{y}'+u^{4}\bar{\omega}h^{y}{}_{t}'=0\,,
\end{equation}
where primes denote derivatives with respect to the radial coordinate, and  $\CG=\sqrt{1+{{\cal S}^{2}}/{u^4}}.$
%
%
%
%

The above equations for the shear perturbations are not gauge-invariant or linearly independent, because choosing the radial gauge is not enough to fix all the gauge redundancy in the system. This problem can be solved by choosing to work with physical, gauge-invariant combination of the variables involved \citep{Kovtun:2005ev}. There are two types of gauge transformations that the system is subject to, $U(1)$ gauge transformations generated by $\lambda(r,x,t)$, and infinitesimal diffeomorphisms  generated by $\xi^{\mu}(r,x,t)$:
\begin{align}
h_{\mu\nu}\rightarrow\widetilde{h}_{\mu\nu}&=h_{\mu\nu}-\left(\bar{\nabla}_{\mu}\xi_{\nu}+\bar{\nabla}_{\nu}\xi_{\mu}\right)\,,\\
a_{\mu}\rightarrow\widetilde{a}_{\mu}&=a_{\mu}-\left(\xi^{\nu}\bar{\nabla}_{\nu}A_{\mu}+A_{\nu}\bar{\nabla}_{\mu}\xi^{\nu}+\bar{\nabla}_{\mu}\lambda\right),
\end{align} 
Here the covariant derivative $\bar{\nabla}_\mu$ is computed with  the background metric $\bar{g}_{\mu\nu}$. In momentum space, the effect on this gauge transformation on the shear perturbations is given by
\begin{equation}
\widetilde{a}_{y}(u)=a_{y}(u),\qquad \widetilde{h}^{y}{}_{t}(u)=h^{y}{}_{t}(u)+i\omega\xi^{y}(u), \qquad \widetilde{h}^{x}{}_{y}(u)=h^{x}{}_{y}(u)-iq\xi^{y}(u).
\end{equation}
This allows us to construct the following dimensionless gauge-invariant variables
\begin{equation}\label{GIXY}
X(u)=h^{y}{}_{t}(u)+\frac{\omega}{q}h^{x}{}_{y}(u), \qquad Y(u)=\mu^{-1}a_{y}(u),	
\end{equation}
which are physical and naturally encode the Ward–Takahashi identities obeyed by the retarded Green's functions of the underlying field theory, up to contact terms \citep{Herzog:2002fn, Kovtun:2005ev}
\begin{equation}
G_{T^{xy}T^{xy}}^{R}=\frac{\omega}{q}G_{T^{ty}T^{xy}}^{R}=\frac{\omega^{2}}{q^{2}}G_{T^{ty}T^{ty}}^{R},\qquad G_{T^{xy}J^{y}}^{R}=\frac{\omega}{q}G_{T^{ty}J^{y}}^{R}.
\end{equation}

The variables (\ref{GIXY}) are not the only gauge-invariant variables that can be constructed. Another common choice is  the Ishibashi-Kodama  ``master fields" \citep{Kodama:2003kk}, which include derivatives of the fluctuations in their definition, and have the advantage of producing decoupled E.o.M \citep{Edalati:2010hk}. However, the price to pay for this is that the Ward-Takahashi identities, as well as the relations between the sources of the $CFT_3$ and the boundary values of the ``master fields" are no longer transparent, and the holographic dictionary becomes cumbersome. For this reason, we have chosen to work with the variables defined in  Eq.~(\ref{GIXY}).

Using the $X$, $Y$ variables, we can construct two coupled, gauge-invariant, linearly independent equations. To obtain the first one, we combine (\ref{A}), (\ref{B}) and (\ref{C}), and for the second one, we use (\ref{B}) and (\ref{D}). This leaves us with the following set of equations
%
\begin{eqnarray}
\left[f\left(u^{4}X' + { \CF \CG}u^{6}f'Y'\right)\right]' &+& \frac{{ \CF^{2}}Q^{2}\left[\bar{\omega}^{2}
-\bar{q}^{2}f\right]}{f}\left[X+f'u^{2}{ \CF \CG}Y\right]  \nonumber \\
&+& {\CP}\frac{\bar{q}^{2}{ \CF^{3}{ \CG'}}f'u^{3}Q^{2}}{2}Y=0\,,\label{GI 1}
\end{eqnarray}
\begin{equation}\label{GI 2}
\left[\frac{f\left(u^{4}X'+{ \CF{N_D}}4Q^{2}Y\right)}{\bar{\omega}^{2}-\bar{q}^{2}f}\right]'+\frac{{ \CF^{2}}Q^{2}}{f}X=0,
\end{equation}
where we used  the  identity
\begin{equation}\label{f'}
\left[u^{4}f'(u)\right]'=\frac{4Q^{2}}{u^{2}}{ \frac{{N_D}}{\CG[u]}}.
\end{equation}

The above equations decouple in two limits: the zero-momentum limit $\bar{q}\rightarrow0$, when all the fluctuation channels become the same \citep{Kovtun:2005ev}, and in the probe limit ${N_D}\rightarrow0$, when the gauge and metric fluctuations decouple. It is also worth noticing that both (\ref{GI 1}) and (\ref{C}) depend explicitly on the parameter ${\cal P}$, which will modify the dispersion relation of the charge diffusion mode
\begin{equation}
\omega(q)=-i{\cal D} q^2-i\Delta({\cal P}) q^4+{\cal O}(q^6).
\end{equation}
Notice that the quadratic part $\CD=\eta/\left(\epsilon+P\right)$ is completely determined by thermodynamic quantities through the $\eta=s/4 \pi$ relation, and therefore cannot depend on $\CP$ since the thermodynamics is completely independent of it. We confirm numerically that the first correction in $\CP$ comes at a quartic order proportional to $\Delta(\CP)q^4$. 

Here we have a clear example in which transport properties can be used to distinguish two thermodynamically identical but microscopically different systems, by looking at their quasi-normal modes\footnote{I would like to thank A. Lucas for this comment.}.
%
We expect this to be the case in higher dimensions as well as long as we have a term of the form $\det (F_{~\nu}^{\mu}) \propto(\vec{B}\cdot\vec{E})^2$ which could be sourced by ``induction" from the shear fluctuations $\delta E\rightarrow\delta B$ or any other term that can source the fluctuations and is zero on-shell. As we already mentioned, this phenomenon is exclusive to the shear channel, and we can safely set the parameter $\CP=0$ in the action when we consider the longitudinal channel, studied in \cite{andy}, before computing the E.o.M. for the fluctuations.


\subsection{Retarded Green's function in the hydrodynamic limit }\label{RGFs}
%
We are finally ready to compute the retarded Green's functions of both $T^{a b}$ and $J^a$ operators of the dual $CFT_3$, 
by using the boundary renormalized on-shell action to quadratic order in the fluctuations, which will act as a generating function for the correlators. The only counterterm we need to include is the vacuum Gibbons-Hawking boundary term, which is consistent with the field theory statement that the vacuum counterterms suffice for renormalization at non-zero $T$ and $\mu$ \citep{Kapusta:2006pm}. Ignoring contact terms, the final result for the boundary action at  $u \rightarrow \infty$ can be written as
\begin{equation}
S^{(2)} =\frac{r_{H}}{2\kappa^{2}}\int_{u\rightarrow\infty}\frac{d\omega dq}{(2\pi)^{2}}\left[-\frac{r_{H}^{2}q^{2}u^{4}f}{2L^{4}\left(\omega^{2}-q^{2}f\right)}X(-\omega)X'(\omega)-2\mu^{2}u^{2}f{ \CG N_{D}}Y(-\omega)Y'(\omega)\right]\,,
\nonumber
\end{equation}
where we omitted the $q$-dependence in the fields $X$ and $Y$ to simplify notation. From the E.o.M., we know that the 
near-boundary expansion of the field is given by
\begin{equation}\label{NBexp}
X\sim X^{(0)}+\frac{\Pi_{X}}{u^{3}}+...,\qquad Y\sim Y^{(0)}+\frac{\Pi_{Y}}{u}+...,
\end{equation}
which can be used to express the renormalized on-shell action to second order in perturbations (up to contact terms) as
\begin{equation}\label{S2}
S^{(2)}=\frac{r_{H}}{2\kappa^{2}}\int_{u\rightarrow\infty}\frac{d\omega dq}{(2\pi)^{2}}\left[\frac{3r_{H}^{2}q^{2}}{2L^{4}\left(\omega^{2}-q^{2}\right)}X^{(0)}(-\omega)\Pi_{X}(\omega)+2\mu^{2}{ \CG N_{D}}Y^{(0)}(-\omega)\Pi_{Y}(\omega)\right]\,.
\nonumber
\end{equation}
Using the standard holographic prescription \citep{Son:2002sd,Iqbal:2008by}, we can determine the dependence of the generalized 
momenta $\Pi_{X,Y}$ on the boundary values $X^{(0)}, Y^{(0)}$.  In general, we can only do this analytically 
in some very special cases, where we can solve exactly the E.o.M. for the fluctuations with ingoing boundary conditions 
and arbitrary parameters $\omega, q$. In our case, however, we can only determine the  retarded Green's functions in the hydrodynamic limit $\omega,q\ll T$. In order to do so, we will be closely following the ``matching procedure" outlined in ref.~\citep{Davison:2013bxa}, and summarized in the Appendix \ref{app-matching}.
%
%

In the hydrodynamic limit, we obtain the following correlators:
\begin{equation}\label{GFS}
G_{T^{ty}T^{ty}}^{R}=\frac{r_{H}^{2}q^{2}}{2\kappa^{2}L^{2}\left(i\omega-{\cal D}q^{2}\right)},\qquad G_{J^{y}J^{y}}^{R}=\frac{8L^{2}Q^{2}q^{2}{ N_{D}^{2}}}{9\kappa^{2}M^{2}\left(i\omega-{\cal D}q^{2}\right)}\,,
\end{equation}
\begin{equation}\label{GFA}
G_{J^{y}T^{ty}}^{R}	=G_{T^{ty}J^{y}}^{R}=\frac{2r_{H}Qq^{2}{ N_{D}}}{3\kappa^{2}M\left(i\omega-{\cal D}q^{2}\right)}\,.
\end{equation}
These expressions  are consistent with the ones obtained in ref.~\citep{Davison:2013bxa}, for $N_D=1$ and $\CS=0$. We observe that in the hydrodynamic limit, all the correlators have the same pole at
\begin{equation}
\omega(q)=-i{\cal D} q^2\equiv-i\frac{Q{ \CF}}{3M \mu} q^{2}=-i\frac{s}{4\pi\left(\epsilon+P\right)}q^{2}\,,
\end{equation}
where he have used the results from section 2. A comparison with the hydrodynamic prediction \eqref{Dhydro} shows that for the dual field theory at strong coupling, ${\eta}/{s}={1}/{4\pi}$ (in natural units $\hbar=k_B=1$), as expected. 
%
%

This raises the question of weather one could reverse the logic to find universal coefficients apart from the ones already known in a given holographic system. For example, if one could use a parameter such as $\CP$ to deform the original Lagrangian by a term that vanishes on-shell and sources the E.o.M. for the fluctuations (without being a boundary term and changing the holographic dictionary), then one could argue that the transport coefficients that are dependent on $\CP$ cannot be universal. However, 
if we could find some higher order transport coefficient (or a combination of them) that are $\CP$-independent, we
 could argue that they are possibly universal. As long as one can find a good candidate for the extra term, one could use this trick to identify either numerically or analytically, possible new universal relations in higher order hydrodynamics, such as  the ones proposed in \citep{Erdmenger:2008rm,Haack:2008xx,Grozdanov:2016fkt}.

In the next section, we will perform a numerical study of the full correlators beyond the hydrodynamic approximation, and determine, among other things, the effect that the non-linearities $\CS$ and $\CP$, finite momenta $\bar{q}$ and the back-reaction $N_D$ have on the quasi-normal modes (QNM) of the system.

\section{Numerical Method and Results}\label{numerics}
\subsection{Leaver's matrix method}
In this section, we will study the correlation functions and quasi-normal modes numerically, using the ``matrix method" also known as Leaver's method \citep{Leaver:1990zz}. It will be more convenient  to use the variable $z=1/u$ in order to make the range of numerical integration finite, $z\in [0,1]$. Computing retarded correlators corresponds to  choosing the  ingoing boundary conditions at the horizon, $X,Y\sim(1-z)^{-{i \omega}/{4 \pi  T }}$, and finding their poles means we have to impose   Dirichlet
 conditions at the boundary $z=0$, $X,Y\sim0$ \cite{Son:2002sd}. According to Eq.~(\ref{NBexp}), this implies
  $X/z^3\sim\Pi_{X}$ and $Y/z\sim \Pi_{Y}$. 
%
%
Correspondingly, it is convenient to redefine the fields as
\begin{equation}
Y(z) = f(z)^{-{i \bar{\omega}}/{4 \pi  \bar{T} }} z \phi (z), \qquad X(z)=f(z)^{-{i \bar{\omega} }/{4 \pi  \bar{T} }} z^3 \psi (z)\,,
\end{equation}
where  the regular part of the fields $\phi(z), \psi(z)$ goes to a constant in both limits\footnote{We have used $f(z)^{-{i \bar{\omega} }/{4 \pi  \bar{T} }}$ instead of $(1-z)^{-{i {\omega} }/{4 \pi  T }}$ since the two expressions are equivalent near the horizon (for $\bar{T}>0$), and we observe that this choice of normalization speeds up the convergence of the numerics.}. We then expand the regular parts in Taylor series around the middle-point $z_m=1/2$ between the two singularities:
\begin{equation}
\phi(z)=\sum_{n=0}^{N_{max}}a_{n}\left(z-z_{m}\right)^{n},\qquad \psi(z)=\sum_{n=0}^{N_{max}}b_{n}\left(z-z_{m}\right)^{n}\,.
\end{equation}
 This guarantees that the series has a radius of convergence of $|z|\leq1/2$, defined by the nearest singularities in the complex plane, and that it will be able to probe both the horizon and the boundary simultaneously. Introducing this Ansatz in the coupled equations  (\ref{GI 1}), (\ref{GI 2}) and expanding  around $z_m$, we obtain a set of $2(N_{max}+1)$ algebraic equations relating the coefficients $a_n$ and $b_n$, which can be put in a matrix form
\begin{equation}
M_{ab}C_b=0,
\end{equation}
where $C_b$ is a vector with the coefficients $a_n$ and $b_n$, and $M_{ab}$ is a $2(N_{max}+1)$-dimensional square matrix relating the coefficients, whose elements are functions of $\bar{\omega}, \bar{q}$ and $\bar{T}$.
We can find  the complex QNMs of the system by demanding that a non-trivial solution to this set of equations exists, which then requires
\begin{equation}
\det M(\bar{\omega}_{_{QNM}})=0.
\end{equation}
The accuracy in the position of the QNM, $\bar{\omega}_{_{QNM}}$, increases as we increase the number of terms $N_{max}$ in the series. A crucial step in constructing the vector $C_b$ and the corresponding  matrix $M_{ab}$ involves inverting the natural order in the coefficients and grouping them together in the following way
\begin{equation}
C_b=(b_{N_{max}},a_{N_{max}},...,b_{_0},a_{_0})^T\,,
\end{equation}
instead of just using the order $C_b= (a_0,...,a_{N_{max}},b_0,...,b_{N_{max}})^T$. This is done with the purpose of 
drastically speeding up the computation of the exact determinant, by making $M_{ab}$ a sparse matrix. 

We proceed by fixing all the desired parameters, except $\bar{\omega}$, to be rational numbers, which will then generate a finite degree rational polynomial in $\bar{\omega}$, given by $\det{M(\bar{\omega})}$. Finally, we solve numerically for all the complex roots of the polynomial with rational coefficients.
This gives as many solutions as the order of $\det{M(\bar{\omega})}$, and therefore we also find some spurious modes located  on a semi-circle centred at $\bar{\omega}=0$ in the complex $\bar{\omega}$ plane. Fortunately, as we increase the degree of the polynomial and the  number of zeroes, the radius of this ``convergence region" becomes larger and the positions of the physical QNMs converge to their actual values, as long as they are well inside this region. Thus the spurious solutions can be safely discarded. This is the case in all the numerics obtained in the current and subsequent sections.

It is also worth noting that we are finding the common QNMs of the coupled fields mixing the components of the metric and gauge field excitations. This means that the correlators of the relevant components of $T^{ab}$ and $J^{a}$ are generically non-trivial, 
and they share the same poles. In showing the QNM results below, we will not specify which correlator we 
have in mind specifically, unless stated otherwise\footnote{The expressions for the decoupled operators  in terms of the dual  Kodama-Ishibashi ``master fields"  can be found in \citep{Edalati:2010hk}, for the case of $AdS_4 RN$.}.

%
%

\subsection{From $AdS_4RN$ to $AdS_4 DBI$ at very high temperatures  $(T \gg \mu)$}
To exemplify the numerical procedure described above, we will start by exploring the effect that the non-linearities, controlled 
by the parameters $\CS$ and $\CP$, have on the  system at very large temperatures and zero momenta. We set the parameters to be
\begin{equation}
0.001\leq {\cal S}\leq1,\quad \CP=1,\quad N_D=1, \quad q/\mu=0,\quad T/\mu = 100\,,
\end{equation}
and follow the behaviour of the QNM spectrum in the complex plane of the dimensionless frequency,  $\omega/2 \pi T$, as we change $\CS$ from $(\CS=0$ or $AdS_4 RN)$ up to $(\CS=1$ or $AdS_4 DBI)$.
\begin{figure}[tt!]
\centering
\includegraphics[width=0.8\textwidth]{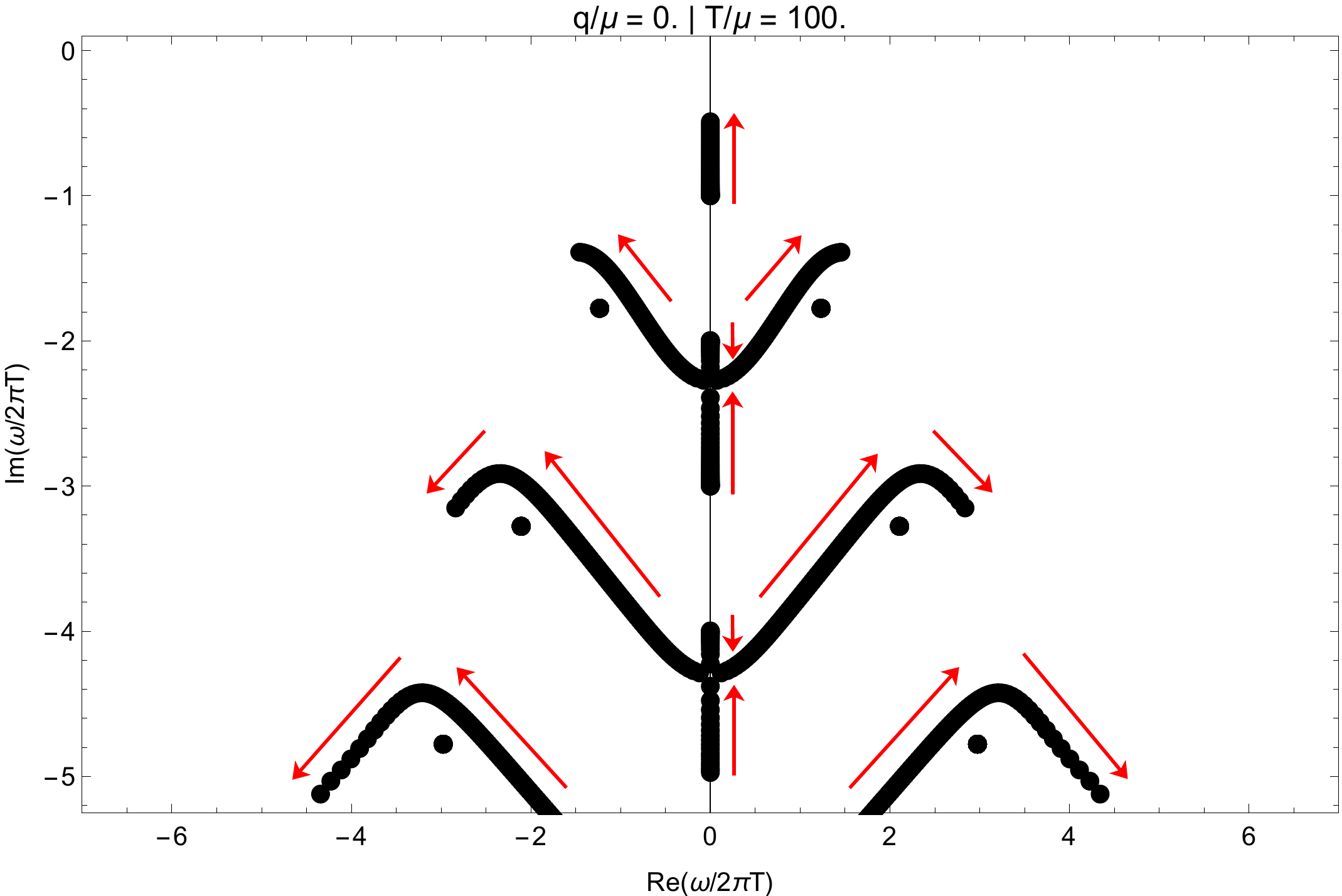}
\caption{\label{F1}Movement of the QNMs  in the complex $\omega/2\pi T$ plane, as we change $\CS$ from $(\CS=0,~N_D=1,~AdS_4RN)$ to $(\CS=1,~\CP=1,~N_D=1,~AdS_4DBI)$, with the rest of the parameters fixed at $T/\mu= 100$ and $~q/\mu=0$. At $\CS=0$, we have the poles on the imaginary axis situated at  $\omega_n = -i 2\pi n T $ with $n=1,2,...$, along with the top part of the ``Christmas tree", represented by the 6 dots in the figure. As we continuously increase ${\cal S}$, the poles on the imaginary axis with $n=5,4$, followed by $n=3,2$ start ``unzipping" by merging in pairs and then attaining a non-zero real part, after which, they go around the static poles, corresponding to the underlying $AdS_4 SC$  geometry that dominates the spectrum of fluctuations at $T/\mu\gg1$. The only exception is the $n=1$ pole, which steadily moves up but has no other pair to merge with and eventually saturates at $\omega=-i\pi T$. (Animated
version of the figure is available on this paper's arXiv page.)}
\end{figure}
At these very high temperatures, the spectrum is dominated by the effects of the underlying AdS-Schwarzschild ($AdS_4 CS$) geometry, whose QNMs form the infinite sequence colloquially known as the ``Christmas tree". We expect the residue 
of the moving modes coming from the charged matter to be suppressed by powers of $\mu/T \ll1$, and therefore negligible at such high temperatures. The results from this smooth interpolation between the two systems is shown in Fig.~\ref{F1}, with red arrows indicated the direction of movement as we increase $\CS$.

Because we are strictly at $q/\mu=0$, none of the shown QNMs is hydrodynamic in nature, since by definition, hydrodynamic 
modes must be gapless. Notice that the effect of non-linearities in the charged matter, can transform purely dissipative modes into propagating modes (the ones with non-vanishing real and imaginary parts).  Fig.~\ref{F1} is the only plot we show
 using the dimensionless $\omega/2\pi T$ normalization, since in the rest of the section we will be interested in exploring 
 the large chemical potential regime, given by $T/\mu \ll1$ and $\omega,q \ll\mu$.

%
\subsection{Diffusion mode at large chemical potential $(T,\omega\ll \mu) $}
\begin{figure}[!tt]
\centering
\includegraphics[width=0.47\textwidth]{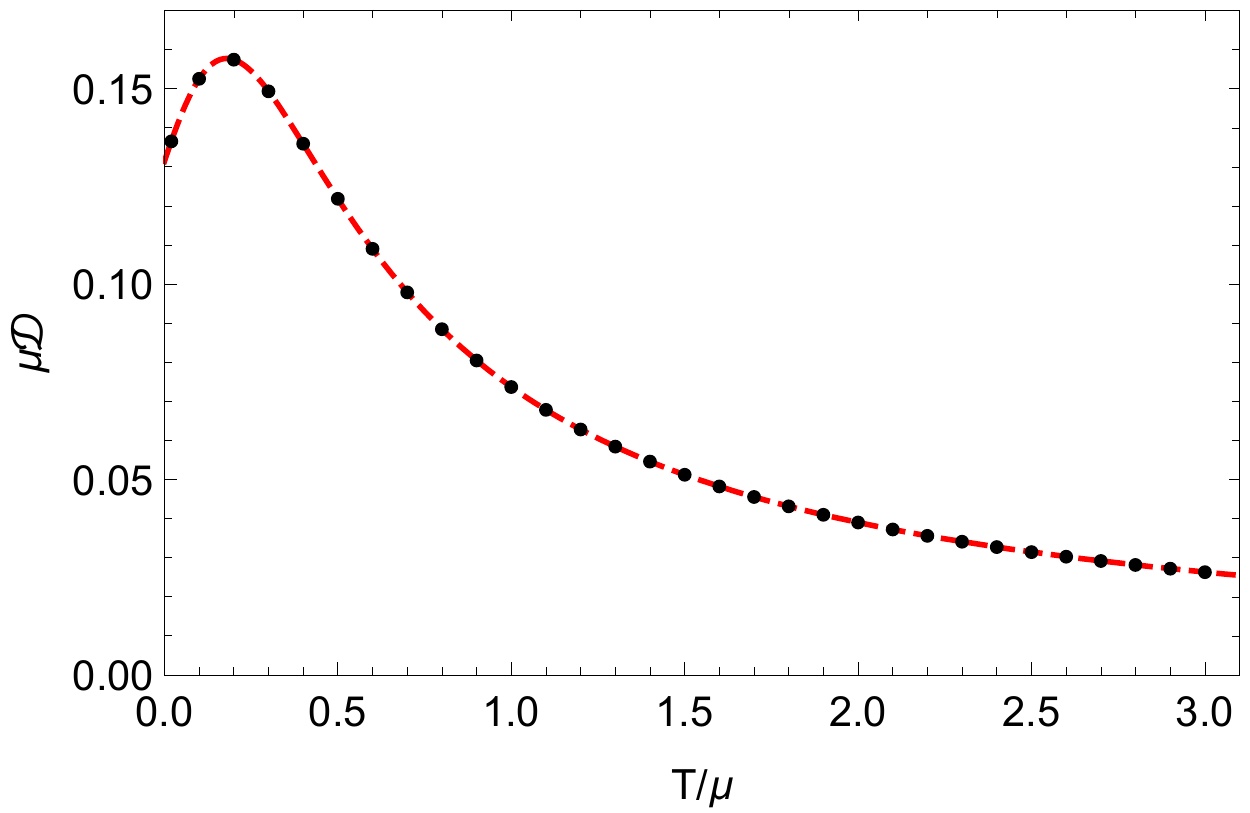}
\includegraphics[width=0.51\textwidth]{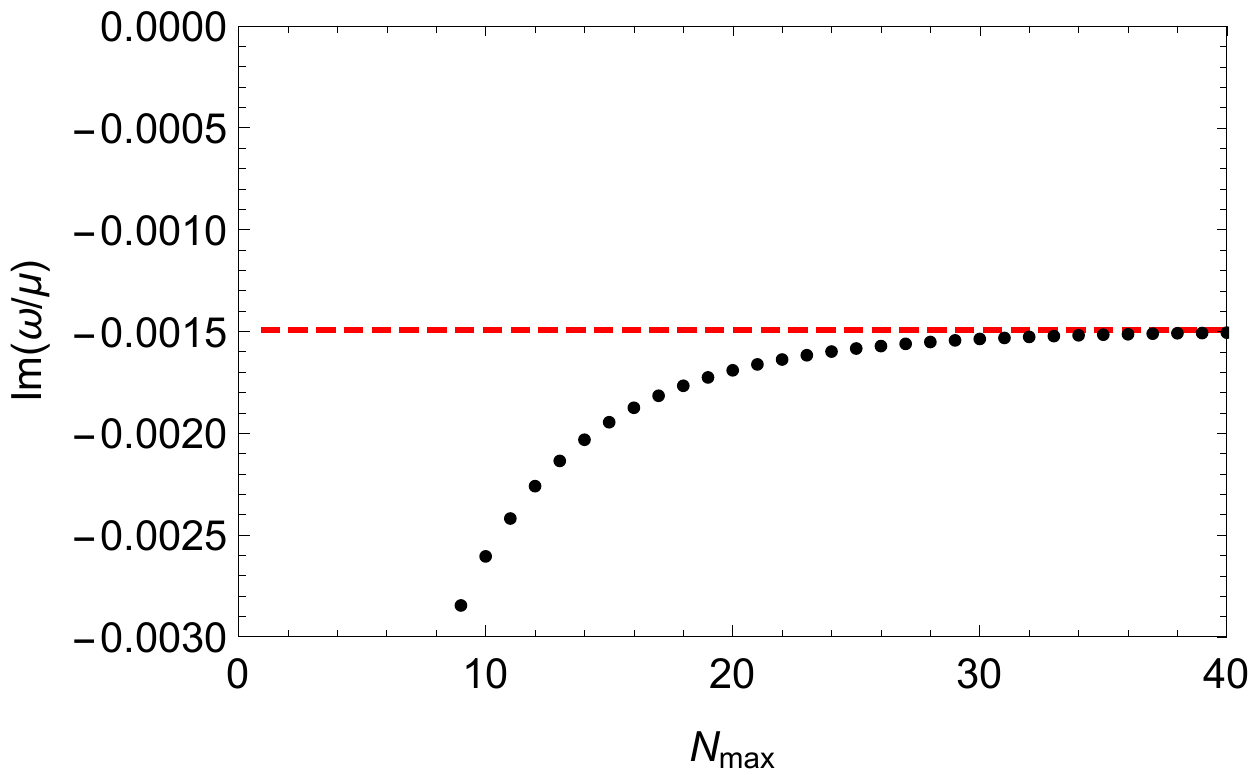}
\caption{\label{F2} The dimensionless diffusion constant $\mu {\cal D}$, determined numerically (black dots), for the case $(\CS=1,~\CP=1,~N_D=1,~AdS_4DBI)$, as a function of normalized temperature $T/\mu$. The red dashed line shows 
the hydrodynamic approximation  (\ref{Dmu}). The numerical values  are extracted by fitting $\omega(q)=-i{\cal D} q^2-i\Delta({\cal P}) q^4$ at $q/\mu  = (0.05, 0.1, 0.15, 0.2, 0.25)$ for different temperatures. For the results shown in the left panel, 
we used Leaver's method truncated at order $N_{max}=30$, except for the small temperature case $T/\mu=0.02$, where we used $N_{max}=40$. An example of the convergence for this point at $q/\mu=0.1$ is shown in the right panel. We observe similar results regardless of the values of $\CS$ and $\CP$.}
\label{fig: MUD}
\end{figure}
In this section, we will explore the effect of non-linearities on the diffusion mode at low temperatures and momenta in the case of $AdS_4 DBI$, where we consider
\begin{equation}
{\cal S}=1,\quad \CP=1,\quad N_D=1,\quad 0.05\le{q/\mu\le0.25,\quad 0.02\le T/\mu\le 3}.
\end{equation}
As discussed in the Introduction, in the hydrodynamic limit $\omega/T \ll1 $, the  momentum diffusion mode has the following dispersion relation
\begin{equation}
\omega(q)=-i{\cal D} q^2+{\cal O}(q^4),
\label{dif-m}
\end{equation}
where the (dimensionless) diffusion constant $\mu \CD$ in theories with non-vanishing chamical potential $\mu$ is  determined by 
\begin{equation}\label{Dmu}
{\mu\CD}=\frac{\mu\eta}{\left(\epsilon+P\right)}=\frac{\mu s}{4\pi\left(\epsilon+P\right)}=\frac{1}{4\pi\left(\frac{T}{\mu}+\frac{{N_D}Q}{\pi}\right)}.
\end{equation}
In the above expression, we have used fact that $\eta / s =1/ 4\pi$ in the (dual) Einstein gravity limit, and  $\epsilon +P=Ts+\mu \rho$.

One may ask a natural question about the applicability regime of the hydrodynamic approximation  (\ref{dif-m})-(\ref{Dmu}). This question was already answered for the case of $AdS_4RN$ in ref.~\citep{Davison:2013bxa}. Surprisingly, in that case, the hydrodynamic approximation turns out to be valid all the way down to zero temperature. The authors of ref.~\citep{Davison:2013bxa} attribute this property to the fact that one has an $AdS_2$ piece, dual to a $CFT_1$, in the low temperature near-horizon geometry. Since $AdS_4 DBI$ has a similar behaviour at low temperature, we expect this to be true also in our system, despite the non-linearities.

To extract the  coefficient $\mu \CD$ from the numerics, we make a quartic polynomial fit of the form\footnote{The reason to only consider even powers of $q$ is that  a diffusion process should be independent of the direction of propagation, i.e invariant under $q\rightarrow-q$.}
\begin{equation}
\omega(q)=-i{\cal D} q^2-i\Delta({\cal P}) q^4
\end{equation}
for a discrete set of small momenta between $0.05\le q/\mu\le0.25$  and for several values of the normalized temperature in the range $0.02\le T/\mu\le3$. We can compare the numerical fit with the hydrodynamics expression for $\mu\CD$ as given in (\ref{Dmu}). The comparison between the numerics and the hydrodynamic prediction is shown in the left panel of Fig.~\ref{F2}.

Notice that if we had instead done the fit using just $\omega=-i {\cal D}{q}^2$, we would have gotten a  result similar 
to the one obtained in ref.~\citep{Edalati:2010hk}, namely that  $\mu\CD$ is a monotonically increasing function at low temperatures, with no  maximum; thus, the extra quartic part in $q^4$ cannot be neglected at small temperatures. We also observe that the deviation from hydrodynamics, proportional to $\Delta(\CP)q^4$, gets smaller 
with increasing $\CP$, and therefore depends on the microscopic data. As an example, in the  $T/\mu=0.159$ case, we find that $\mu^3\Delta(\CP=0)\approx 0.02>\mu^3\Delta(\CP=1)\approx 0.01$. Here we provide a whole family of solutions with $AdS_2$ near horizon geometries, in which the hydrodynamic result of $\mu\CD$, derived in the $T \gg \omega$ limit, is still valid even for $T \lesssim \omega$, as long as we are in the large charge density limit $\omega, T \ll\mu$. This confirms and extends
 the results of ref.~\citep{Davison:2013bxa}.

The temperature $\bar{T}_{m}$ for which the maximum in $\mu\CD$ occurs (see Fig.~\ref{F2}) provides
 a convenient measure for the separation of scales between the high temperature hydrodynamic regime ($T/\mu \gg \bar{T}_{m}$), and the low temperature regime ($T/\mu \ll \bar{T}_{m}$). The position of the maximum 
  can be determined analytically using (\ref{Dmu}). To find the maximum, we first need to invert (\ref{thermo1}) to 
  determine $Q(\bar{T})$
\begin{eqnarray}
\label{QRN}
Q N_D &=& \frac{\sqrt{\frac{6 N_D}{\CS^2}\left[\sqrt{1+{\cal S}^{2}}-1\right] + (2\pi \bar{T} \CF)^2}-2\pi \bar{T} \CF}{\frac{2 }{\CS^2}\left[\sqrt{1+{\CS}^{2}}-1\right]} \nonumber 
\\
&=&
 {\sqrt{3 N_D + (2\pi \bar{T})^2}-2\pi \bar{T}}+\CO(\CS^2)\,,
\end{eqnarray}
and then require $\partial_{\bar{T}}\CD(\bar{T}_{m})\propto (\pi+N_D Q'(\bar{T}_{m}))=0$. If we use the linear approximation assuming
 $\CS\ll1$, we find the position and the height of the maximum to be, respectively
\begin{eqnarray}
\label{tauM}
\bar{T}_{m} &\approx&	\frac{\sqrt{N_D}}{2\pi}\left(1+\frac{7 \CS^2}{40}\right)+\CO(\CS^4)\,, \\
{\mu\CD(\bar{T}_{m})} &\approx& \frac{1}{6\sqrt{{N_D}}}\left(1-\frac{3 \CS^2}{40}\right)+\CO(\CS^4)\,. \label{tauM-1}
\end{eqnarray}
Comparing these results at $N_D=1$ with the  results shown in Fig.~\ref{F2}, we observe that the relative error  in both quantities is around 3\% for $\CS=1$ and around 6\% for $\CS=0$. In this respect, the non-linearities do not have a big impact on the value of 
 $\bar{T}_{m}$, while the back-reaction will be the dominant effect. In particular, as $N_D\rightarrow0$, we observe that the position of the maximum in Fig.~\ref{F2} shifts to the left, and its height increases (this is also evident from Eqs.~\ref{tauM}). Thus, for 
 $N_D\rightarrow0$,  the hydrodynamic regime becomes valid for all range of finite temperatures, as expected in the probe limit. 

\subsection{Spectral functions at large chemical potential $(T,\omega \ll \mu)$}

Another example of how accurate  hydrodynamics describes the large chemical potential regime, can be given by considering spectral functions. 
%
%
For a general field theory operator $\CO$, the spectral function is defined by
\begin{equation}
\chi_{_{\CO \CO}}(\bar{\omega},\bar{q}) \equiv -2\text{Im}\left[G_{\CO \CO}^{R}(\bar{\omega},\bar{q})\right].
\end{equation}
A simple pole of the correlator  $G_{\CO \CO}^{R}$ located at $\omega = \bar{\omega}_{_{QNM}}$ , produces a peak in $\chi_{_{\CO \CO}}$, 
with
\begin{equation}
\text{Peak width} \propto 2|\text{Im}(\bar{\omega}_{_{QNM}})|,\qquad
\text{Peak height} \propto \frac{\text{Res}[G_{\CO \CO}^{R}(\bar{\omega}_{_{QNM}})]}{|\text{Im}(\bar{\omega}_{_{QNM}})|}\,.
\end{equation}
This implies that the spectral function will be dominated by the the excitations of $G_{\CO \CO}^{R}$ that are both close to the real axis (long lived modes) and have large residues. We will use the appropriately normalized\footnote{ We will choose the same normalization as in ref.~ \citep{Davison:2013bxa}, where $\chi_{_{T^{ty}T^{ty}}}$ is given in units of $r_{H}^3/2\kappa^2 L^4$ and $\chi_{_{J^{y}J^{y}}}$ in units of $r_{H}/2\kappa^2$.} hydrodynamic spectral functions associate with $G_{T^{ty} T^{ty}}^{R}$ and $G_{J^{y} J^{y}}^{R}$ as given in (\ref{GFS}).
%
We will fix the (normalized) momentum  $q/\mu$ and consider  both high and low temperature behaviour, as well as different values of the charged matter parameters, including $\CS$ and $N_D$, in the following range
\begin{equation}
0\le\CS\le1,\quad \CP=1,\quad 0.5\le N_D\le1,\quad {q/\mu=0.1,\quad 0.006\le T/\mu\le 5}.
\end{equation}
\begin{figure}[h!tt]
\includegraphics[width=0.495\textwidth]{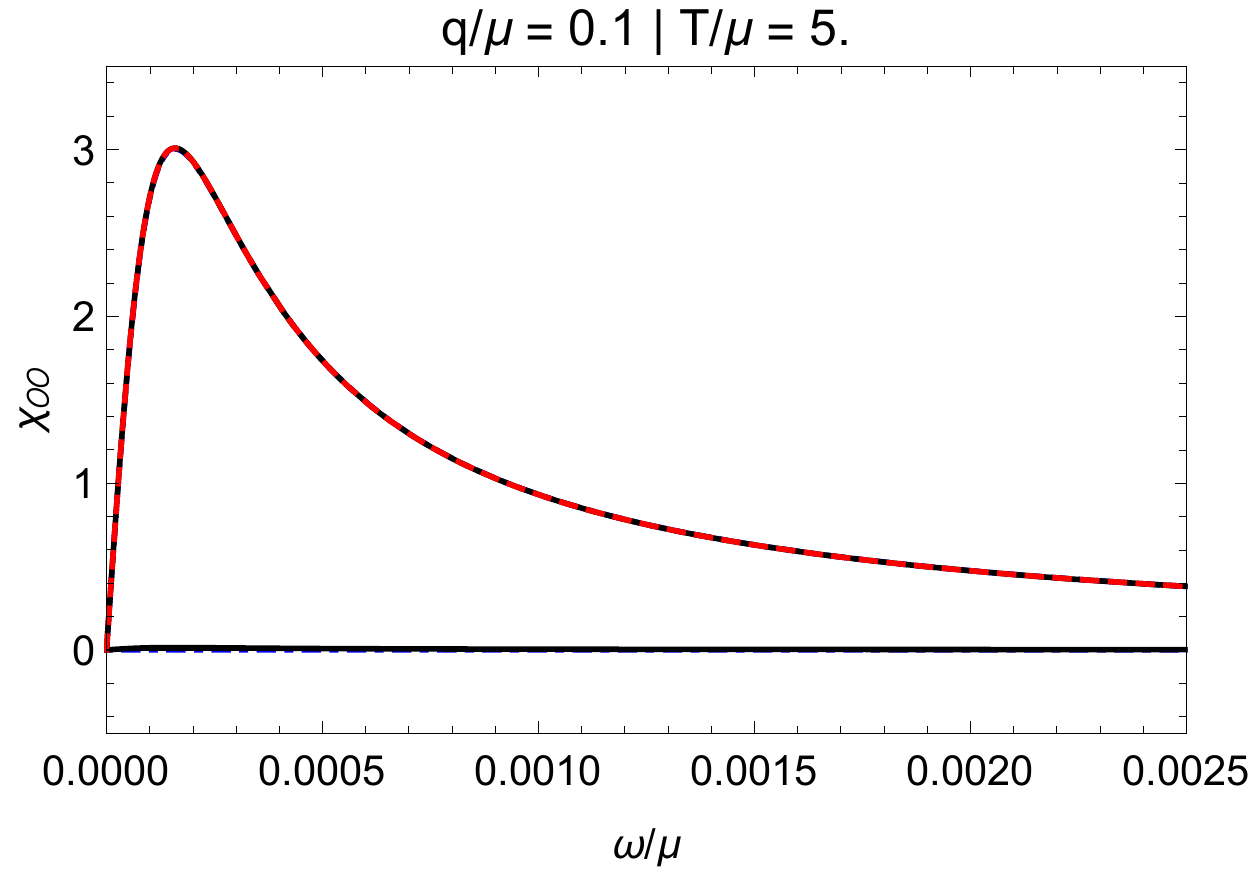}
\includegraphics[width=0.495\textwidth]{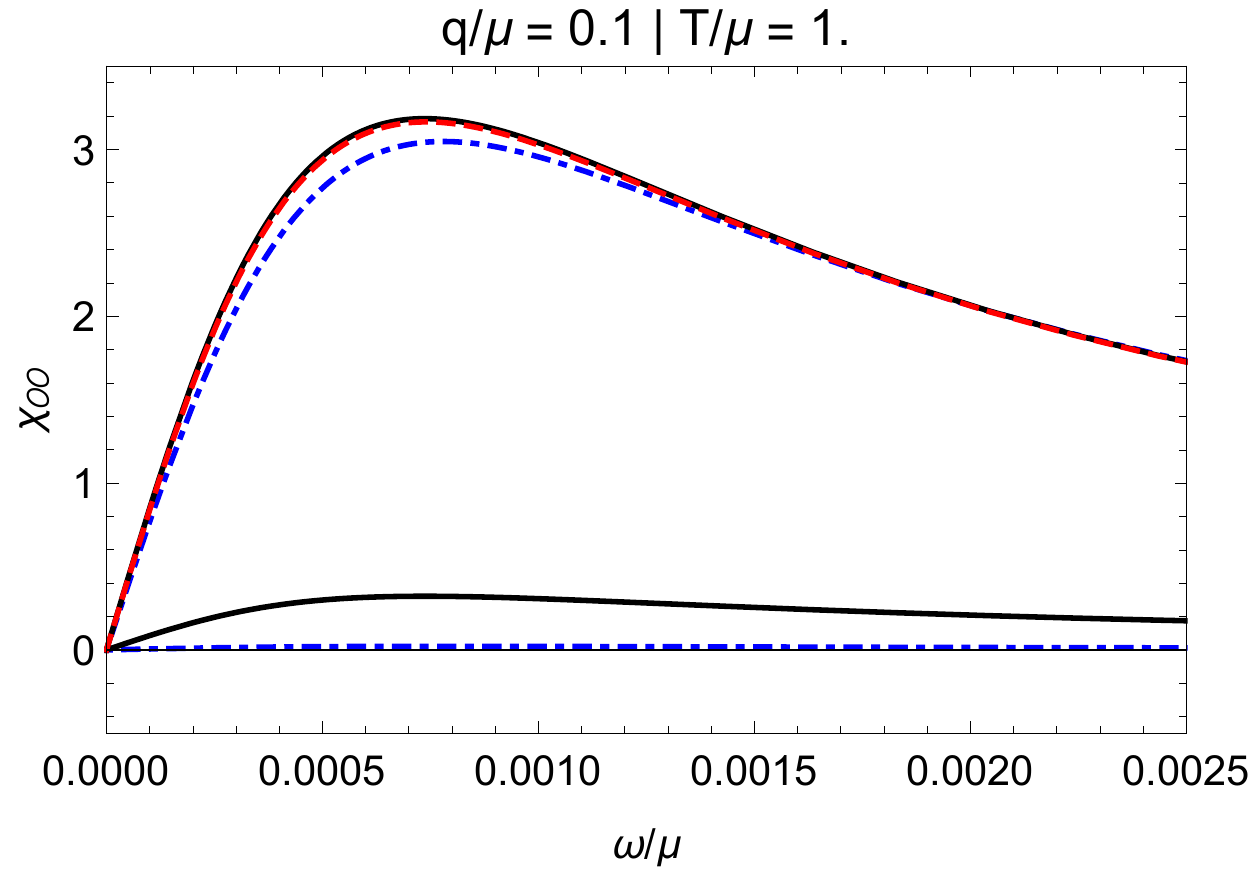}
\includegraphics[width=0.495\textwidth]{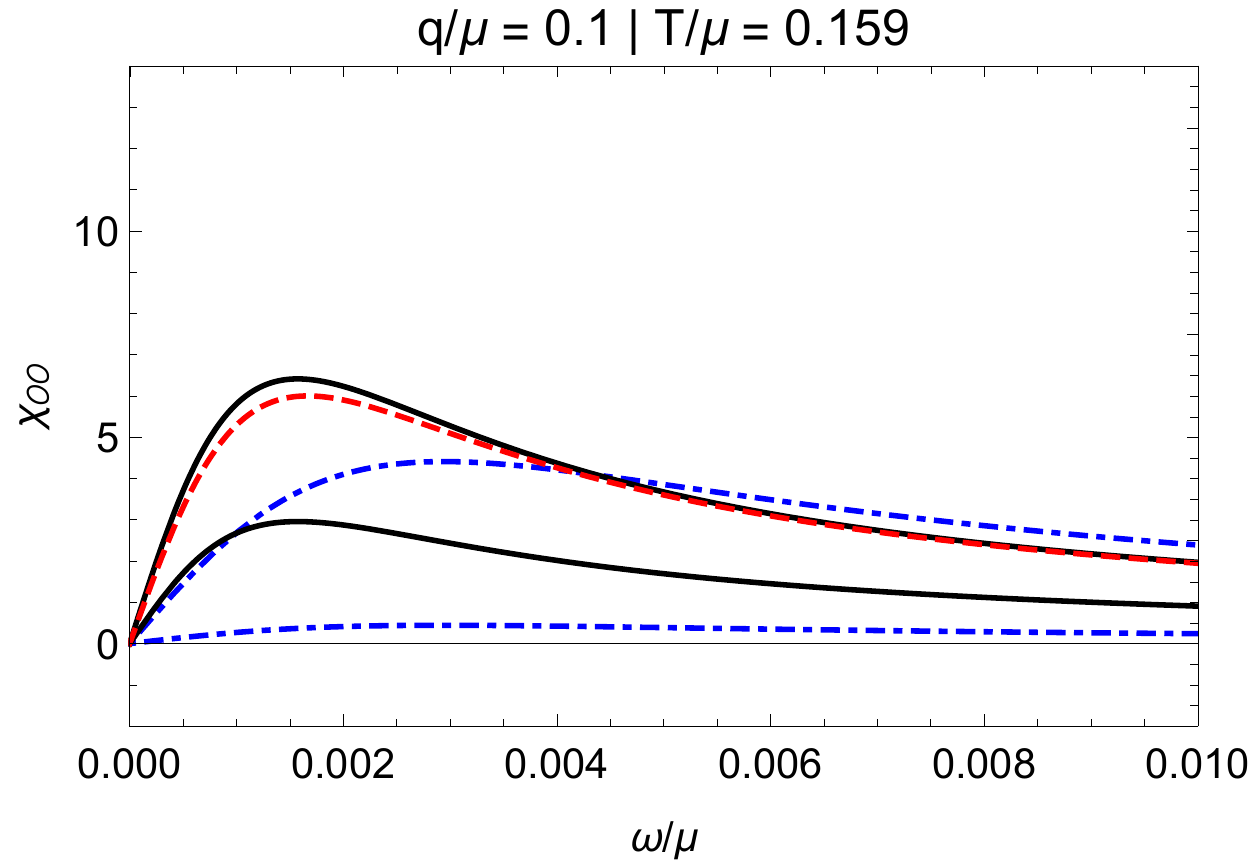}
\includegraphics[width=0.495\textwidth]{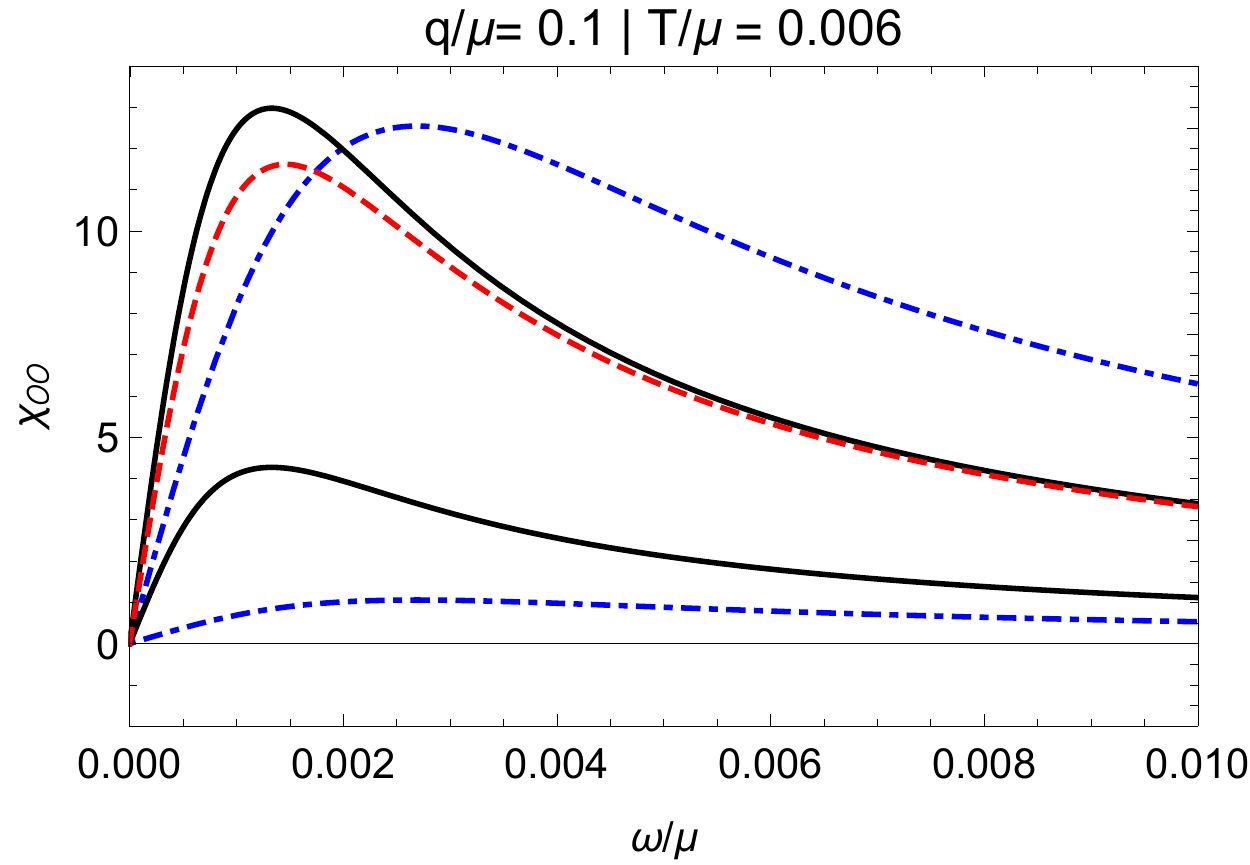}
\caption{\label{F3}The low frequency behaviour $\omega/\mu$ of the appropriately normalized hydrodynamic spectral functions $\chi_{_{T^{ty}T^{ty}}}$ (top curves) and $\chi_{_{J^{y}J^{y}}}$ (bottom curves), as the temperature $T/\mu$ is decreased 
at fixed momentum $q/\mu=0.1$. For each spectral function, we show a black solid curve ($\CP=1,{\cal S}=1 ,N_D=1$, $AdS_4 DBI$), a red dashed curve (${\cal S}=0 ,N_D=1$, $AdS_4 RN$) and blue dot-dashed curve ($\CP=1,{\cal S}=1 ,N_D=0.5$). 
We observe that the  non-linearities play an important role only at low enough temperatures 
$T/\mu\leq \bar{T}_{m}\approx 0.159$, in agreement with the separation of scales given by Eq.~(\ref{tauM}).}
\end{figure}

The spectral functions are shown in Fig.~\ref{F3}.  In each of the 4 panels (each corresponding to a fixed value of $T/\mu$), 
the three  top curves correspond to  $\chi_{_{T^{ty}T^{ty}}}$, 
and the 
three bottom curves to $\chi_{_{J^{y}J^{y}}}$, plotted at different values of the parameters, as described in the caption.
Since both spectral functions are dominated by the same pole, the relative strength of the response is completely determined by the relative size of the residues of their respective Green's functions.

In the high temperature regime (top left panel in Fig.~\ref{F3}, where $T/\mu=5$ and $q/\mu =0.1$), we observe a single peak in $\chi_{_{T^{ty}T^{ty}}}$ 
and no response from the current spectral function, which is suppressed compared to the spectral function of the 
energy-momentum tensor, $\chi_{J^{y} J^{y}}\ll\chi_{T^{ty} T^{ty}}$. 
At  high temperature, the influence  of the charged matter is negligible, and we are effectively in the probe limit. This is the 
reason why the three curves  that correspond to different values of $\CP ,{\cal S}$ and $N_D$  are essentially on top of each other.

As we decrease the temperature (top right panel in Fig.~\ref{F3}, where $T/\mu=1$ and $q/\mu =0.1$), we observe that the current spectral functions are no longer negligible, and that the transport peak broadens. We also observe that the three curves showing each of the  spectral functions at  different values of $\CP ,{\cal S} ,N_D$ now differ from each other. This regime still corresponds to the high temperature hydrodynamic regime, i.e. the region situated to the right of the maximum  $\bar{T}_{m}$ in the $\mu\CD$ vs $T/\mu$ plot shown in Fig.~\ref{F2}. 

As we continue to lower  the temperature and we enter the regime where $T/\mu\leq0.159$ (two bottom panels in Fig.~\ref{F3}), we observe that the effect of non-linearities becomes significant. In particular, we observe that the height of the peaks grows as they become narrower, which implies that the diffusive pole is moving closer to the real axis  and that both residues are increasing.

\subsection{Non-linearities and the QNMs for large momenta  $(q\gg\mu)$}
We have seen in the previous sections that the hydrodynamic approximation describes the regime $T/\mu \ll 1$ remarkably well.
\begin{figure}[!tt]
\centering
\includegraphics[width=0.495\textwidth]{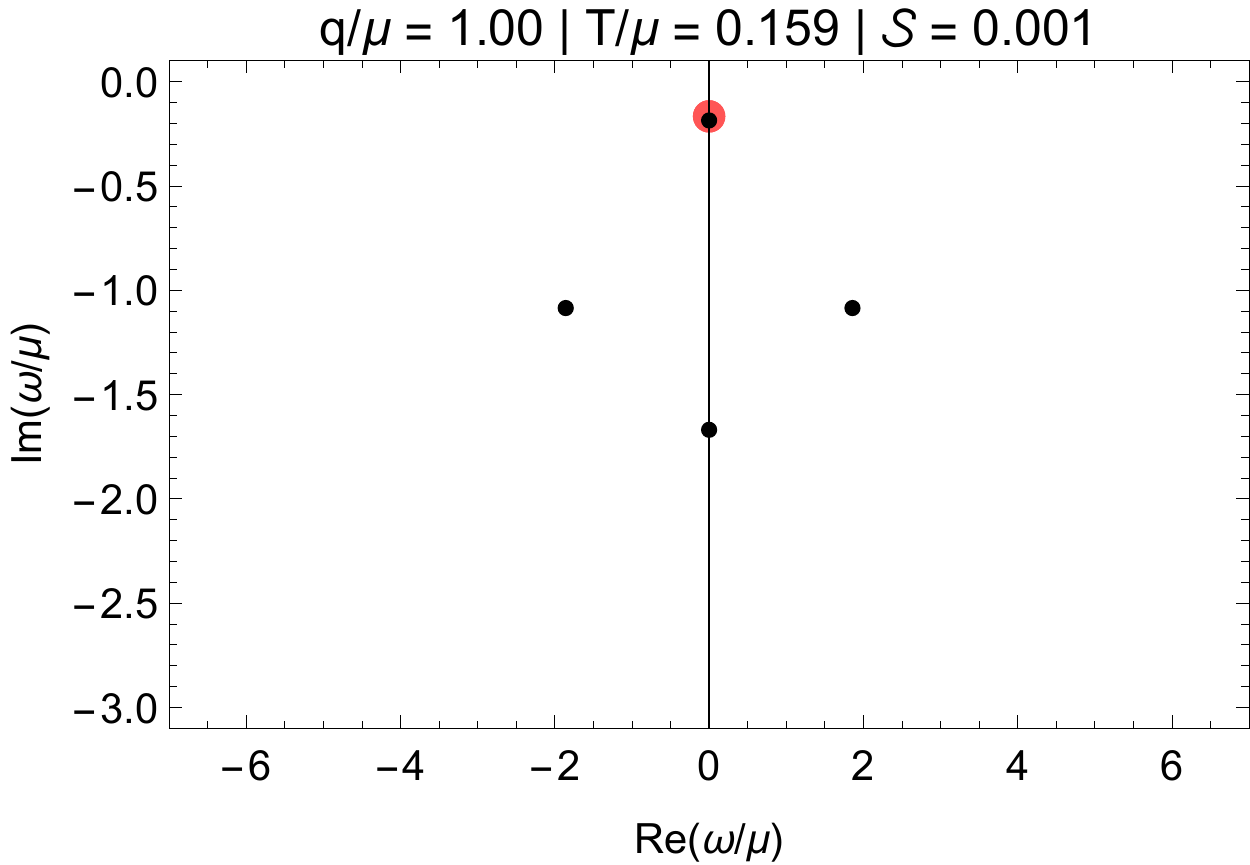}
\includegraphics[width=0.495\textwidth]{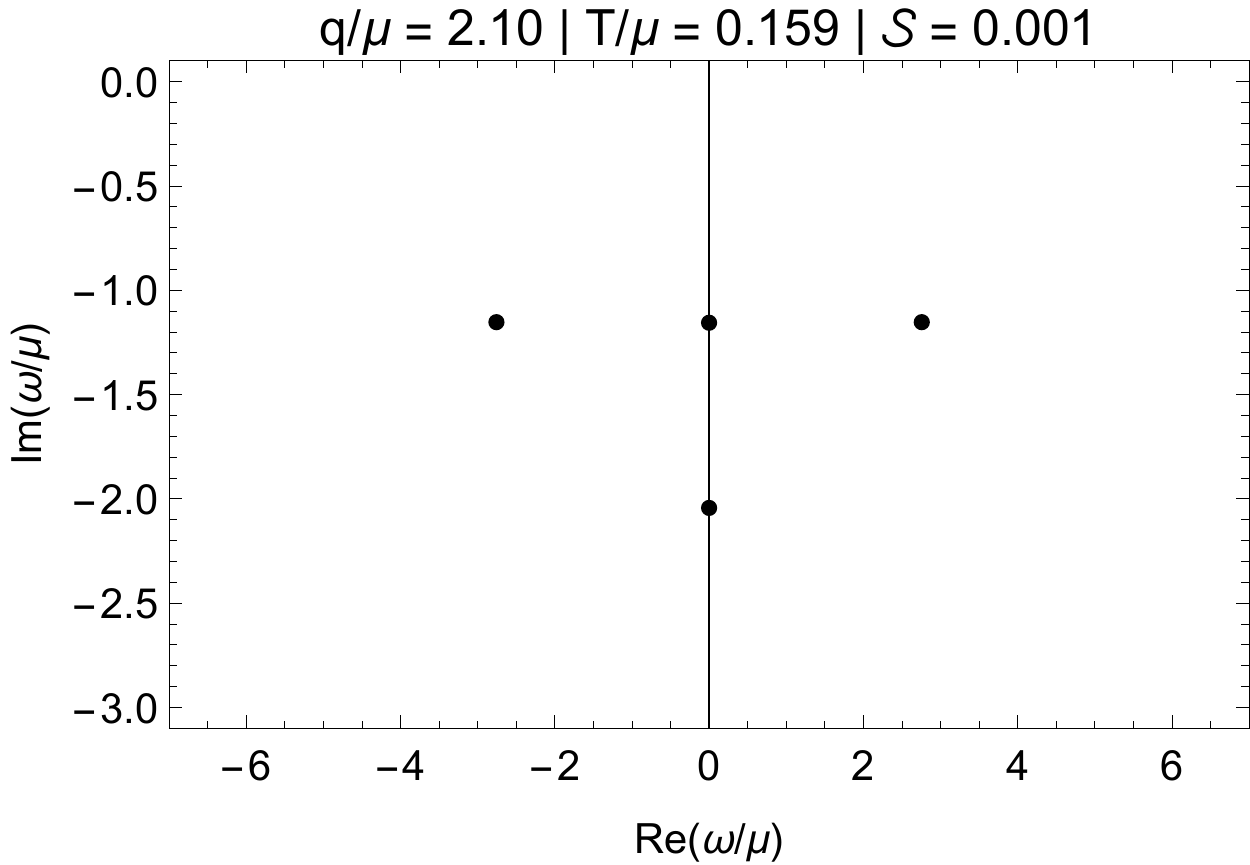}
\includegraphics[width=0.495\textwidth]{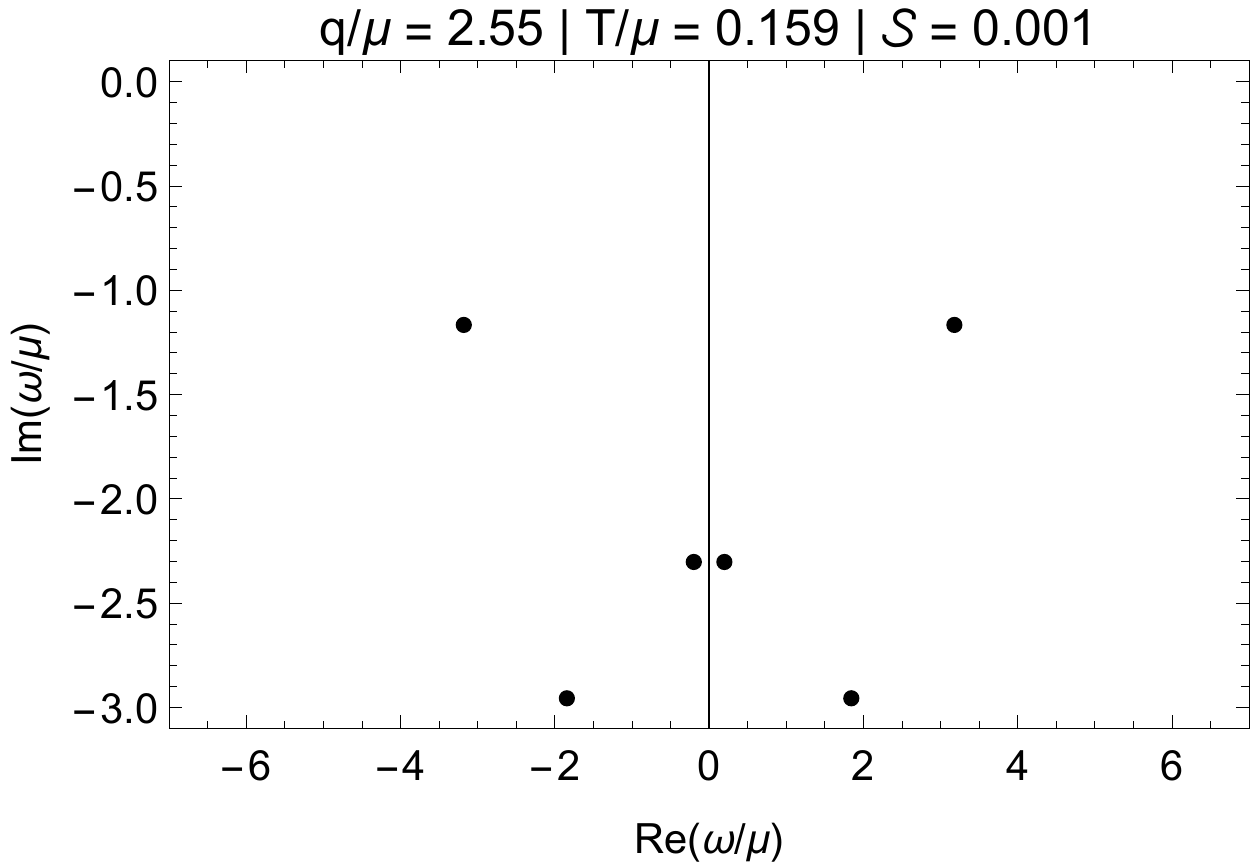}
\includegraphics[width=0.495\textwidth]{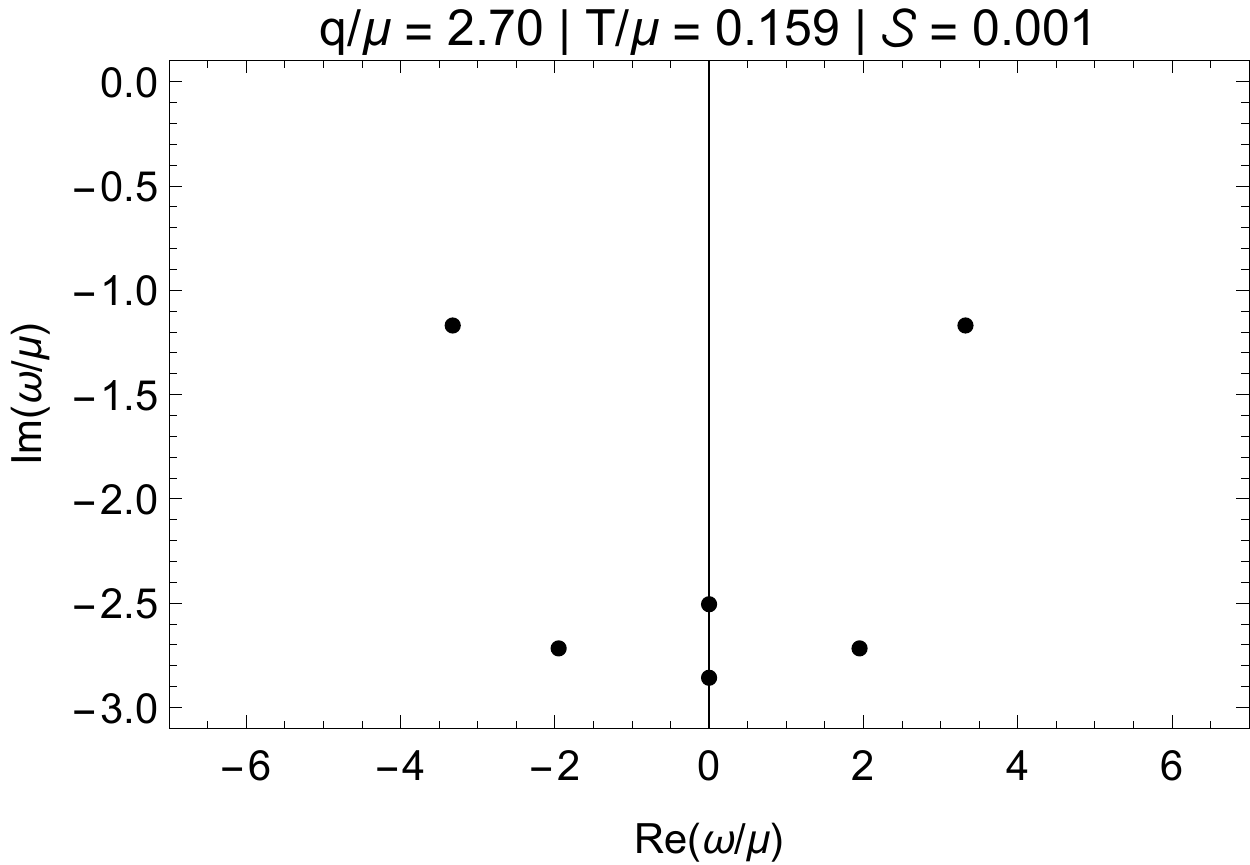}
\includegraphics[width=0.495\textwidth]{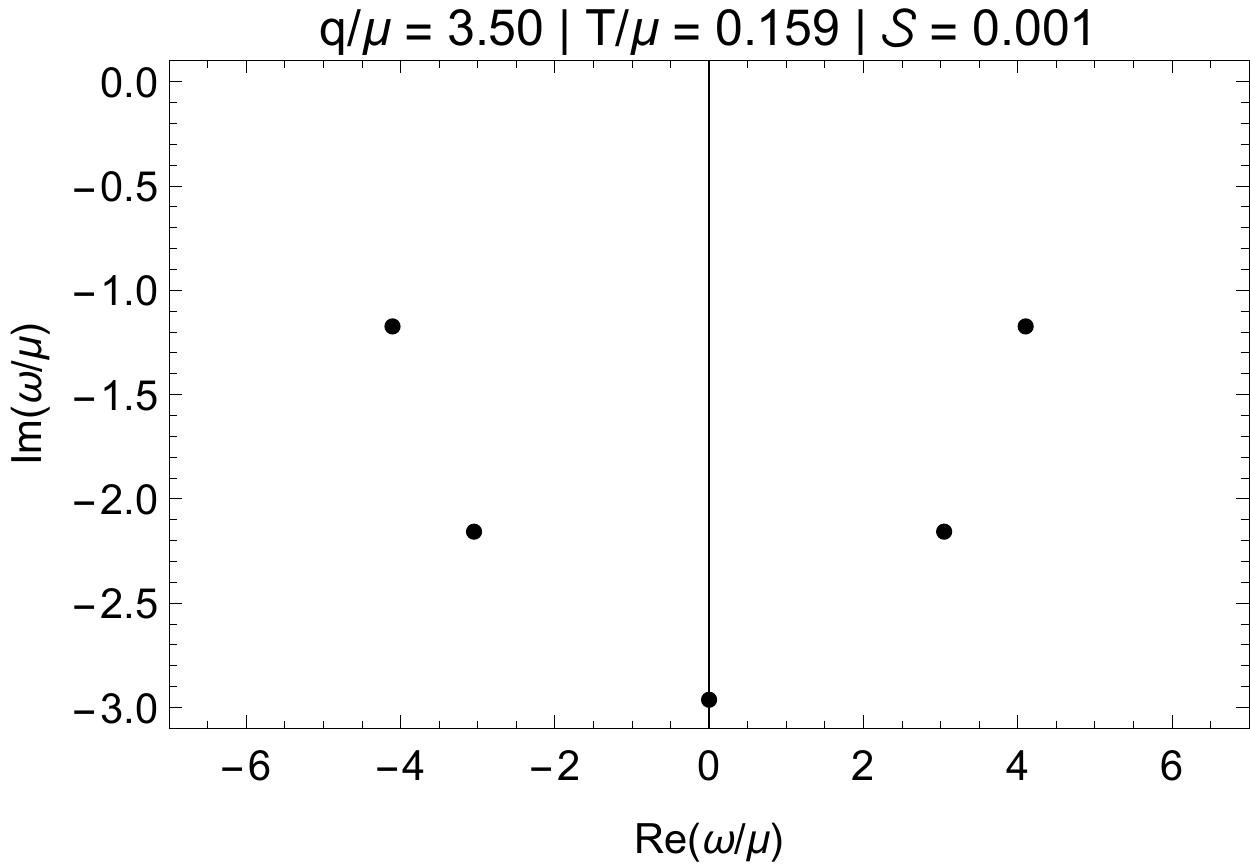}
\includegraphics[width=0.495\textwidth]{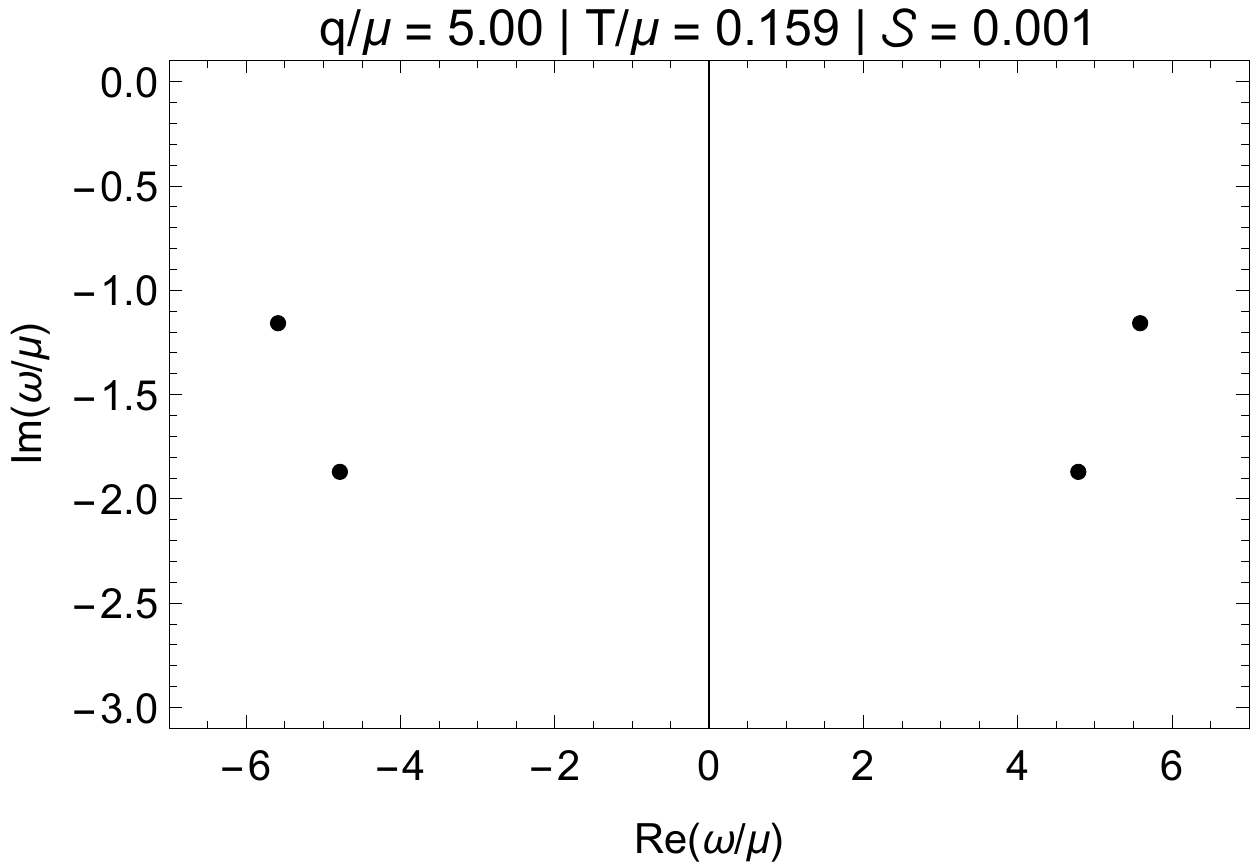}
\caption{\label{F4}The QNMs closest to the real axis in the complex $\omega/\mu$ plane, at various values of the momentum $q/\mu$, fixed $T/\mu = 0.159$, and fixed parameters $(\CS=0.001,~\CP=1,~N_D=1,~AdS_4RN)$. For small momenta, 
the purely imaginary mode closest to the origin corresponds to the hydrodynamic diffusion pole of the correlators (top left panel). 
As we increase $q/\mu$, the two purely imaginary modes ``attract" each other and coalesce, after which they attain  finite real parts  for a range of momenta $2.45\le q/\mu\le 2.7$. Eventually, after reaching their maximum separation at $q/\mu\approx2.55$, they come back to the imaginary axis and disappear down the complex plane. Simultaneously, the other QNMs are joined with two new QNM modes entering from below and drift together away from the imaginary axis (bottom right panel). (Animated
version of the figure is available on this paper's arXiv page.)}
\end{figure}
In this section, we will consider the QNM spectrum of the system in the regime of large momenta,  $q/\mu \gg1$,  with $T/\mu$ fixed at  $T/\mu = \bar{T}_{m}\approx 0.159$.

Alongside this, we will be also interested in how the non-linearities, characterized by $\CS$ and $\CP$, change the large momentum behaviour of the spectrum. In particular, we will show how the presence of $\CP$ seems to create an effective repulsion between some of the modes in the mixed correlator, which we will explore for the following range of parameters
\begin{equation}
0.001\le\CS\le1,\quad 0\le\CP\le1,\quad N_D =1,\quad 1\le q/\mu\le 5,\quad T/\mu=0.159.
\end{equation}

In Fig.~\ref{F4}, we show the position of the six least damped  modes, as we increase the momentum $q/\mu$ from $q/\mu=1$ 
(top left panel) to $q/\mu =5$ (bottom right panel), at fixed temperature $T/\mu=0.159$, with full backreaction $N_D=1$ and for very small non-linearities $\CS=0.001,~\CP=1$. In this limit, we expect the system to be described by the $AdS_4 RN$ geometry, since we are in the linear regime.
We start with $q/\mu=1$, where the purely imaginary mode closest to the real axis is well approximated by the hydrodynamic diffusion mode.  We observe that this is a gappless mode, since it approaches the origin with the momentum approaching zero. 
The other QNMs  are gapped and remain at a finite distance from the origin. 
As we increase the momentum and move further away from the hydrodynamic regime, this diffusive mode starts moving down and eventually is ``attracted" and coalesce with another QNM on the imaginary axes at $q/\mu \approx 2.45$ (not shown), after which they both attain finite real parts and move off the imaginary axis, but not for long. They move down and reach their 
maximum relative separation at $q/\mu\approx 2.55$ (middle left panel in Fig.~\ref{F4}), before they return back to
 the imaginary axis, spitting into two new imaginary modes at $q/\mu\approx2.7$ (middle right panel in Fig.~\ref{F4}). For larger momenta, the two modes move down. Meanwhile, the QNMs located off the imaginary axis, start moving away from the
  axis and are joined by two new QNMs, that emerge from below. Eventually, they all start drifting away  from the imaginary axis, while remaining at a fixed distance from each other. A similar merging behaviour was observed in the 
  shear correlator in ref.~\citep{Brattan:2010pq}.

\begin{figure}[!tt]
\centering
\includegraphics[width=0.495\textwidth]{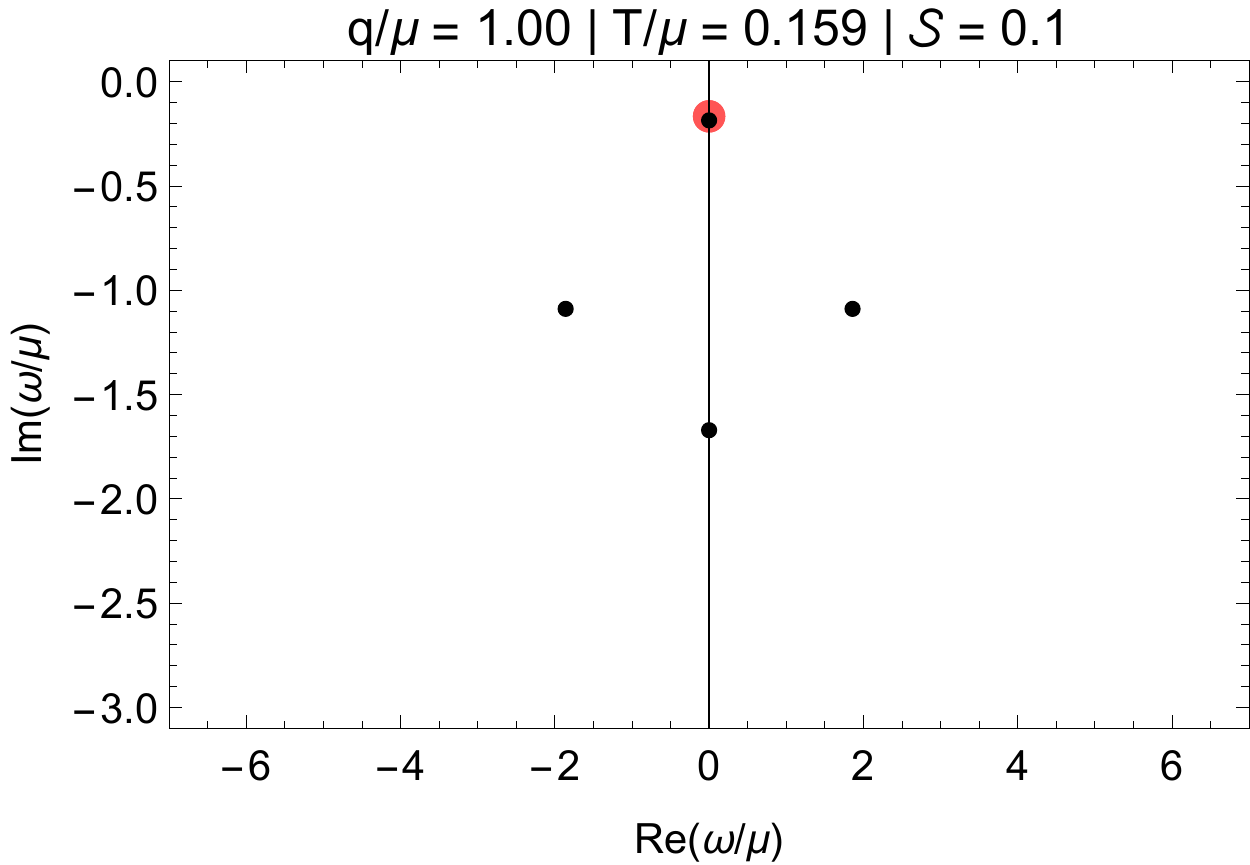}
\includegraphics[width=0.495\textwidth]{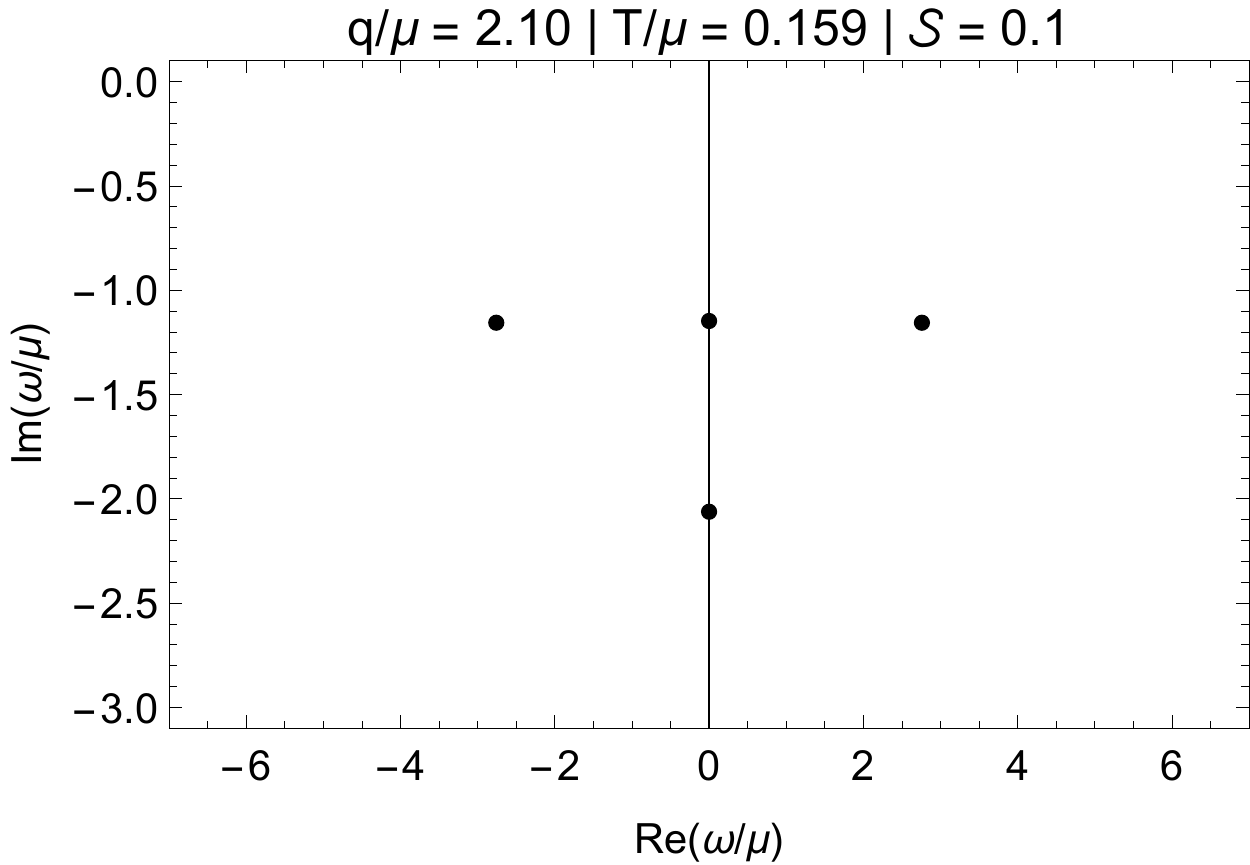}
\includegraphics[width=0.495\textwidth]{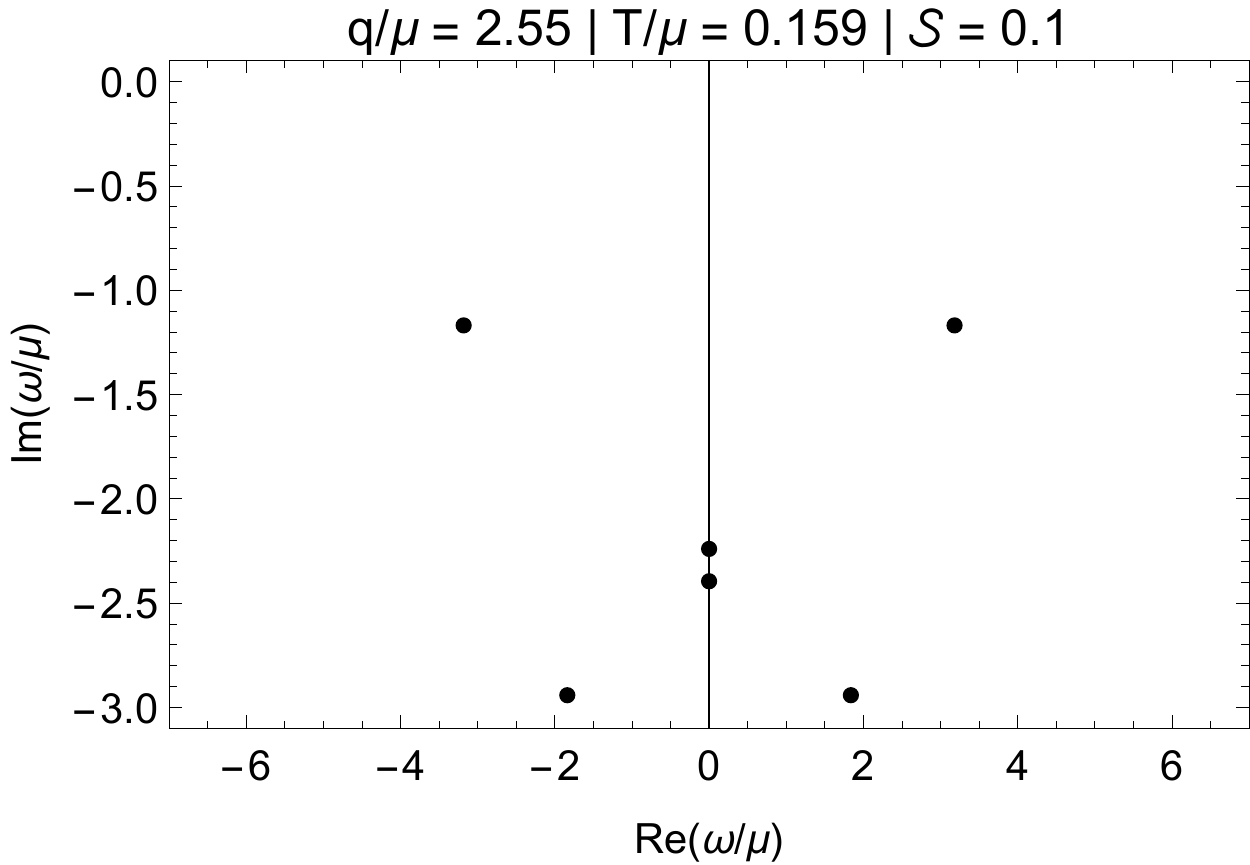}
\includegraphics[width=0.495\textwidth]{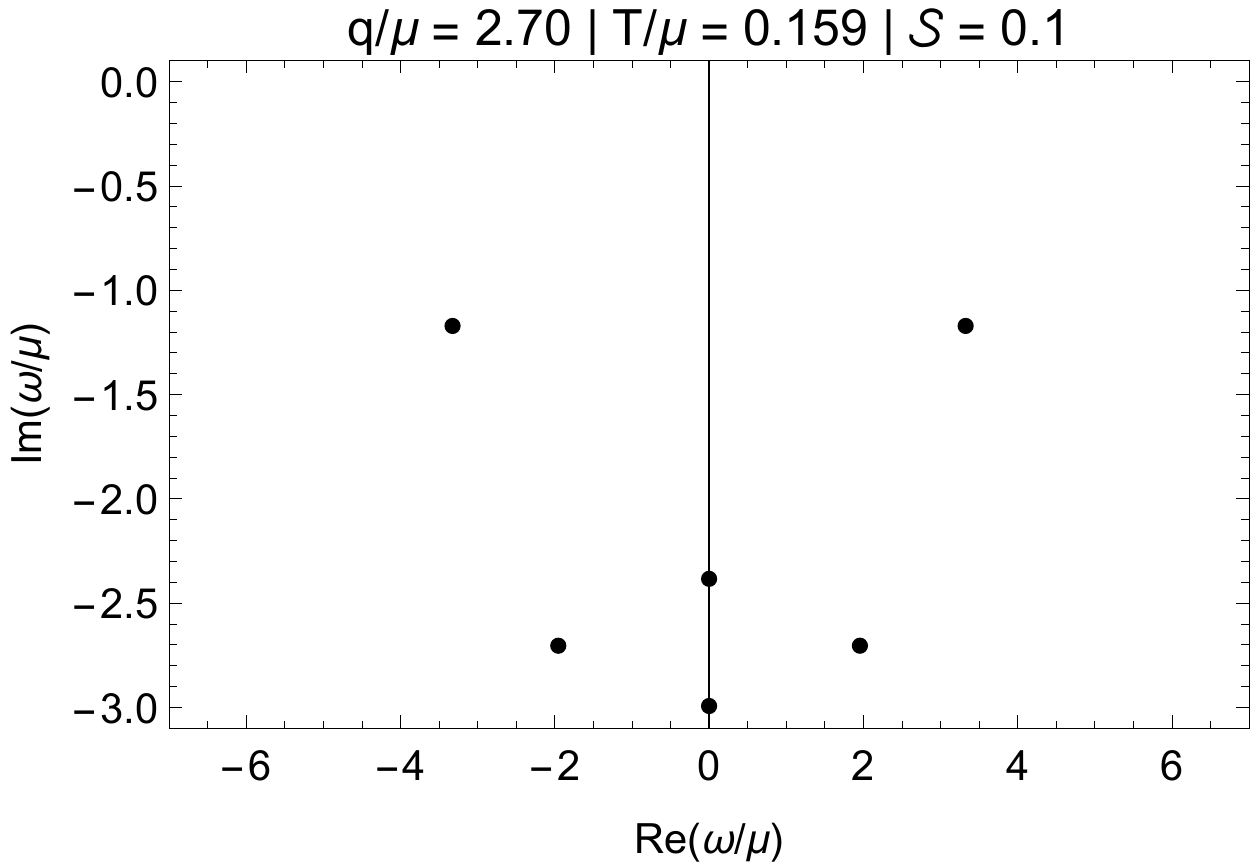}
\includegraphics[width=0.495\textwidth]{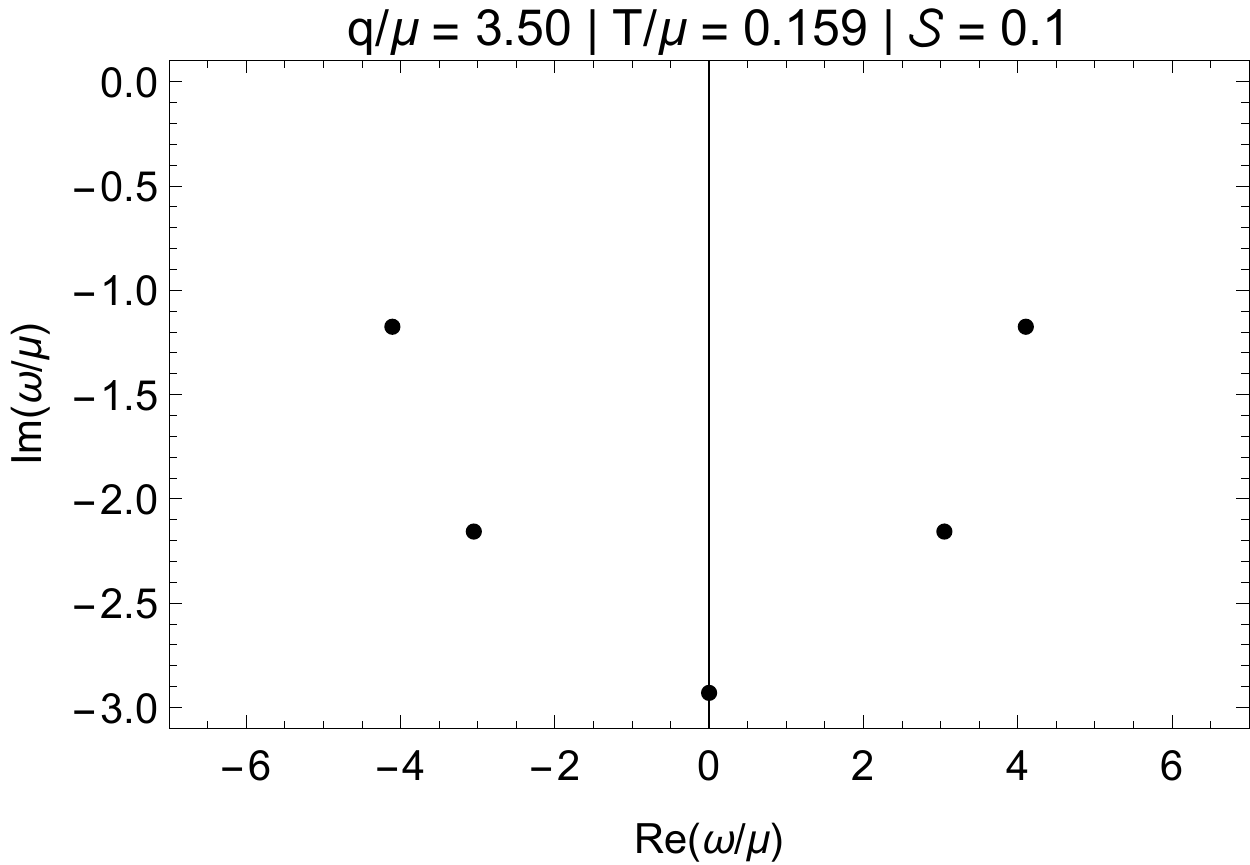}
\includegraphics[width=0.495\textwidth]{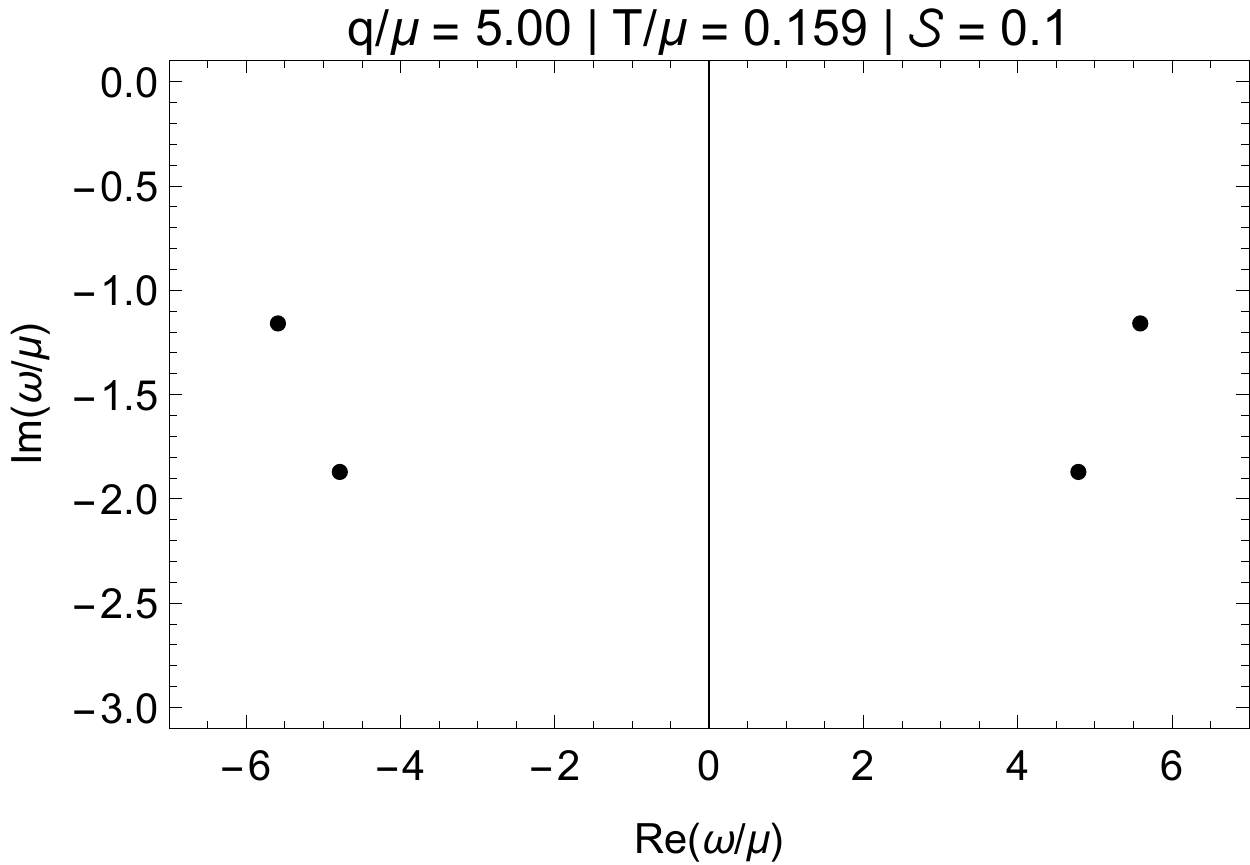}
\caption{\label{F5}The QNMs closest to the real axis in the complex $\omega/\mu$ plane, at various values of the momentum $q/\mu$, fixed $T/\mu = 0.159$, and fixed parameters $(\CS=0.1,~\CP=1,~N_D=1)$. As we increase $q/\mu$, the two purely imaginary modes ``repel" each other for a range of momenta $2.45\le q/\mu\le 2.7$. After reaching their minimum separation at $q/\mu\approx2.55$ (middle left panel), the lowest mode is repelled down into the complex plane, leaving the diffusive mode behind (bottom left panel). The diffusive mode  eventually also disappears into the complex plane (bottom right panel). Other QNMs are not visibly affected. (Animated
version of the figure is available on this paper's arXiv page.)}
\end{figure}

Whith the increase of  non-linearity parameters, new phenomena emerge. In Fig.~\ref{F5}, we show 
the same range of parameters as in Fig.~\ref{F4}, but now with larger non-linearity effects ($\CS=0.1,~\CP=1$). 
We observe that the regimes of small ($q/\mu\leq 2.45$) and large ($q/\mu\geq2.7$) momenta remain essentially unaffected
 by  non-linearities. However, as we increase the momentum $q/\mu$ and the diffusive QNM starts moving down the imaginary axis, we now observe a picture very different from Fig.~\ref{F4}: there, we had  ``attraction" and merging of poles, whereas in 
 Fig.~\ref{F5} the non-linearities create  ``repulsion" between the modes. Once the diffusive mode gets close enough to the other QNM on the imaginary axes at $q/\mu \approx 2.45$ (not shown), it starts pushing down the other mode until at $q/\mu\approx 2.55$ (middle left panel  in Fig.~\ref{F5}), they now reach the point of closest approach, and then the lowest mode is repelled away at $q/\mu\approx 2.7$ (middle right panel in Fig.~\ref{F5}), leaving the diffusive mode behind. With even large $q/\mu$, the diffusive mode eventually moves down into the complex plane as well.

%
%

\begin{figure}[!tt]
\centering
\includegraphics[width=0.495\textwidth]{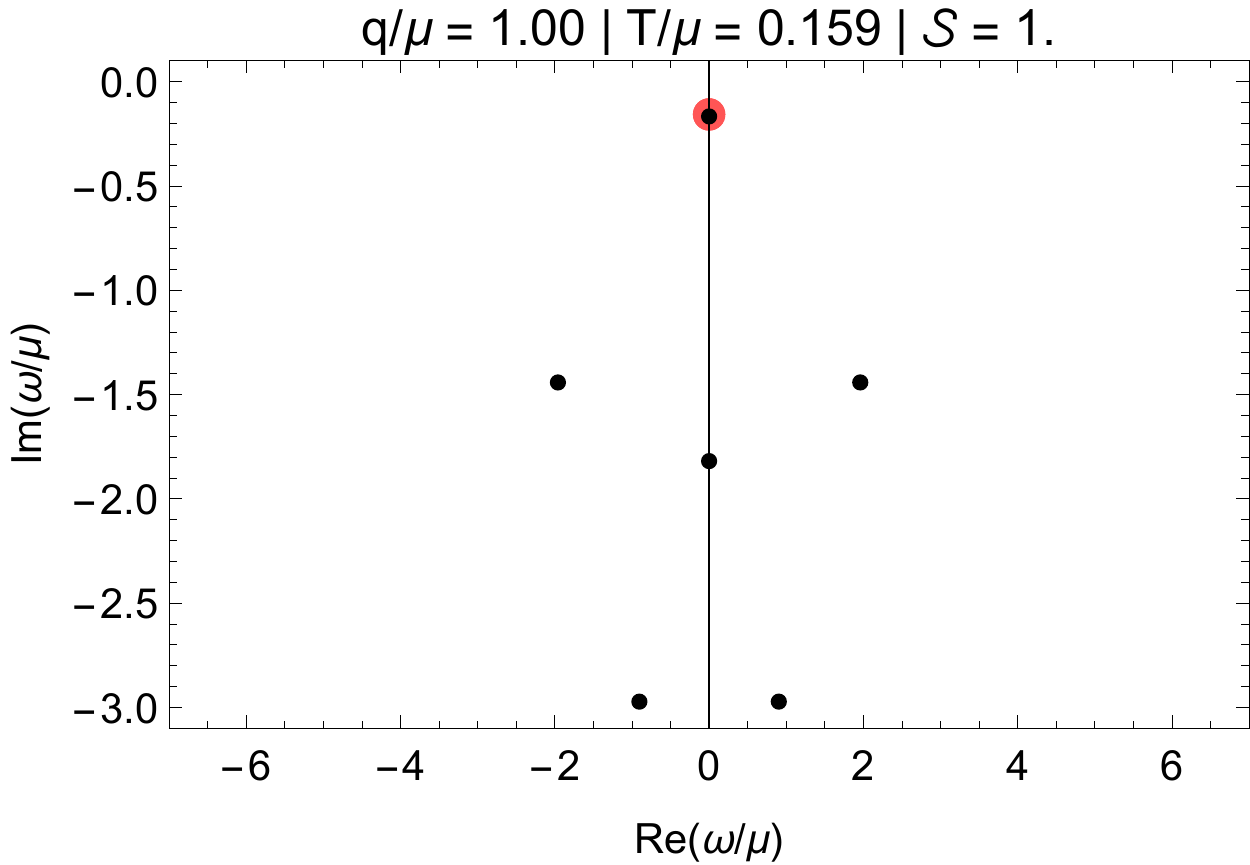}
\includegraphics[width=0.495\textwidth]{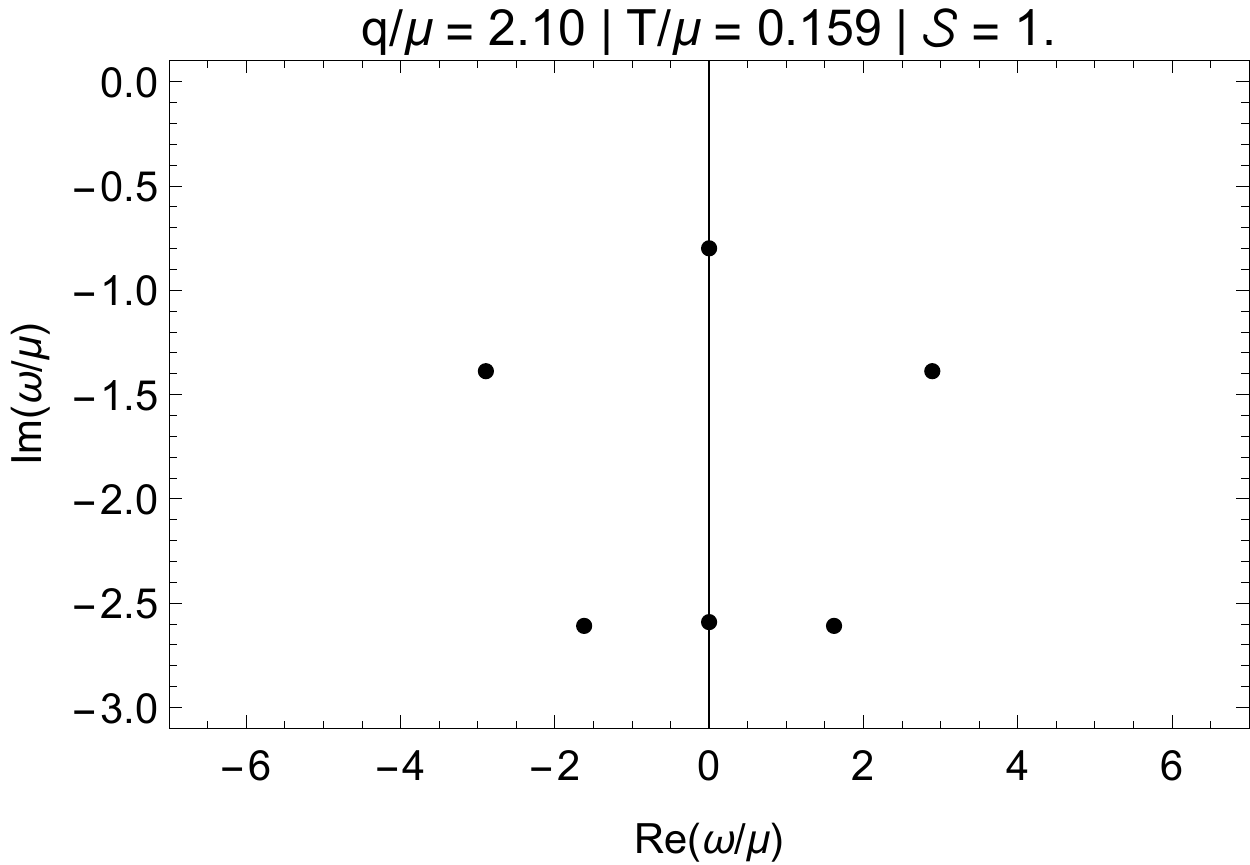}
\includegraphics[width=0.495\textwidth]{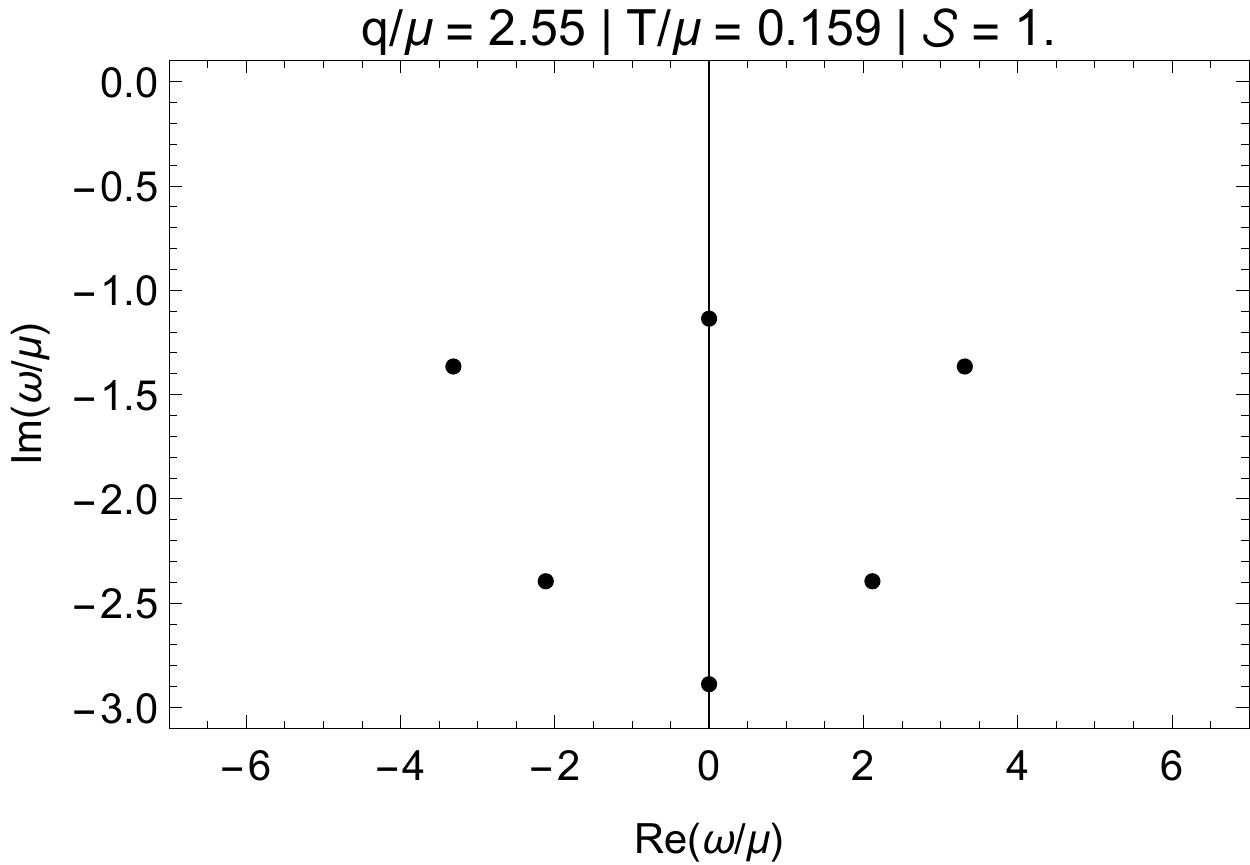}
\includegraphics[width=0.495\textwidth]{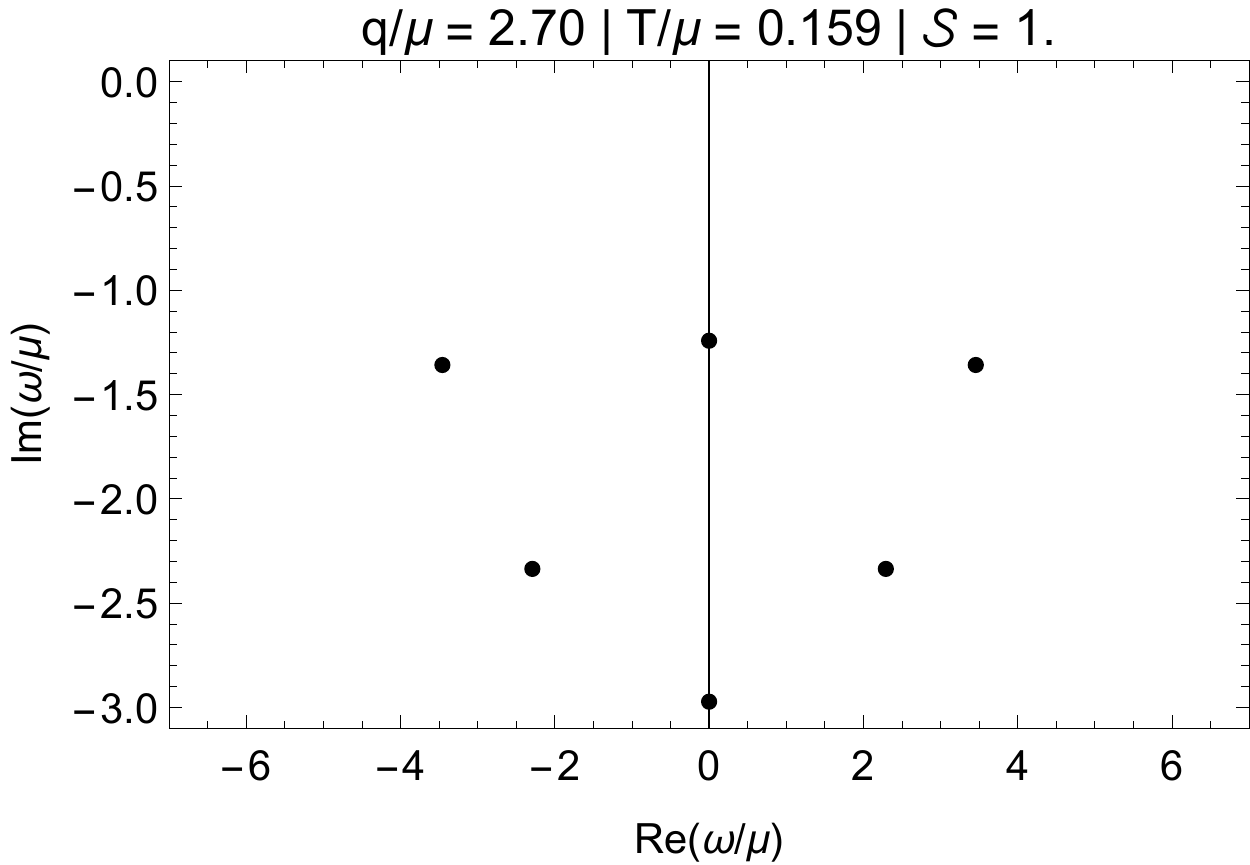}
\includegraphics[width=0.495\textwidth]{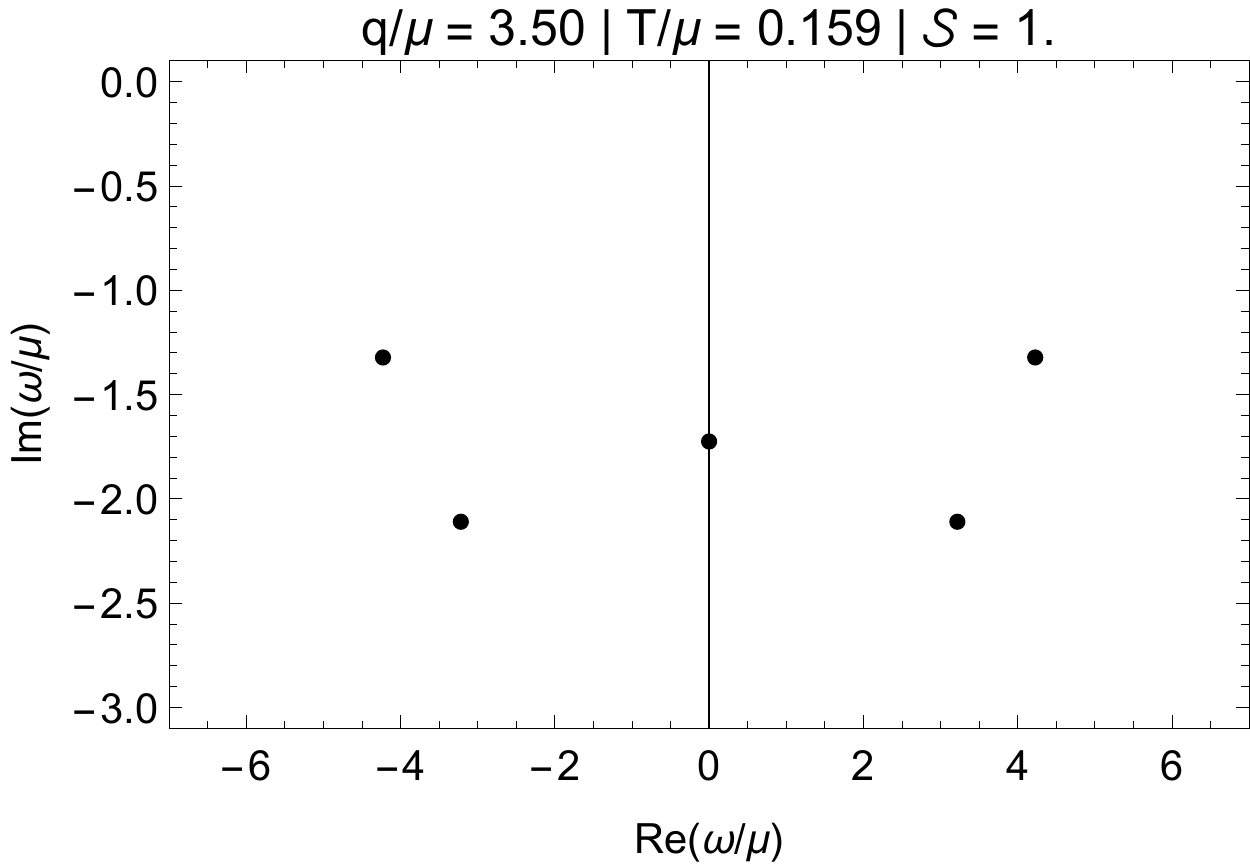}
\includegraphics[width=0.495\textwidth]{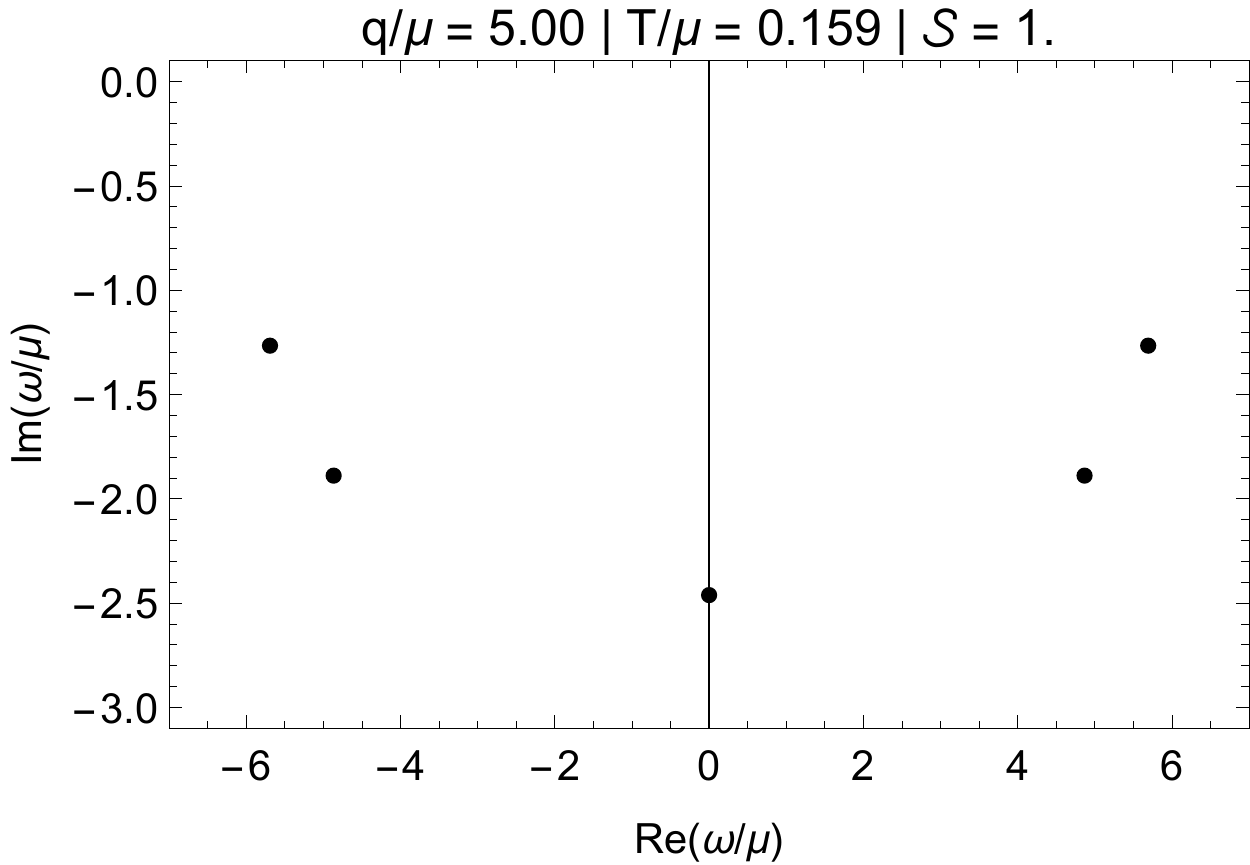}
\caption{\label{F6}The QNMs closest to the real axis in the complex $\omega/\mu$ plane, at various values of the momentum $q/\mu$, fixed $T/\mu = 0.159$, and fixed parameters $(\CS=1,~\CP=1,~N_D=1,~AdS_4DBI)$. As we increase $q/\mu$, the two purely imaginary modes  ``repel" each other for the range of momenta $1\le q/\mu\le 5$. They stay at a fixed ``equilibrium" distance and move down together, before eventually disappearing down into the complex plane. (Animated
version of the figure is available on this paper's arXiv page.)}
\end{figure}

Finally, we explore the $\CS=1$, $\CP=1$ regime, where the non-linearities are prevalent, in the 
same range of parameters as in the previous two cases. The spectrum of QNMs is shown in Fig.~\ref{F6}.
Now, as we  increase $q/\mu$, we observe that the non-linearities affect the whole range of momenta we have explored, $1\le q/\mu\le 5$. The effect of the ``repulsion'' also increases, insomuch that the two modes cannot even get close together any more, so there is no notion of minimal approach distance. Instead, they stay at a fixed ``equilibrium" distance and move down together.

\begin{figure}[!tt]
\centering
\includegraphics[width=0.495\textwidth]{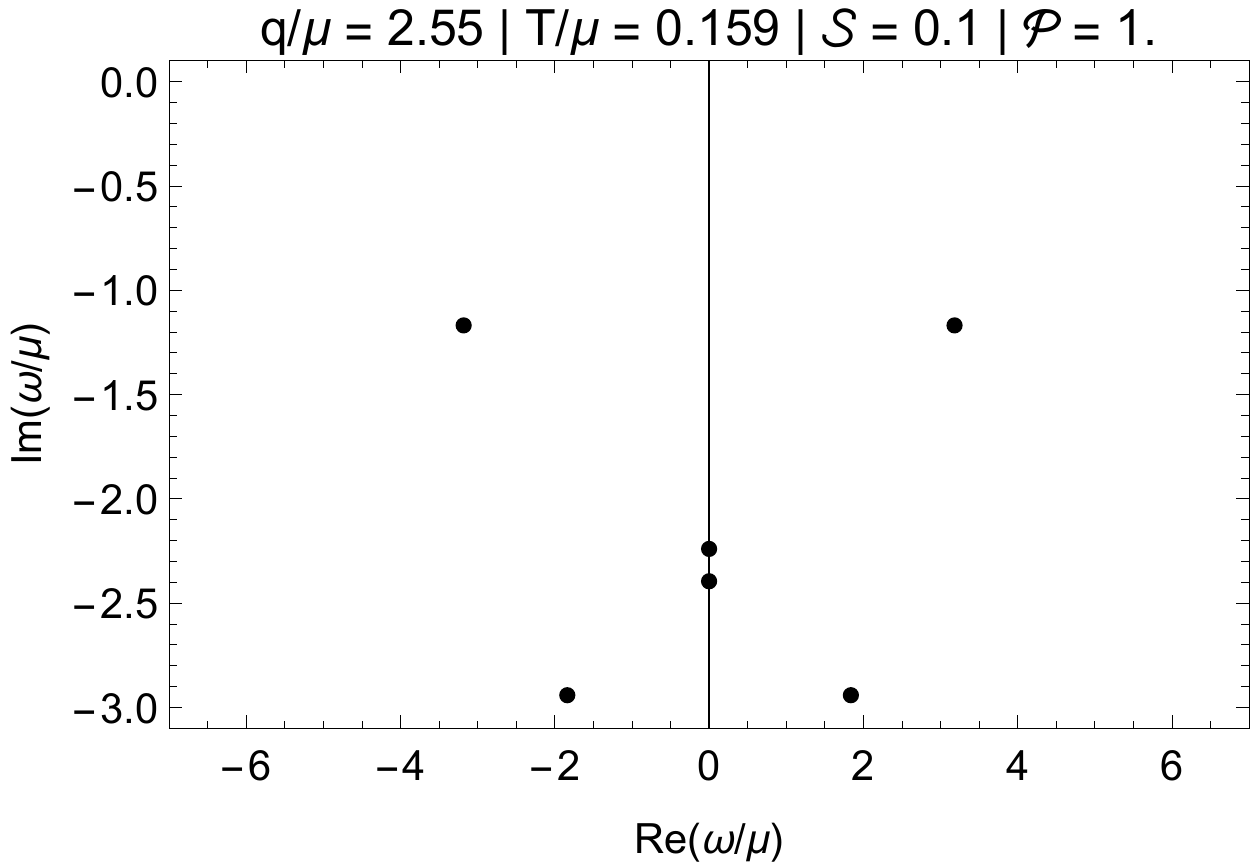}
\includegraphics[width=0.495\textwidth]{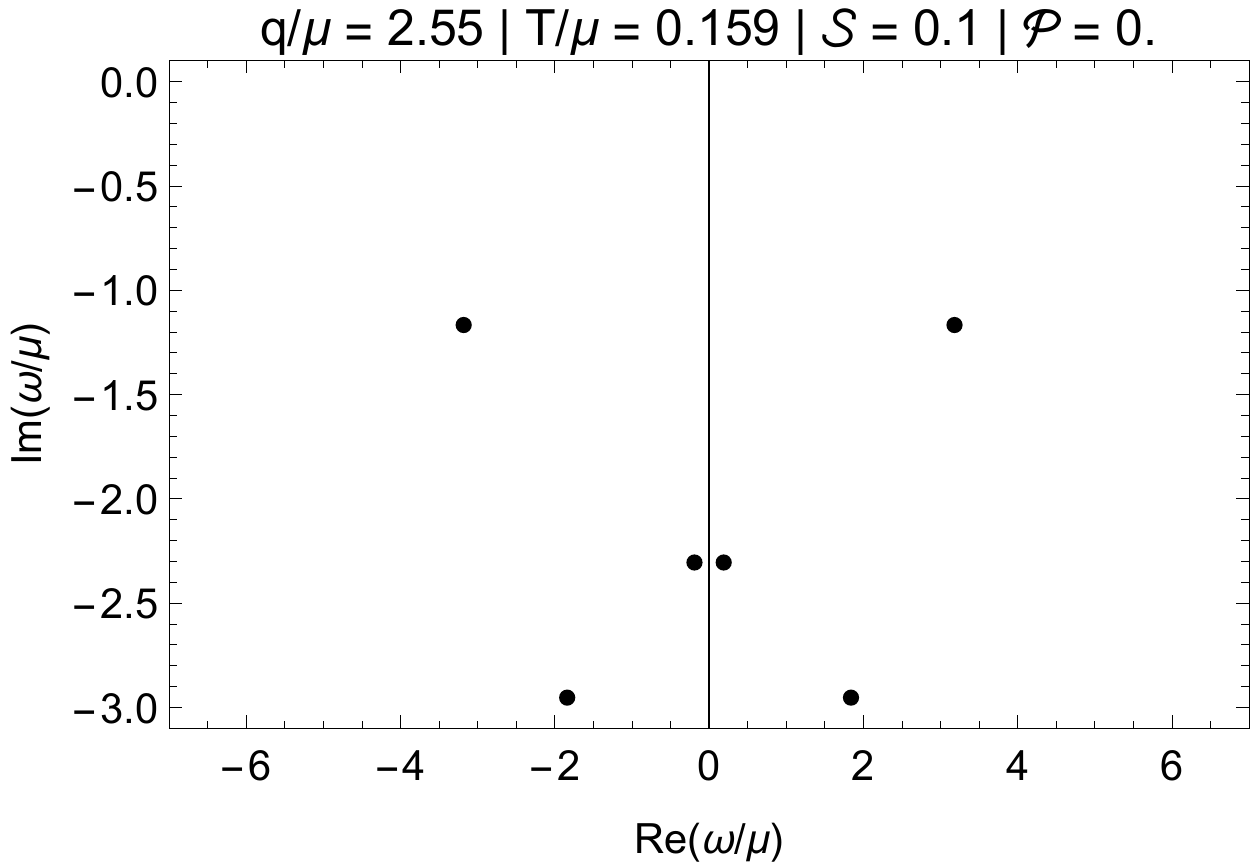}
\includegraphics[width=0.495\textwidth]{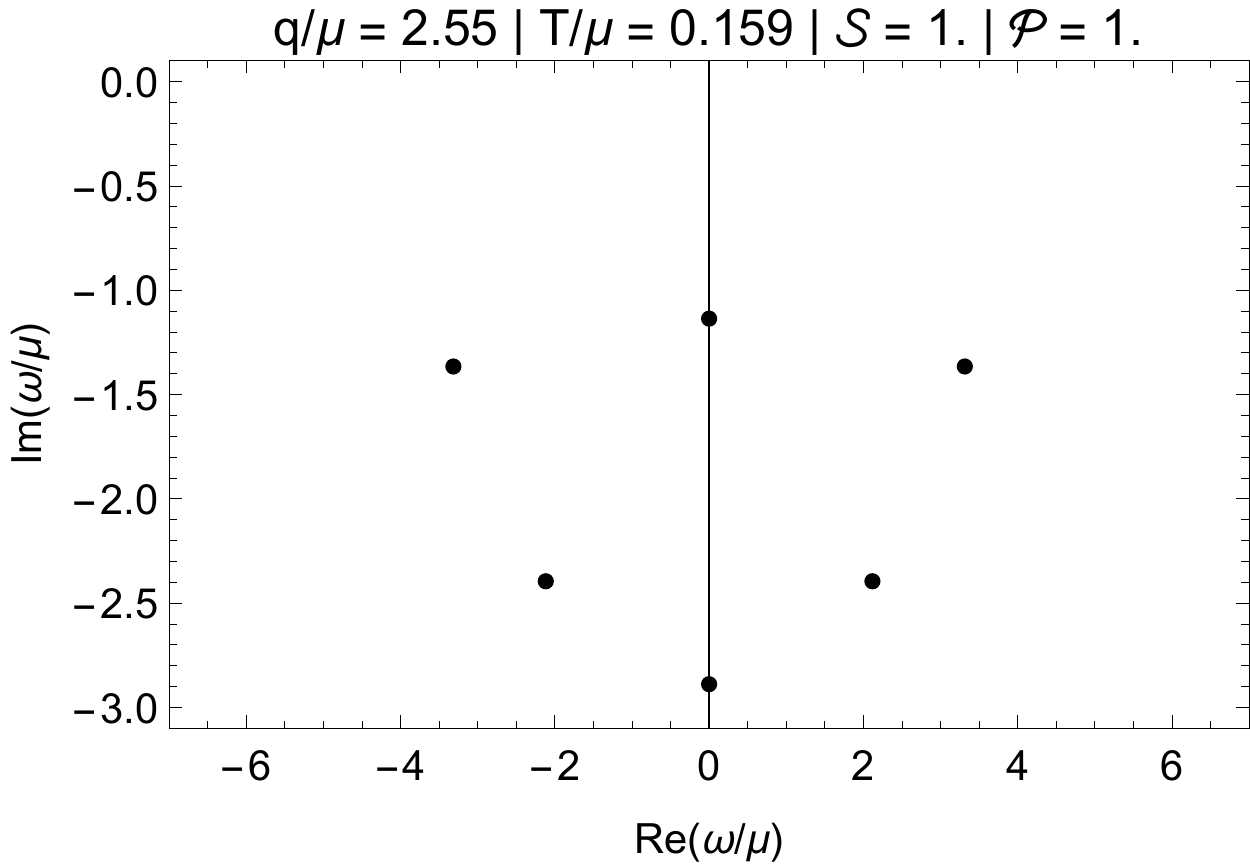}
\includegraphics[width=0.495\textwidth]{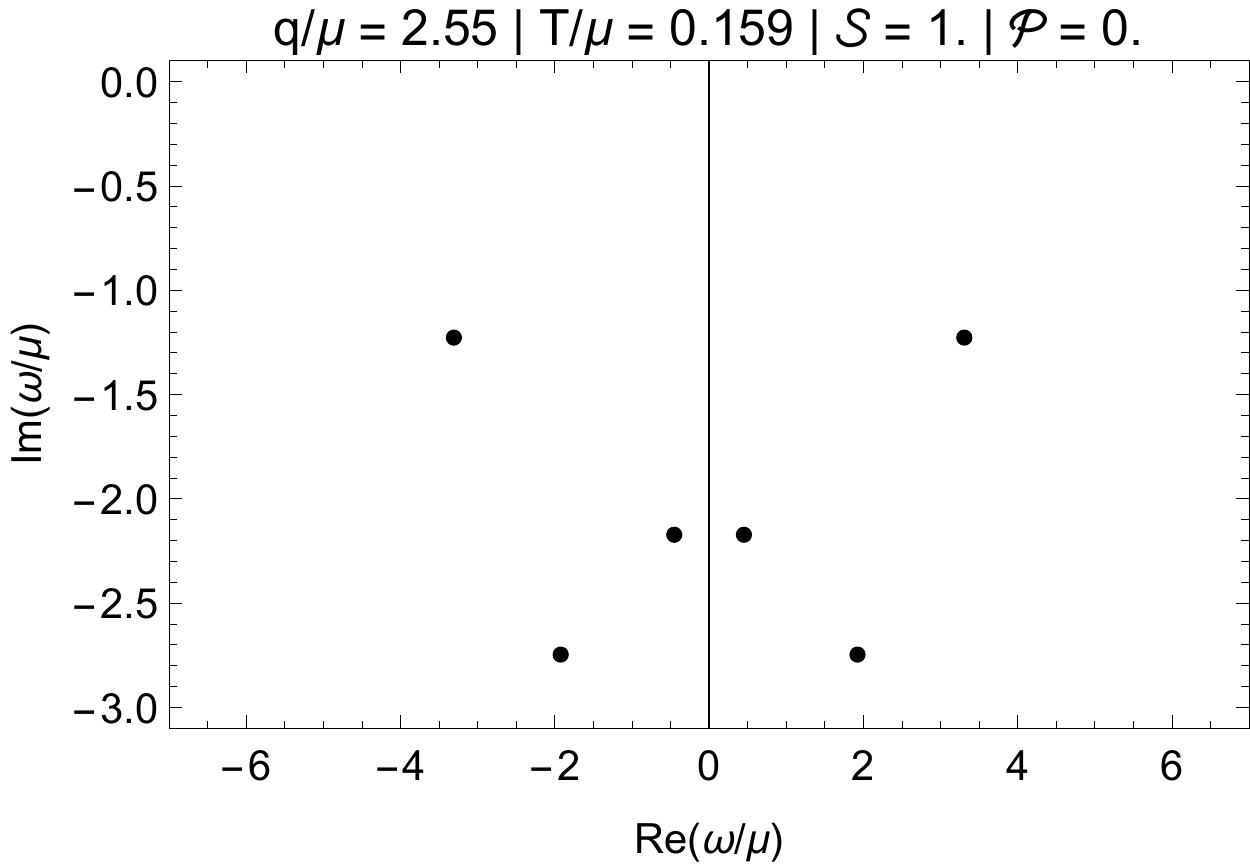}
\caption{\label{F7} The QNMs  for different values of the non-linearity parameter ${\CP=(0,1)}$,  fixed $q/\mu=2.55$, $T/\mu = 0.159$, ${N_D}=1$, and for ${\CS}=(0.1, 1)$. Setting $\CP=0$ seems to completely remove the ``repulsion" effect between the imaginary QNMs. Instead, we obtain an ``attraction" and splitting of the modes effect, similar to the linear case.}
\end{figure}

As already mentioned, the origin of this repulsion between the diffusive mode and the other imaginary QNM can be traced back to the presence of the $\CP \det (F_{~\nu}^{\mu})$ term in the Lagrangian density (\ref{LDBI}). Setting $\CP=0$, the behavior of the QNM spectrum  for $\CS=1$ and $\CS=0.1$ is very similar to the $\CS=0.001$ case, where the poles ``attract" and coalesce for a range of momenta. To illustrate this, we compare the $\CP=0$ and $\CP=1$ cases in Fig.~\ref{F7}, where we have fixed the momentum at $q/\mu =2.55$. Interestingly, for $\CP=0$, this value of the momentum corresponds to the point of maximal separation, whereas for $\CP=1$, it corresponds to the minimal approach (whenever the latter can be defined).

\subsection{Back-reaction, non-linearities and the  branch cut  $(T\ll \mu)$} 
In this section, we focus on understanding the effect that both the non-linearities $\CS$ and the finite backreaction $N_D$, 
have on the QNMs  at very low temperatures $T/\mu\ll1$.

As  already mentioned in the Introduction, the presence of the $AdS_2$ piece in the near-horizon geometry at $T/\mu=0$ indicates the existence of light modes described by an effective $CFT_1$ or ``semi-local quantum liquid" \citep{Iqbal:2011in}, and therefore we expect the retarded correlators of both $T^{a b}$ and $J^a$ to exhibit a continuous spectrum with a branch cut along the negative imaginary axis \citep{Edalati:2010hk,Brattan:2010pq,Denef:2009yy}. In our finite temperature case, we will observe the formation of a branch cut, as an infinite set of purely imaginary poles  that approach the origin and become denser as we lower the temperature (this is reminiscent of the situation described in ref.~\cite{Moore:2018mma}). We will explore the effects  the back-reaction parameter $N_D$ and the non-linearity parameter $\CS$
have on the formation of the branch cut  in the following range of parameters:
\begin{figure}[!tt]
\centering
\includegraphics[width=0.495\textwidth]{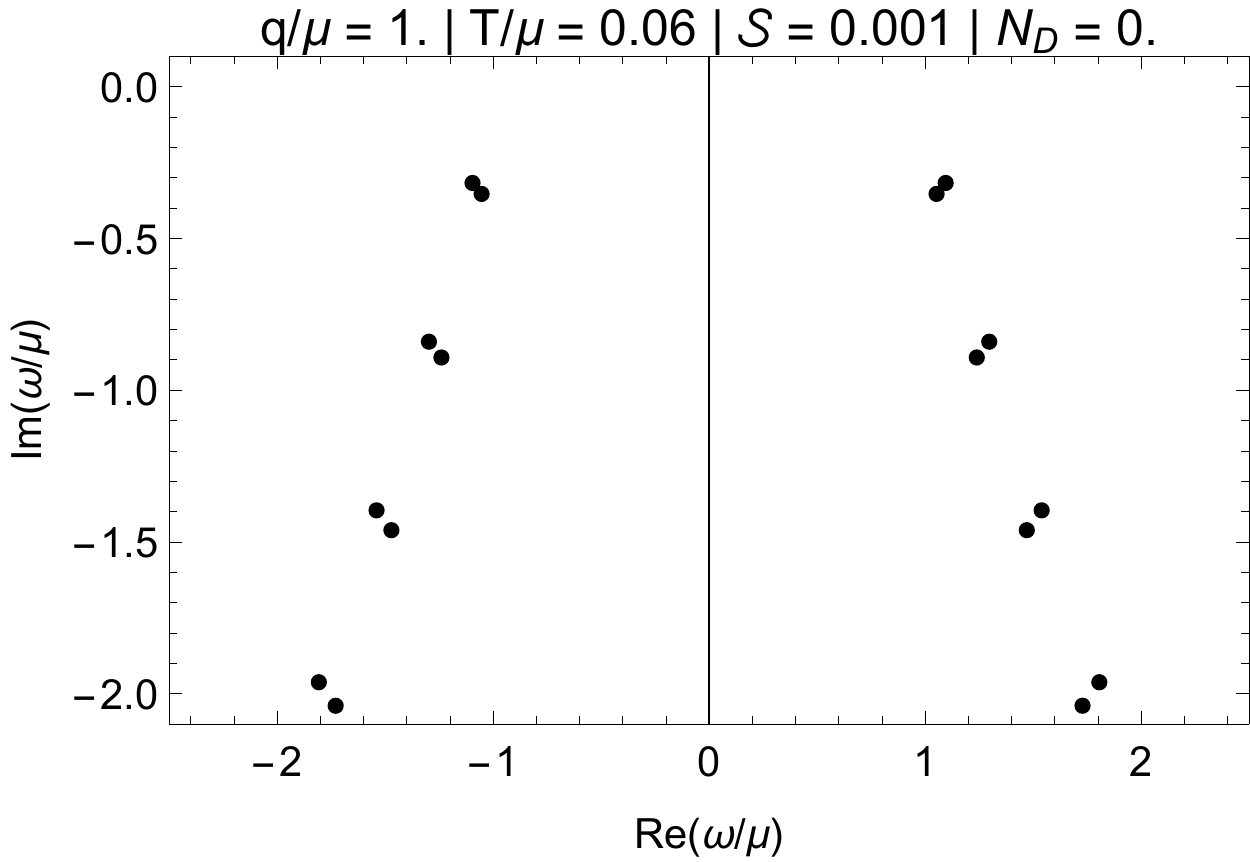}
\includegraphics[width=0.495\textwidth]{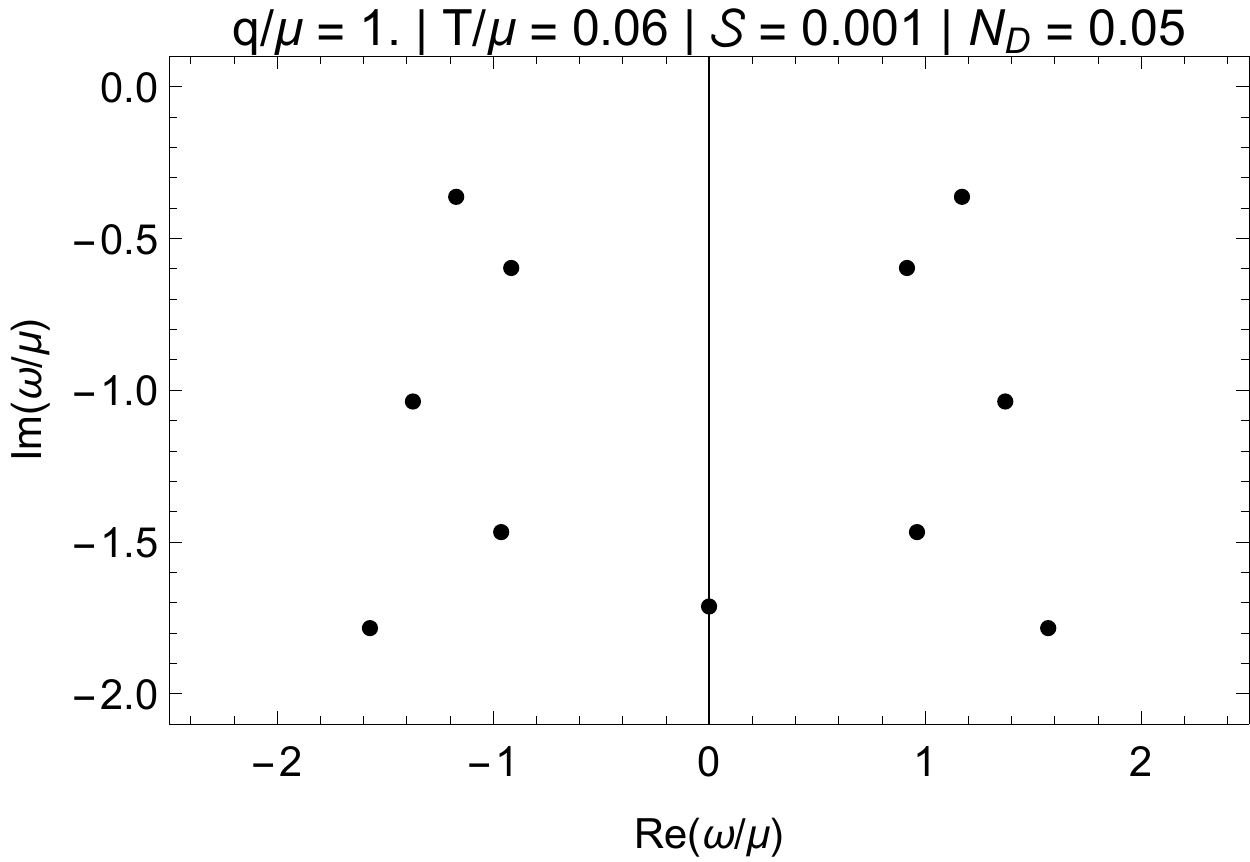}
\includegraphics[width=0.495\textwidth]{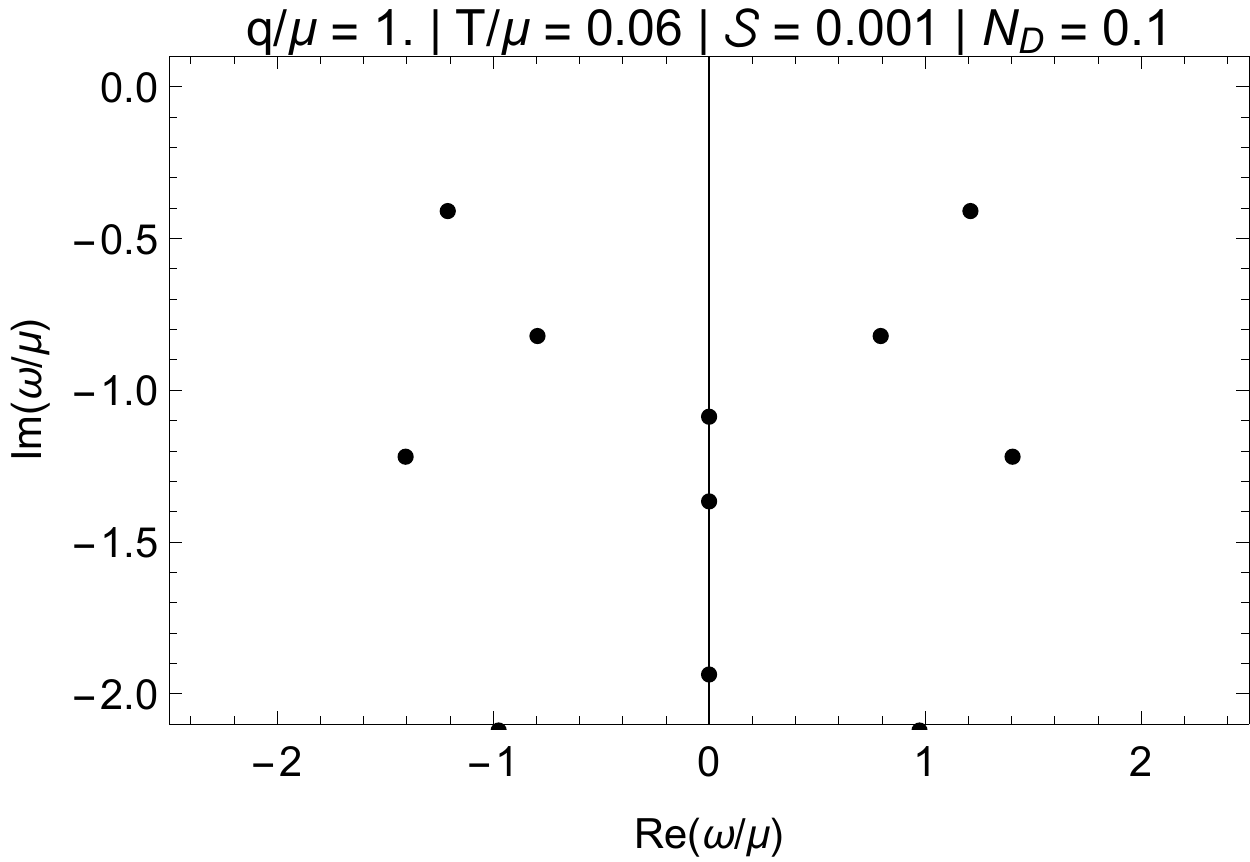}
\includegraphics[width=0.495\textwidth]{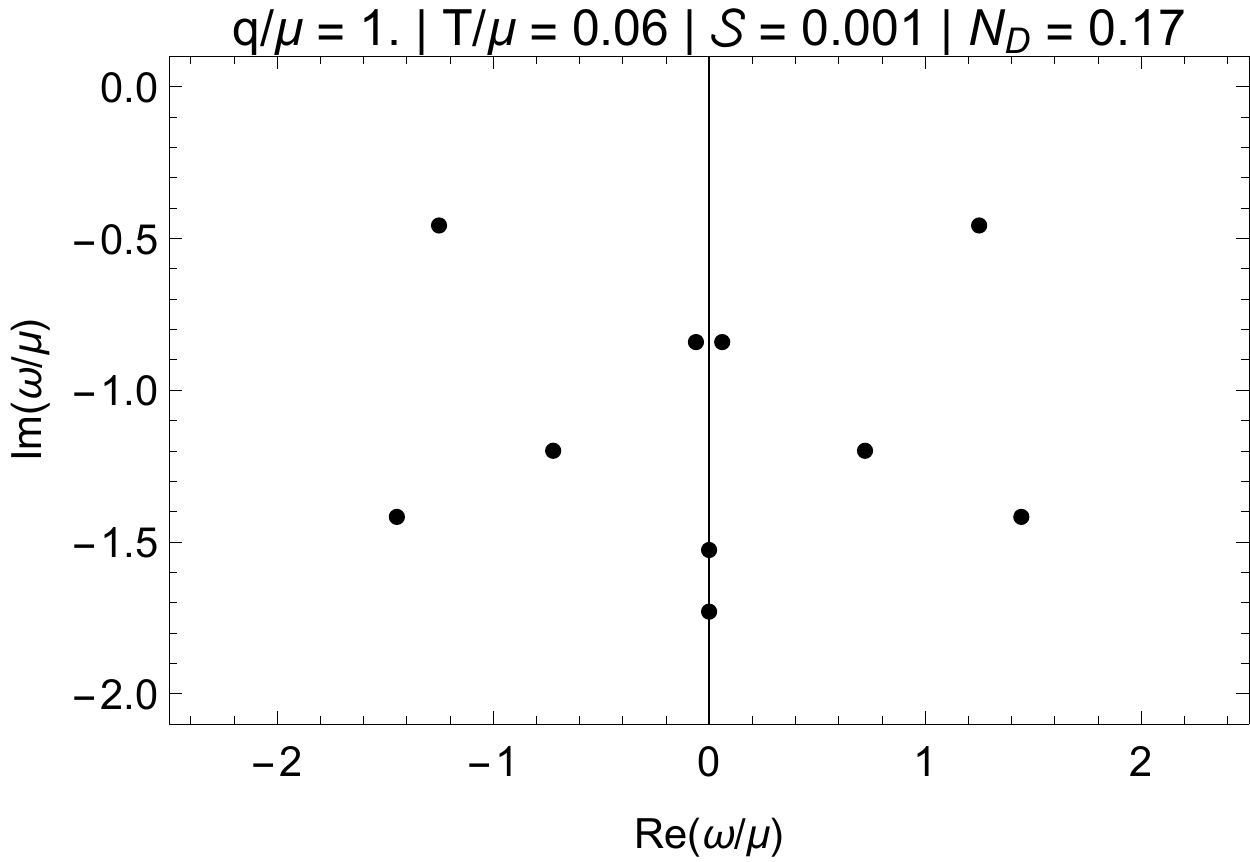}
\includegraphics[width=0.495\textwidth]{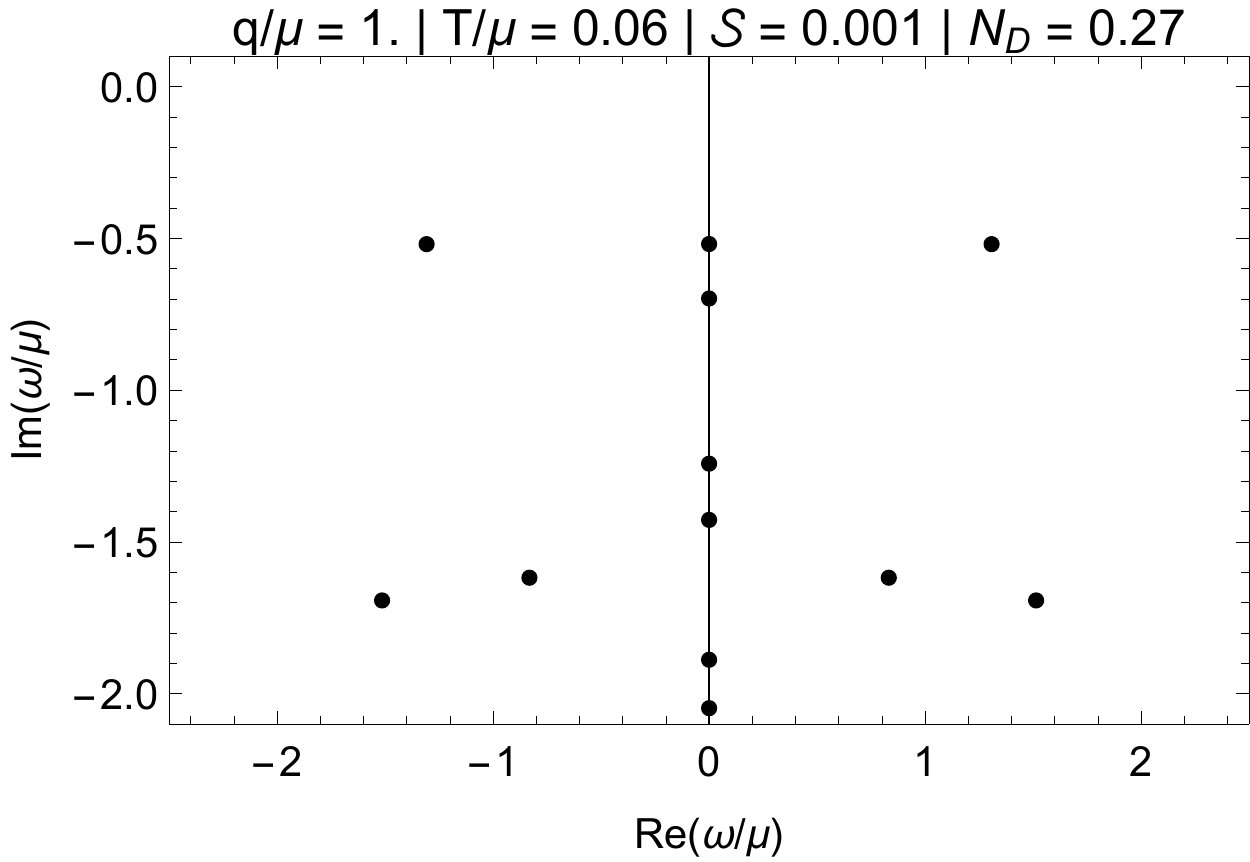}
\includegraphics[width=0.495\textwidth]{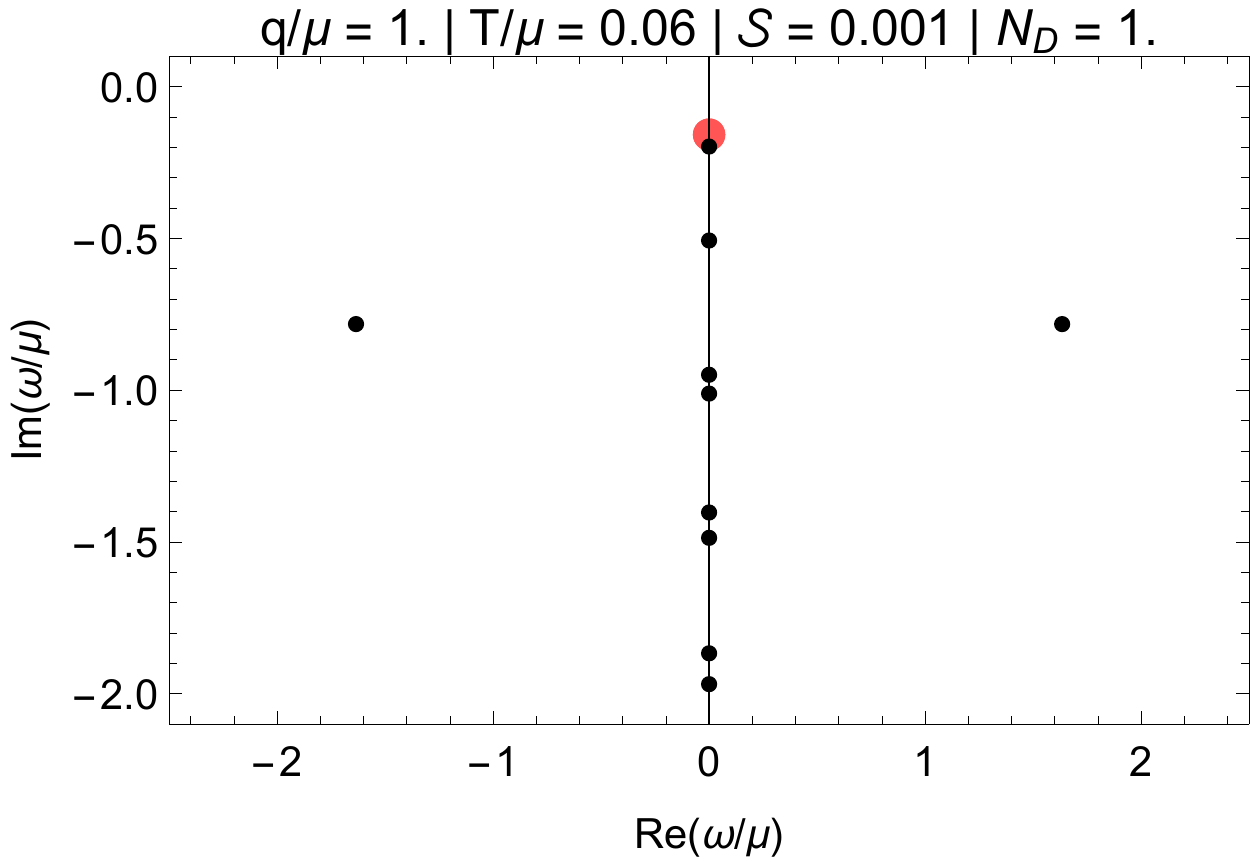}
\caption{\label{F8} The QNMs in the complex $\omega/\mu$ plane, at different values of the back-reaction parameter 
$N_D$, with fixed $T/\mu = 0.06$, $q/\mu=1$, and for $(\CS=0.001,~\CP=1,~\text{``Probe"}~ AdS_4RN)$. 
As we increase $N_D$, new QNMs appear  on the imaginary axis, getting closer together and moving up 
to the real axis. We also observe the ``attraction" and coalescence of the pairs of the imaginary modes 
for $N_D$ in the  range  $0.15\le N_D\le0.2$, with maximal separation between them at $N_D=0.17$. In the end, the
 QNMs on the imaginary axis form a dense structure, consistent with the formation of a branch 
 cut in the limit  $T/\mu \rightarrow 0$ (bottom right panel). (Animated
version of the figure is available on this paper's arXiv page.)}
\end{figure}
\begin{figure}[!tt]
\centering
\includegraphics[width=0.495\textwidth]{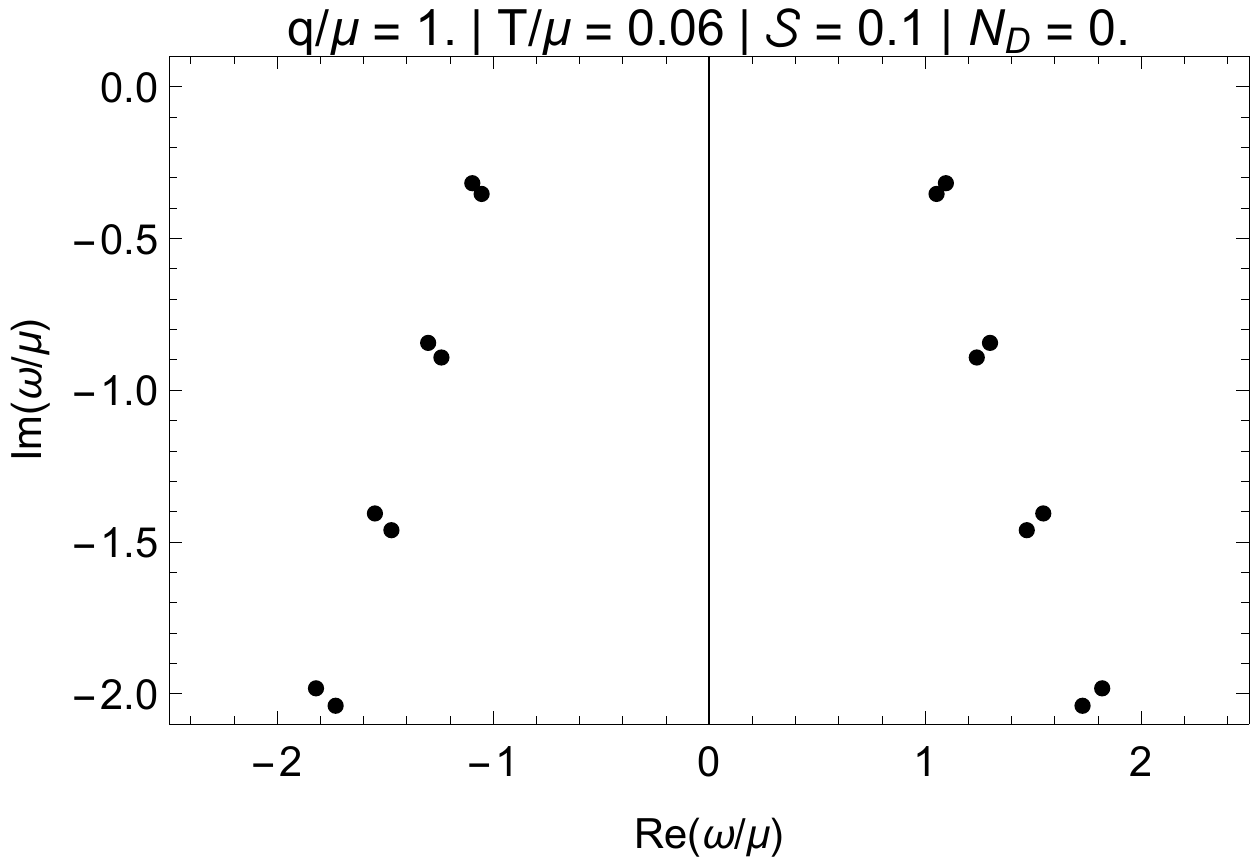}
\includegraphics[width=0.495\textwidth]{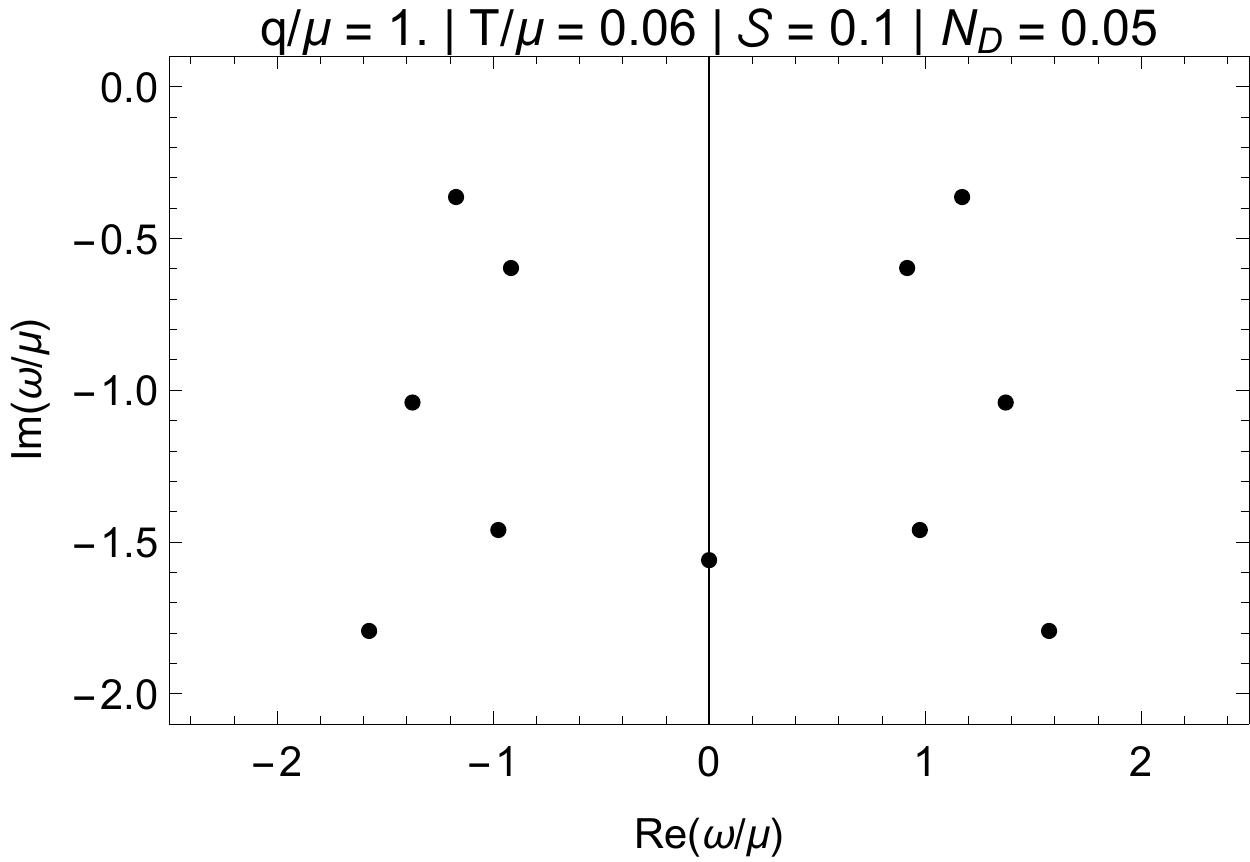}
\includegraphics[width=0.495\textwidth]{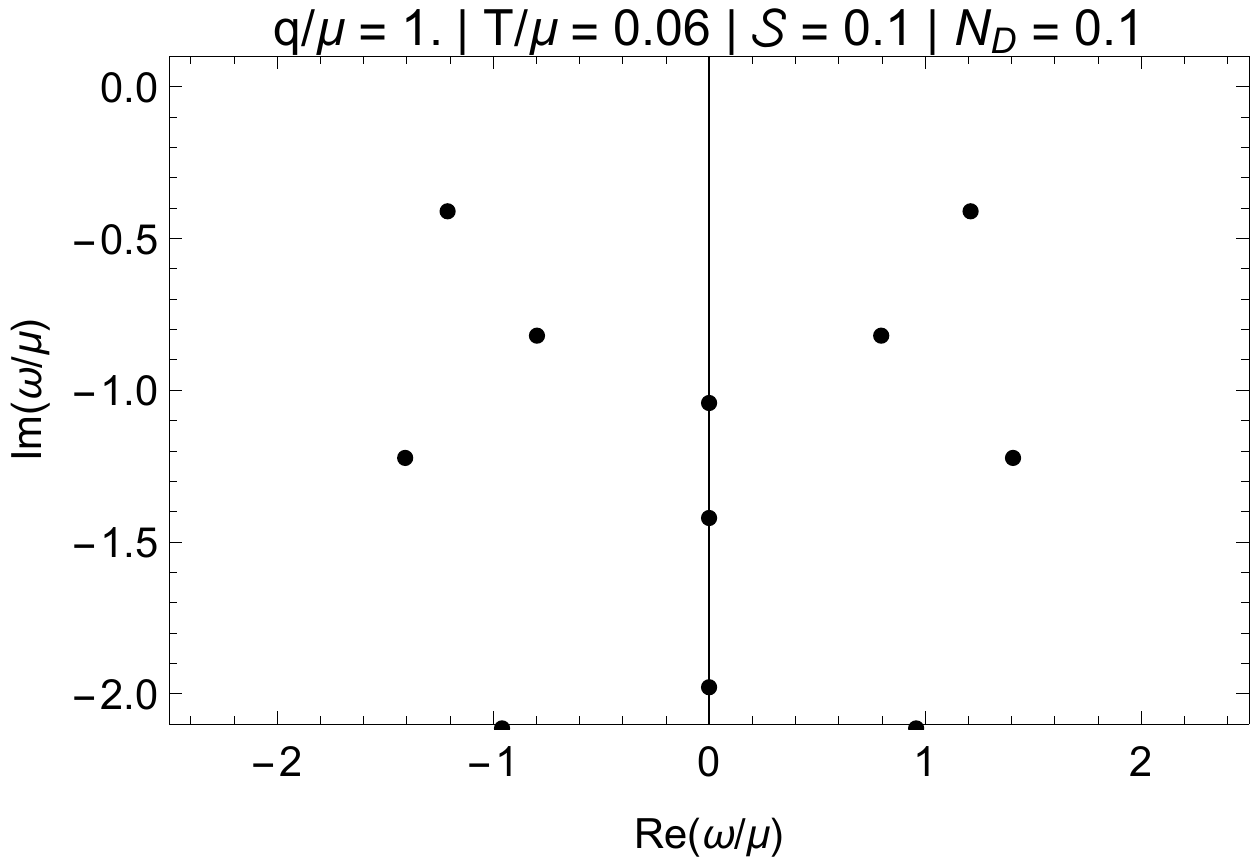}
\includegraphics[width=0.495\textwidth]{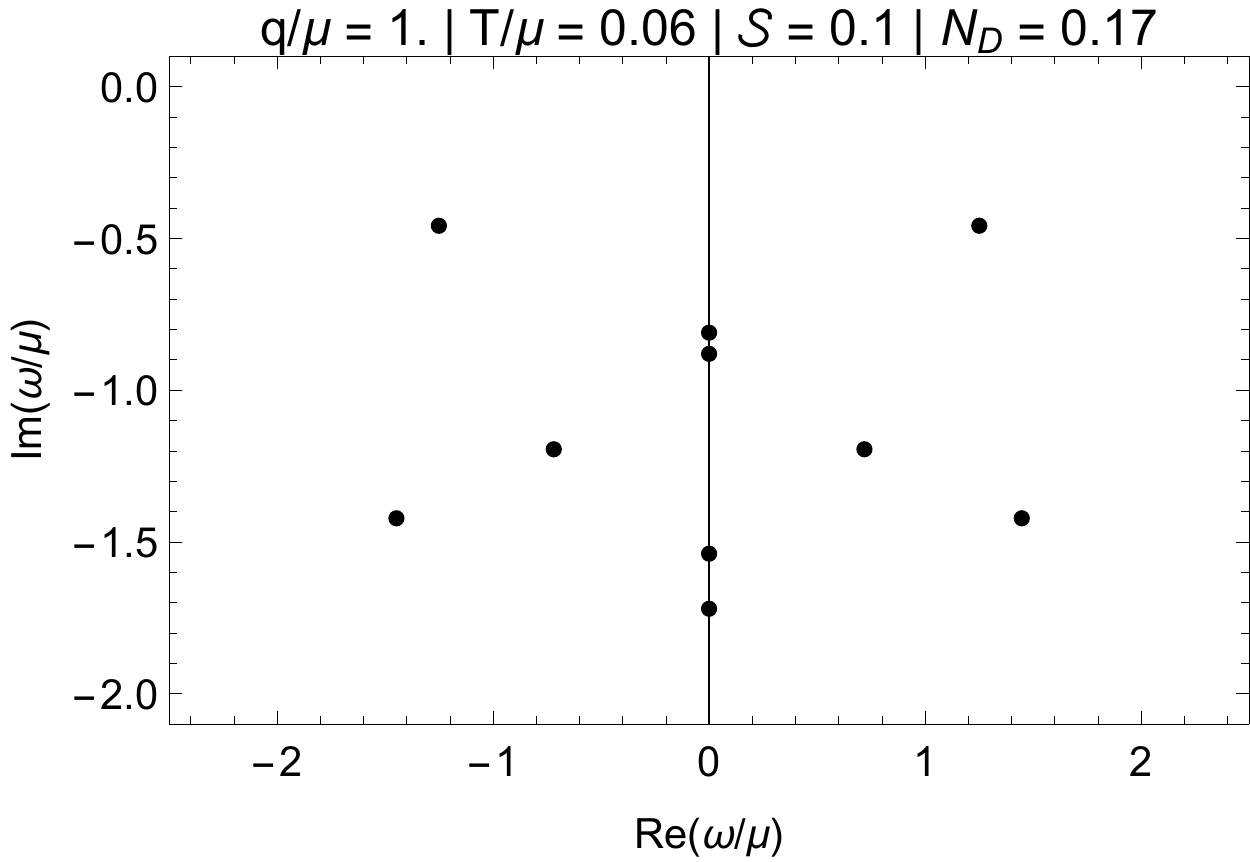}
\includegraphics[width=0.495\textwidth]{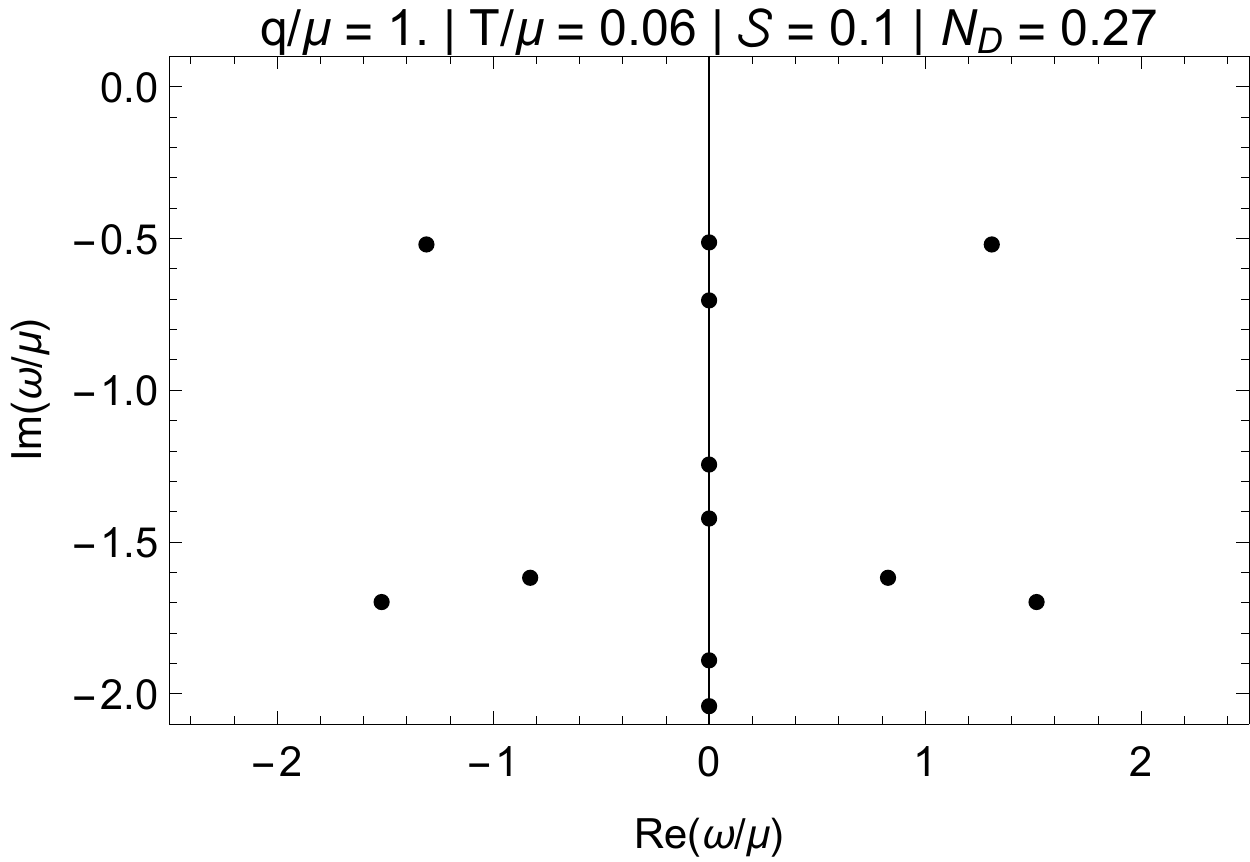}
\includegraphics[width=0.495\textwidth]{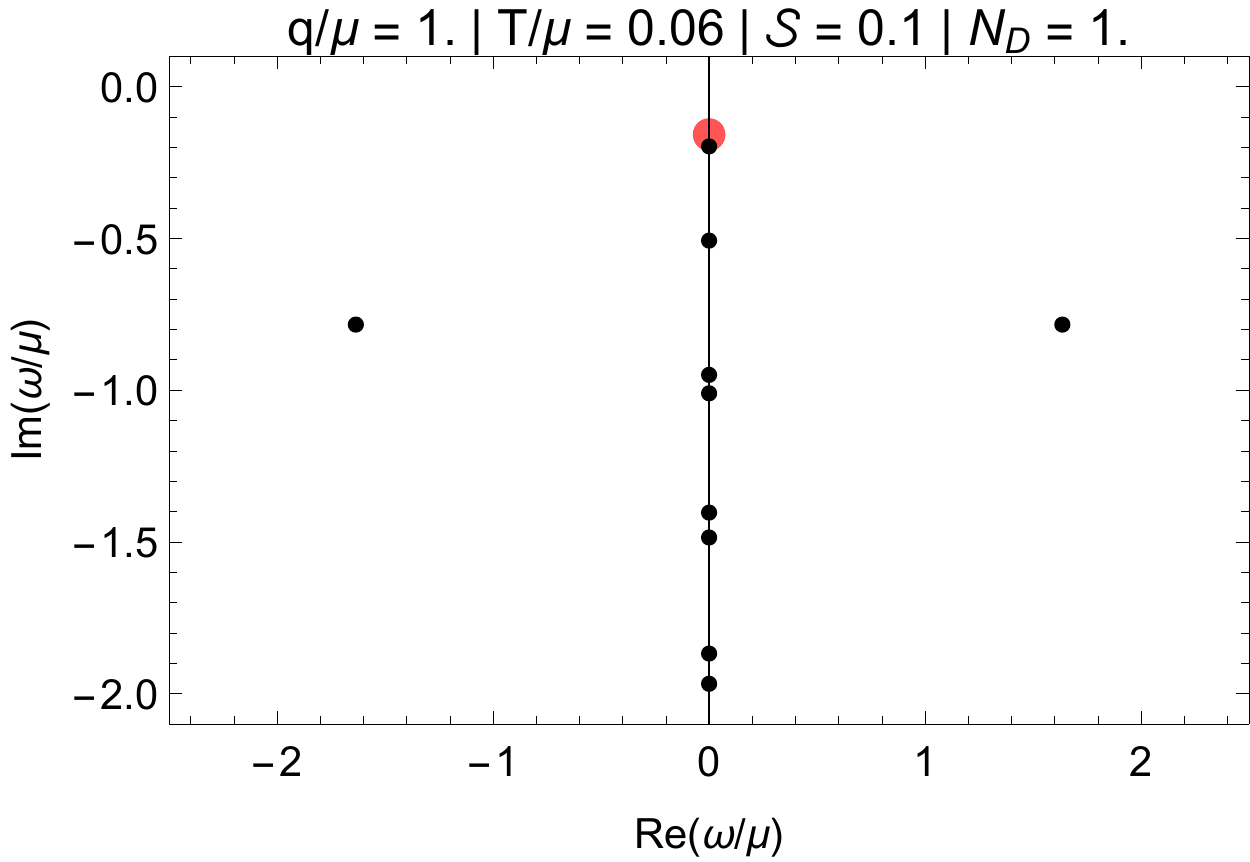}
\caption{\label{F9} The QNMs in the complex $\omega/\mu$ plane, at different values of the back-reaction parameter 
$N_D$, with fixed $T/\mu = 0.06$, $q/\mu=1$, and for $(\CS=0.1,~\CP=1)$. As we increase $N_D$, new QNMs appear  on the imaginary axis, getting closer together and moving up 
to the real axis. We also observe a ``repulsion" of the two imaginary modes in the range corresponding to $0.15\le N_D\le0.2$, with minimal separation between them at $N_D=0.17$ (middle right panel). In the end, the
 QNMs on the imaginary axis form a dense structure, consistent with the formation of a branch 
 cut in the limit  $T/\mu \rightarrow 0$ (bottom right panel). (Animated
version of the figure is available on this paper's arXiv page.)} 
\end{figure}
\begin{equation}
0.001\le\CS\le1,\quad 0\le\CP\le1,\quad 0\le N_D\le1,\quad q/\mu=1,\quad T/\mu=0.06.
\end{equation}

In Fig.~\ref{F8}, we fix $q/\mu=1$, $T/\mu=0.06$, $\CS=0.001,~\CP=1$, and vary the back-reaction parameter $N_D$ from in the interval $[0,1]$. This results in a rather complicated movement of QNMs in the complex plane, as a result of which most QNMs move deeper into the complex plane leaving two modes off the imaginary axis (bottom right panel in Fig.~\ref{F8}). At the same time, new QNMs appear on the imaginary axis (top right panel in Fig.~\ref{F8}, corresponding to $N_D=0.05$), and at even higher value of $N_D$,  the imaginary QNMs  move up until the two highest original modes ``feel the attraction" and coalesce. 
They split off the axis for a range $0.15\le N_D\le0.2$, and then they merge again on the imaginary axes. The point of maximum separation between them is shown in the right middle panel in  Fig.~\ref{F8}, corresponding to $N_D=0.17$. 
After they merge again, the two modes separate and the one closest to the origin becomes the diffusion mode, 
while simultaneously many pairs of new QNMs  move up the imaginary axis, starting a formation of what becomes a branch cut at $T/\mu=0$ (bottom right panel  in  Fig.~\ref{F8}). 

In Fig.~\ref{F9}, we repeat the same analysis, but now  with higher value of the non-linearity parameter 
$\CS=0.1,~\CP=1$. For small back-reaction ($N_D<0.15$), the picture is very similar to the linear case. The only major difference comes, again, for the range of the back-reaction parameter  $0.15\le N_D\le0.2$. While in the linear case we saw the ``attraction" and merging of the imaginary QNMs, we now observe a ``repulsion" between the same modes. The point of closest approach on the imaginary axis is shown in the middle right panel of Fig.~\ref{F8}(corresponding to $N_D=0.17$). 
Finally, once this interaction has ended at $N_D>0.2$, the picture is again very similar to the linear case.
The rest of the off-axis QNMs  seem to be unaffected by this change in non-linearity parameters.

\begin{figure}[!tt]
\centering
\includegraphics[width=0.495\textwidth]{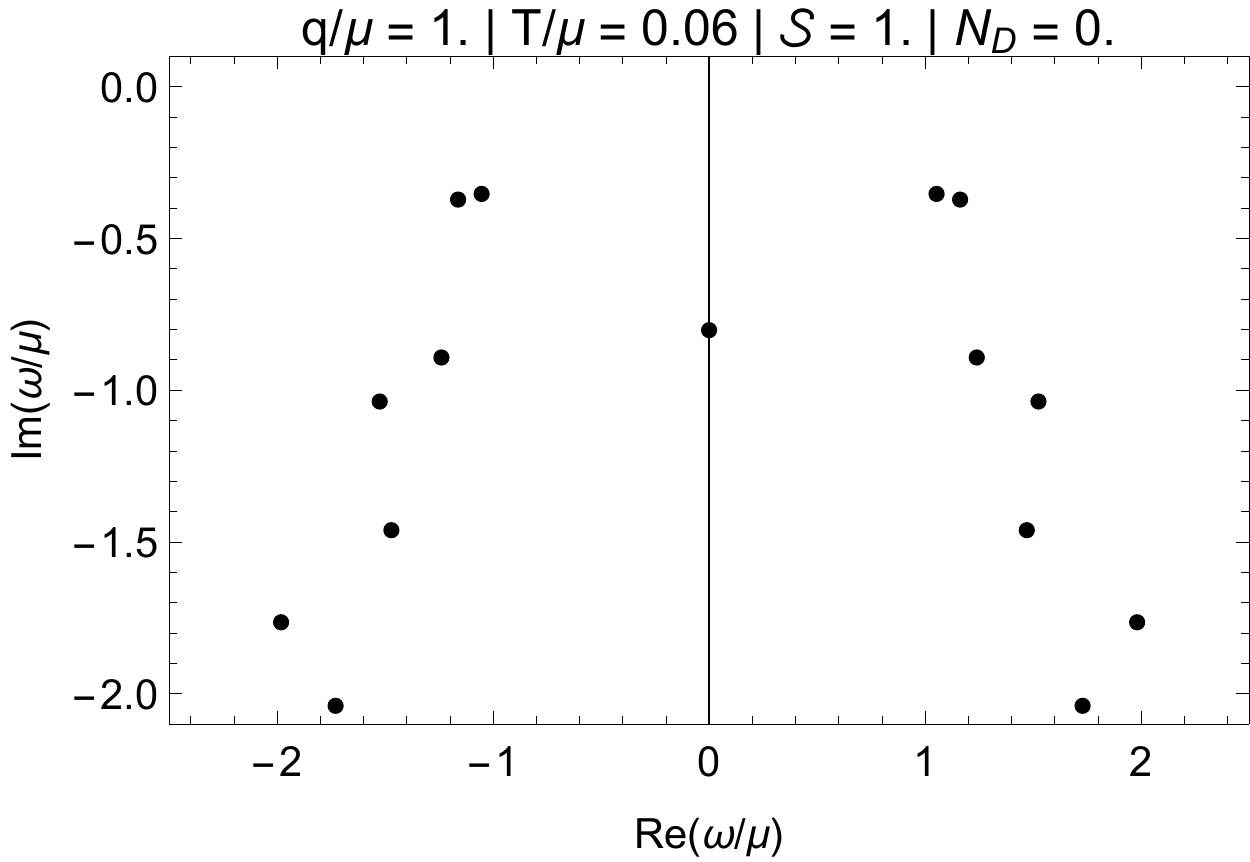}
\includegraphics[width=0.495\textwidth]{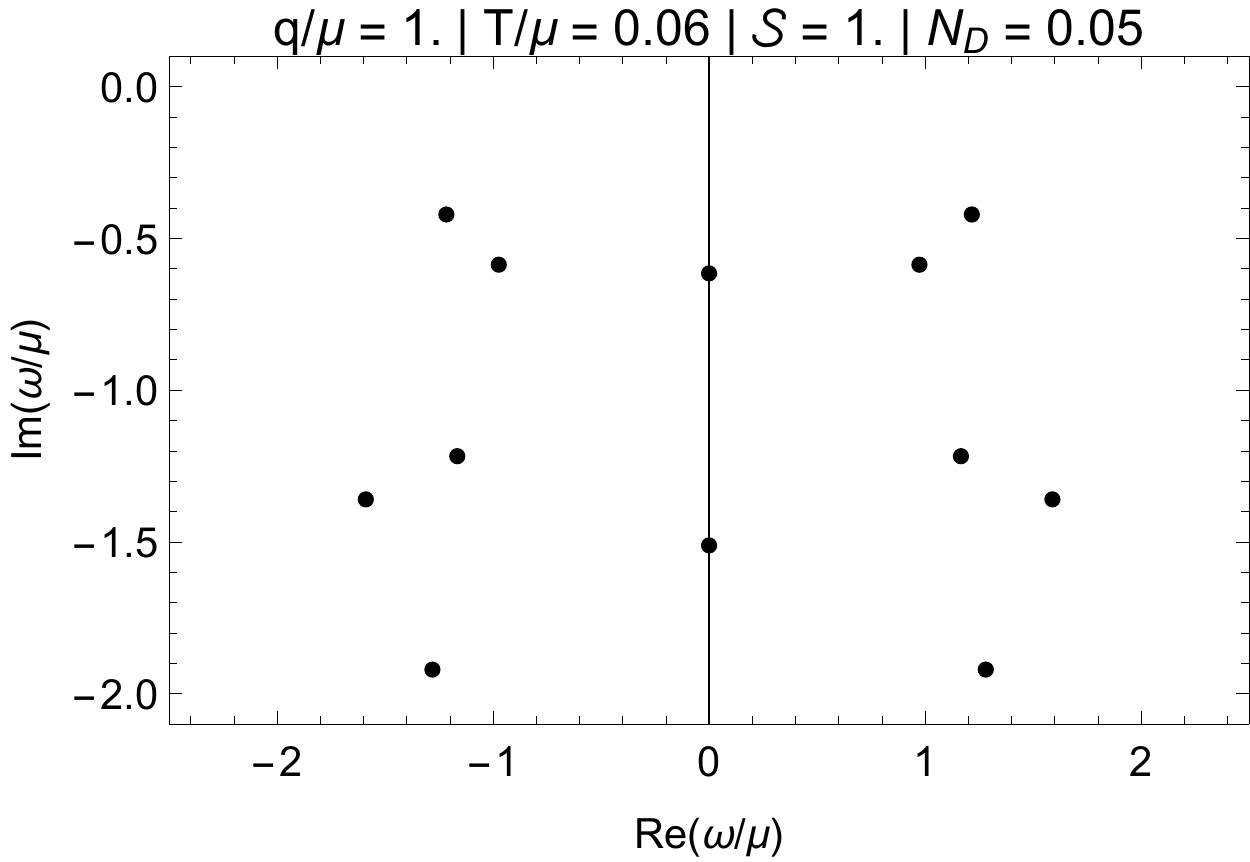}
\includegraphics[width=0.495\textwidth]{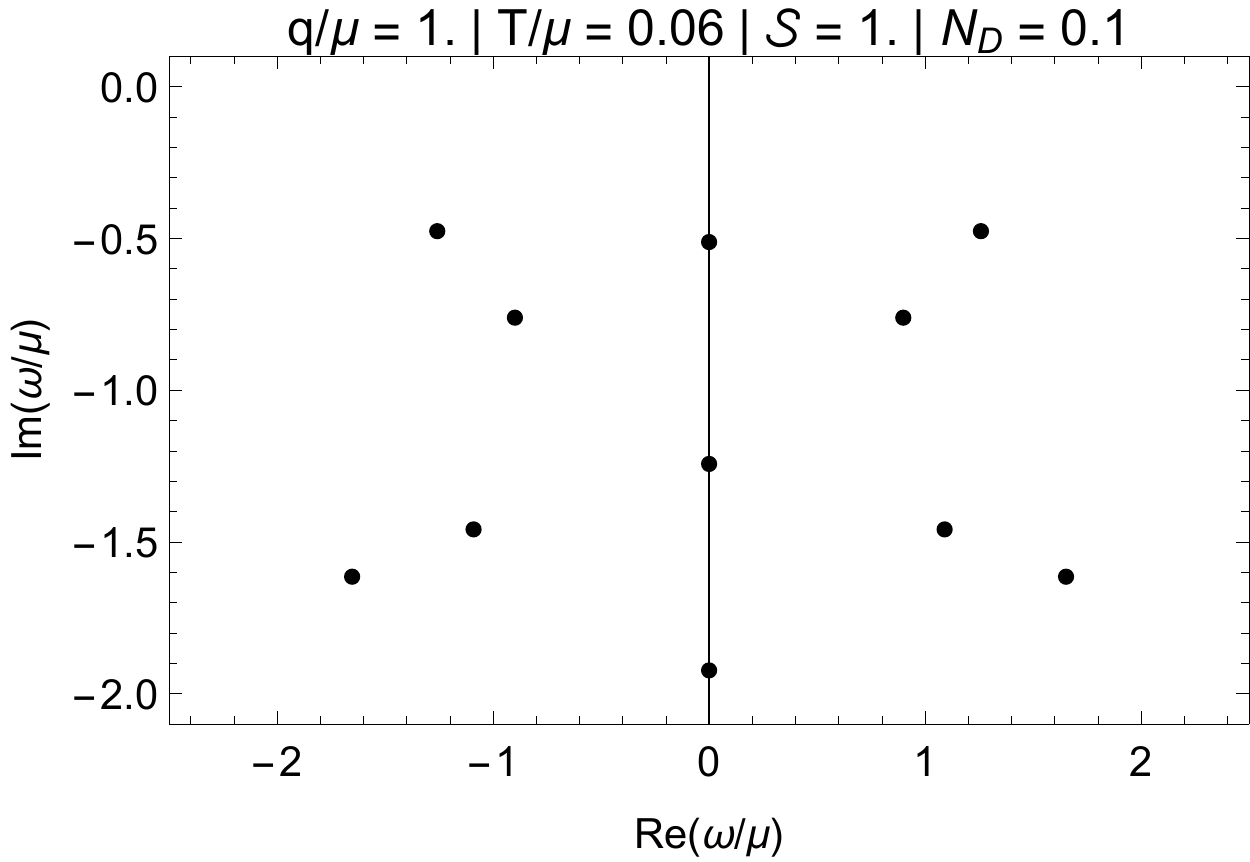}
\includegraphics[width=0.495\textwidth]{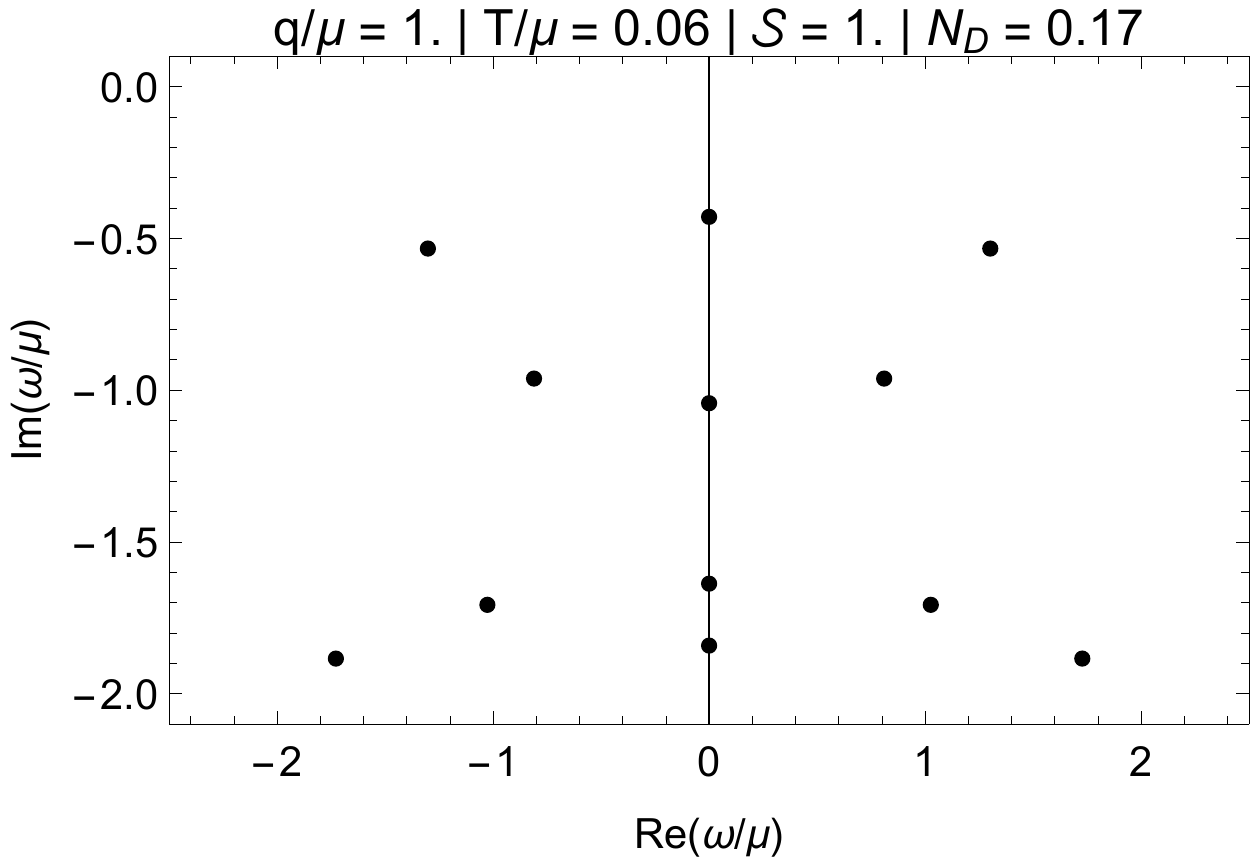}
\includegraphics[width=0.495\textwidth]{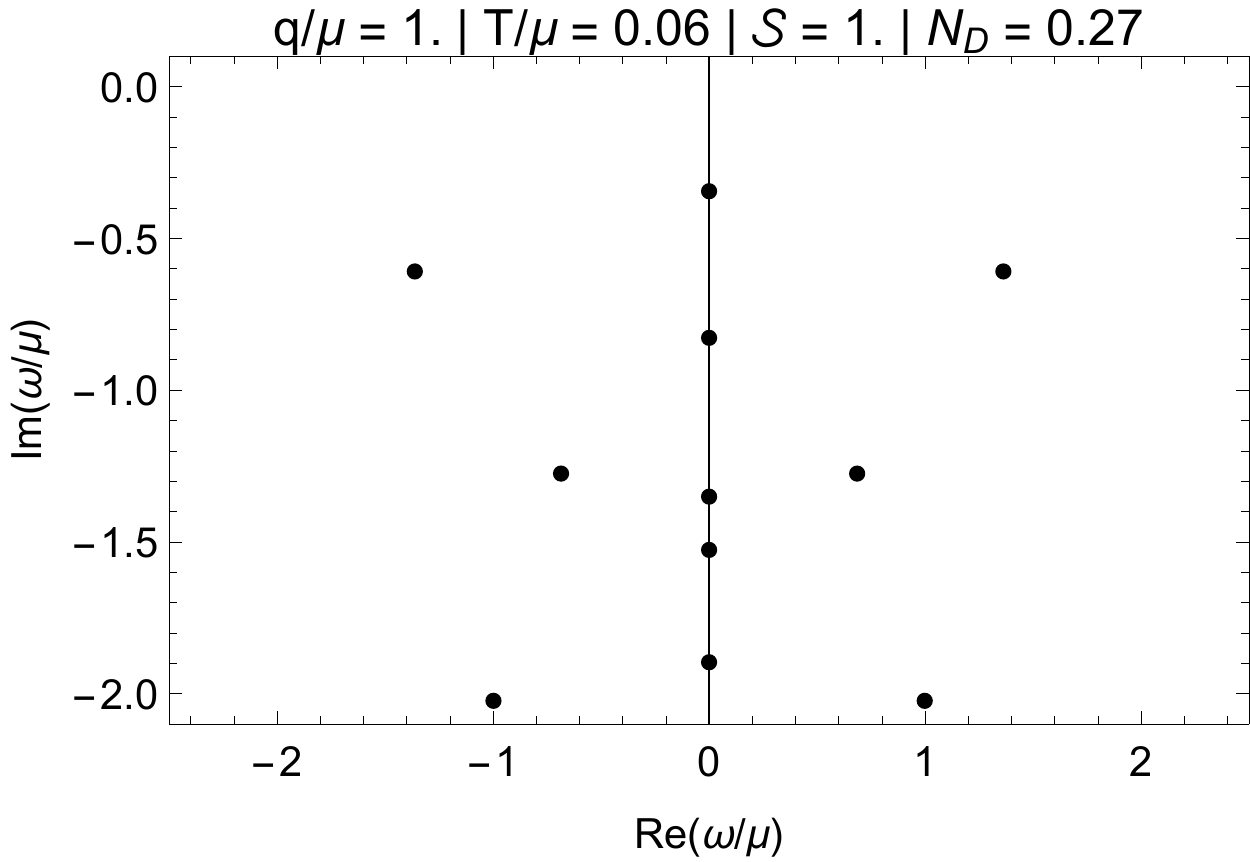}
\includegraphics[width=0.495\textwidth]{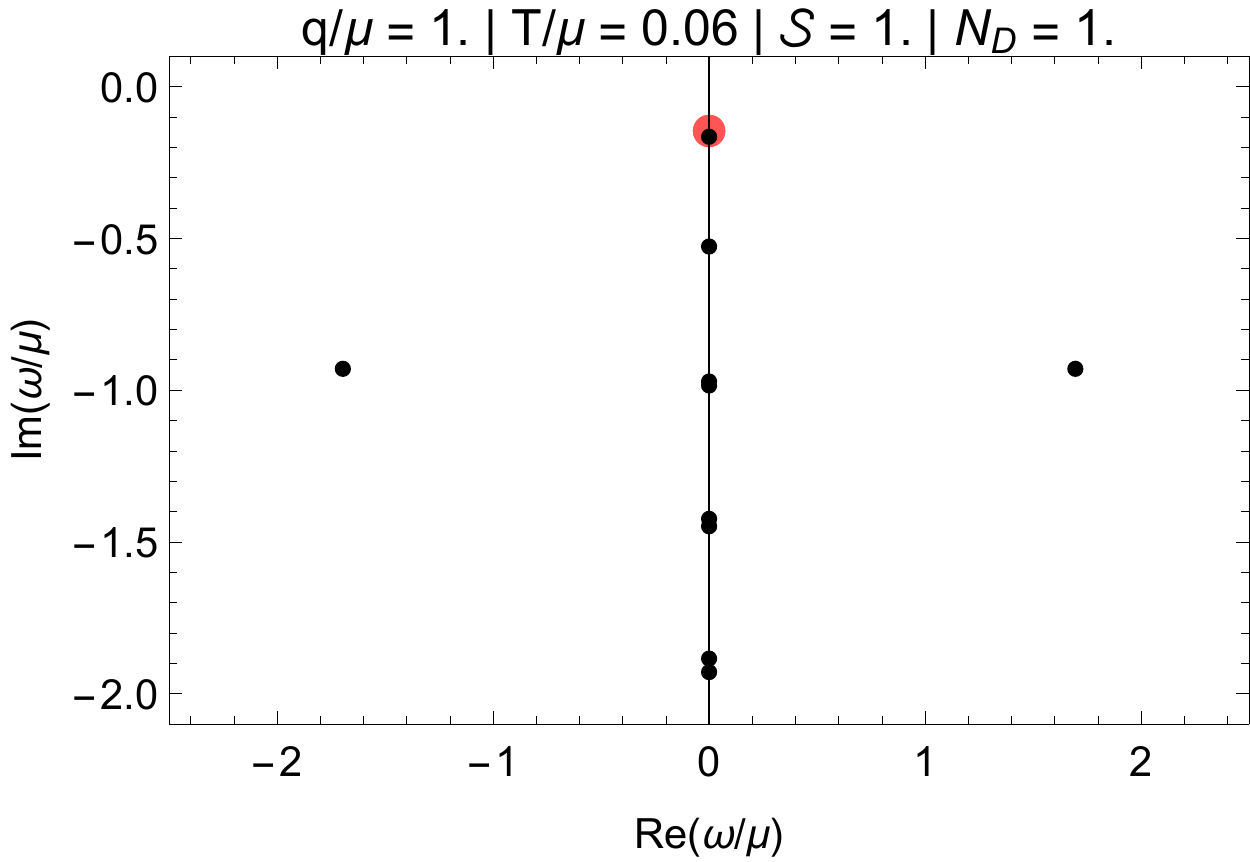}
\caption{\label{F10} The QNMs in the complex $\omega/\mu$ plane, at different values of the back-reaction parameter 
$N_D$, with fixed $T/\mu = 0.06$, $q/\mu=1$, and for $(\CS=1,~\CP=1,~\text{``Probe"}~ AdS_4DBI)$. A a purely 
imaginary QNM is present already for $N_D=0$. As we increase $N_D$, 
new QNMs appear  on the imaginary axis, getting closer together and moving up 
to the real axis. 
 We also observe a ``repulsion" of the two imaginary modes for essentially all range of $N_D$ we are exploring. We observe that eventually they stay at a fixed ``equilibrium" distance while the rest of the imaginary QNMs pair up and 
 form a dense structure, consistent with the formation of a branch 
 cut in the limit  $T/\mu \rightarrow 0$ (bottom right panel). (Animated
version of the figure is available on this paper's arXiv page.)}
\end{figure}

Finally, in Fig.~\ref{F10} we explore the back-reaction effect on the fully non-linear system with $\CS=1,~\CP=1$ in the same range of parameters as before. Now, the structure of the QNMs is completely changed by the non-linearities, for all the range of $N_D$ we have explored.

Starting at $N_D=0$ (top left panel in Fig.~\ref{F10}), we observe that the strong non-linearities change the off-axis QNMs considerably, in comparison with  the linear case. Very importantly, we observe a purely imaginary QNM, which would correspond to the diffusive mode in the hydrodynamic limit. As we  increase $N_D$, we observe that the diffusive mode and the next 
imaginary mode right below it get closer together, but the ``repulsion'' keeps the distance between them finite. Meanwhile, several other QNMs appear  on the imaginary axis in pairs; this time, they  get closer together than in the linear case. 

\begin{figure}[!tt]
\centering
\includegraphics[width=0.495\textwidth]{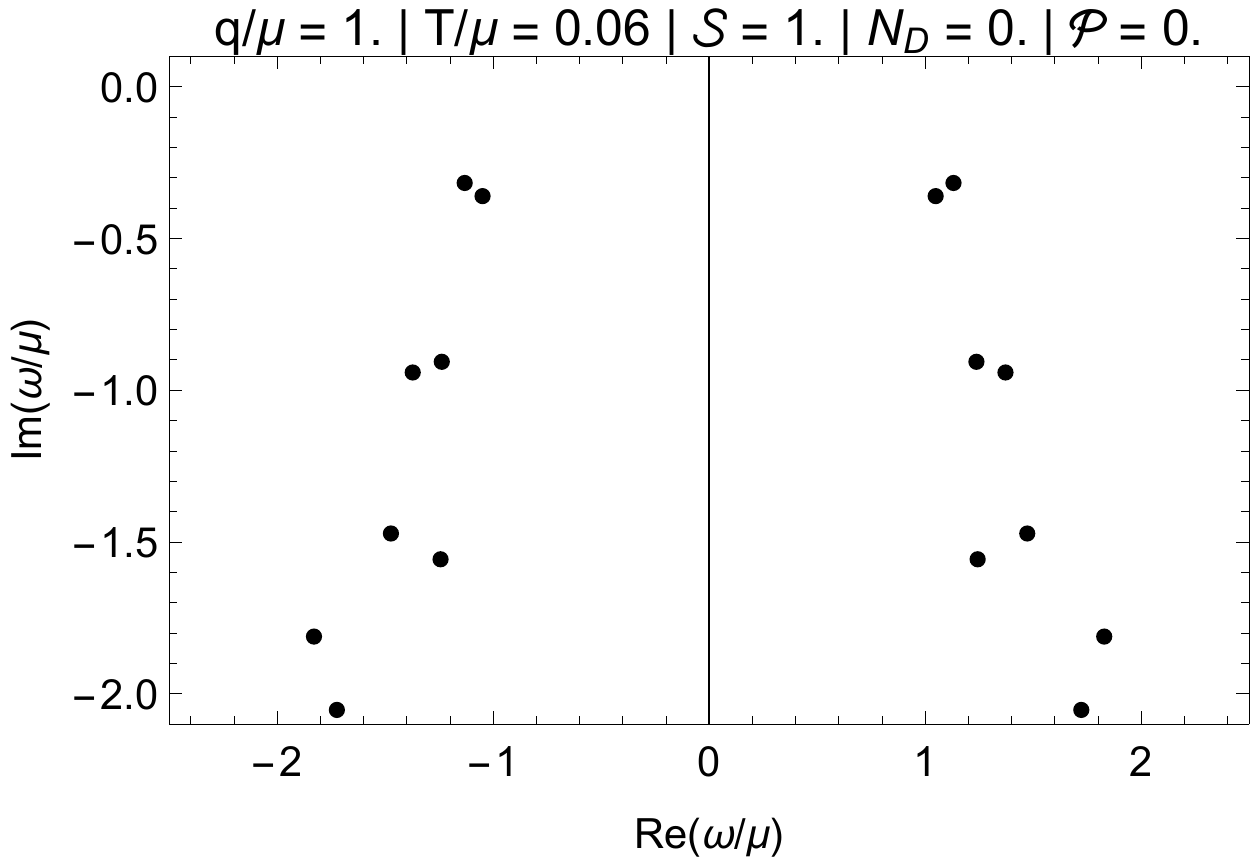}
\includegraphics[width=0.495\textwidth]{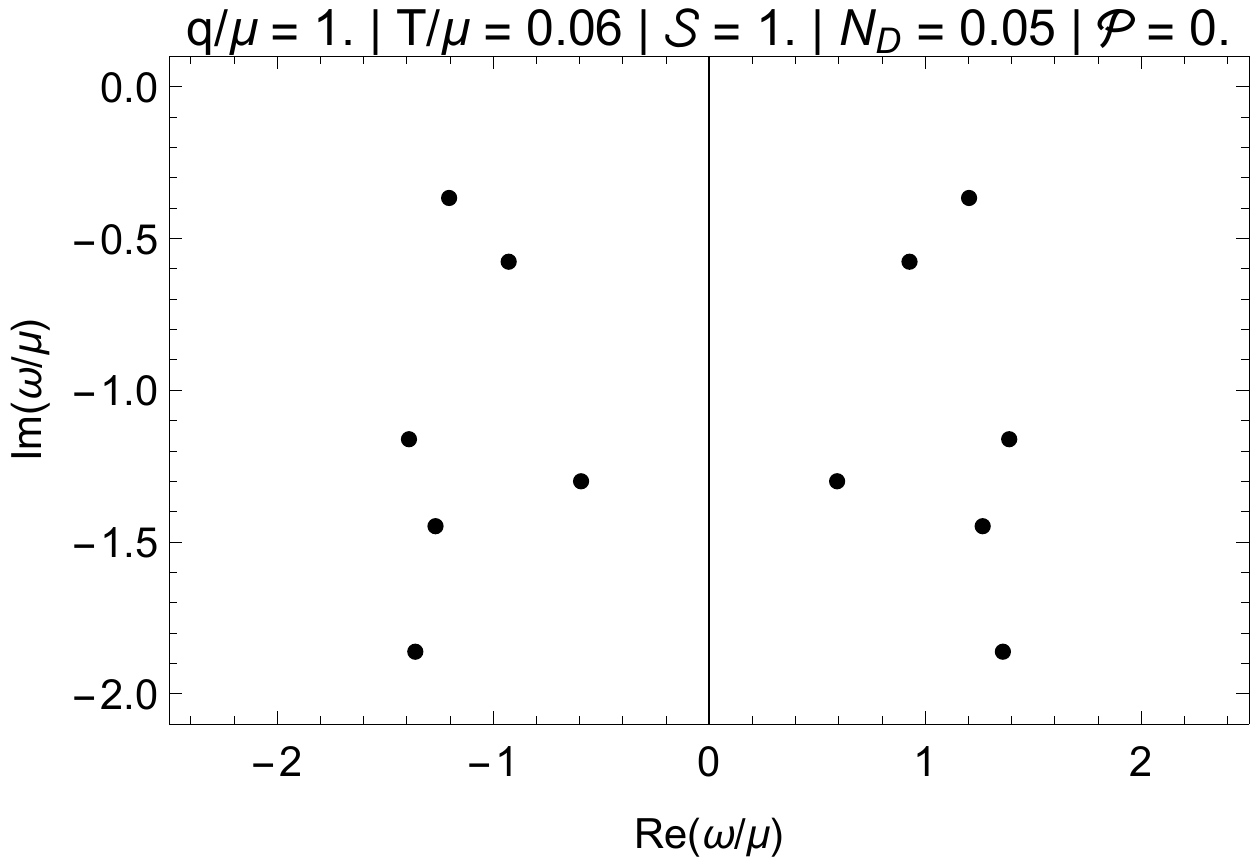}
\includegraphics[width=0.495\textwidth]{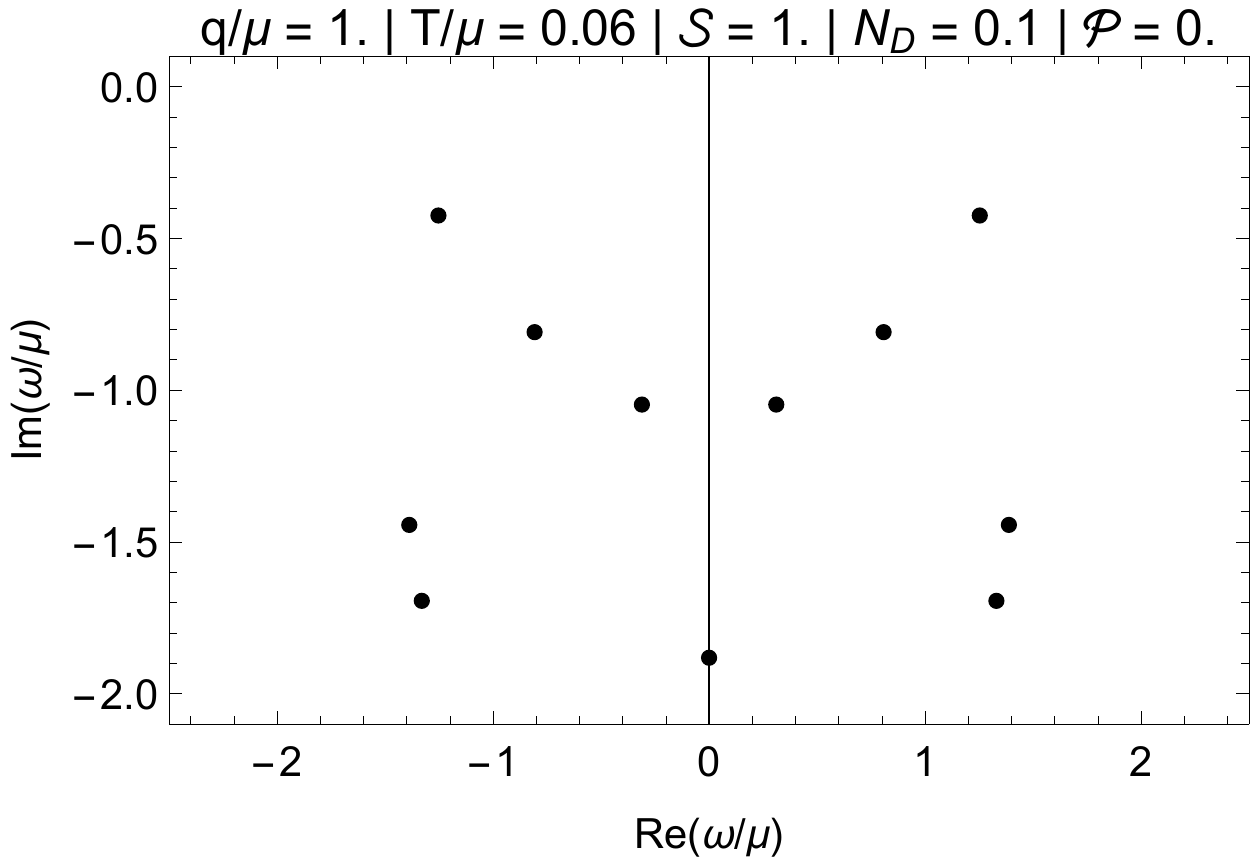}
\includegraphics[width=0.495\textwidth]{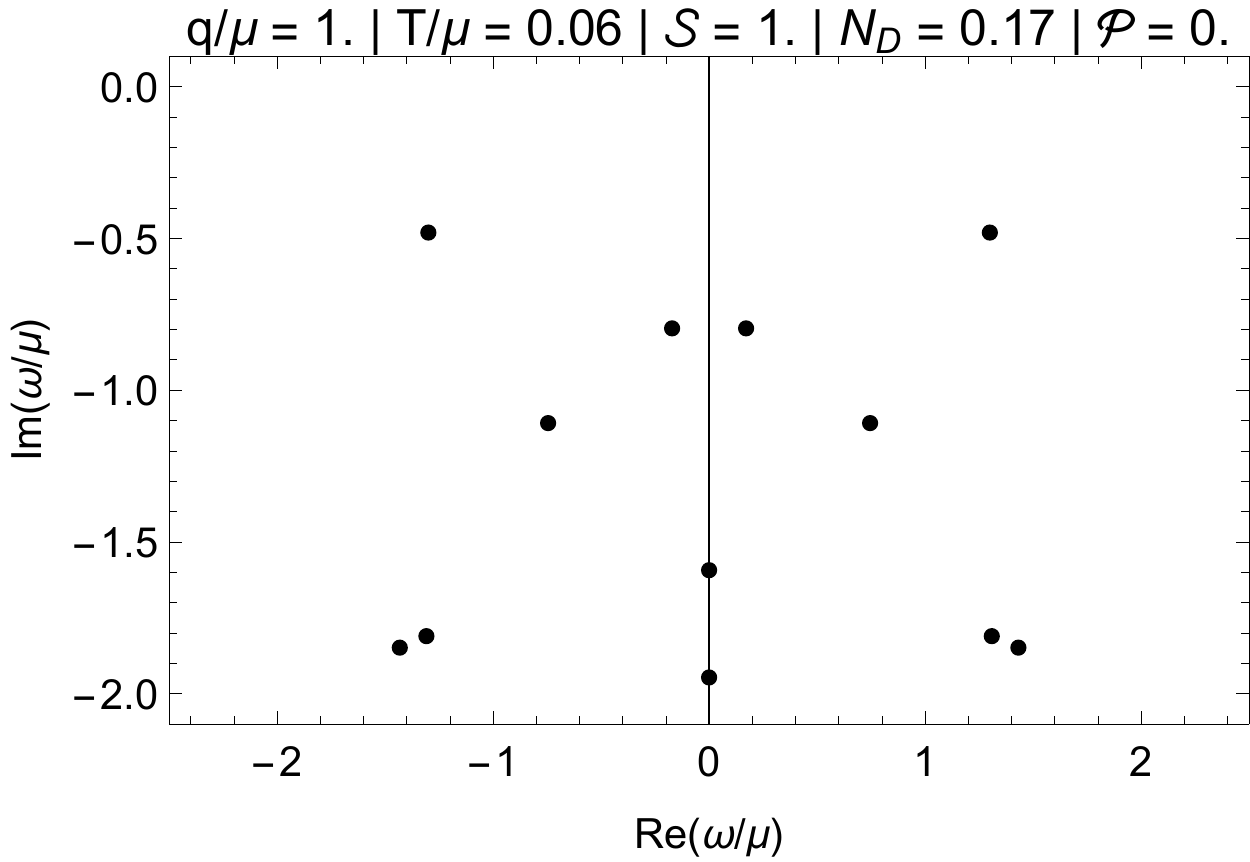}
\includegraphics[width=0.495\textwidth]{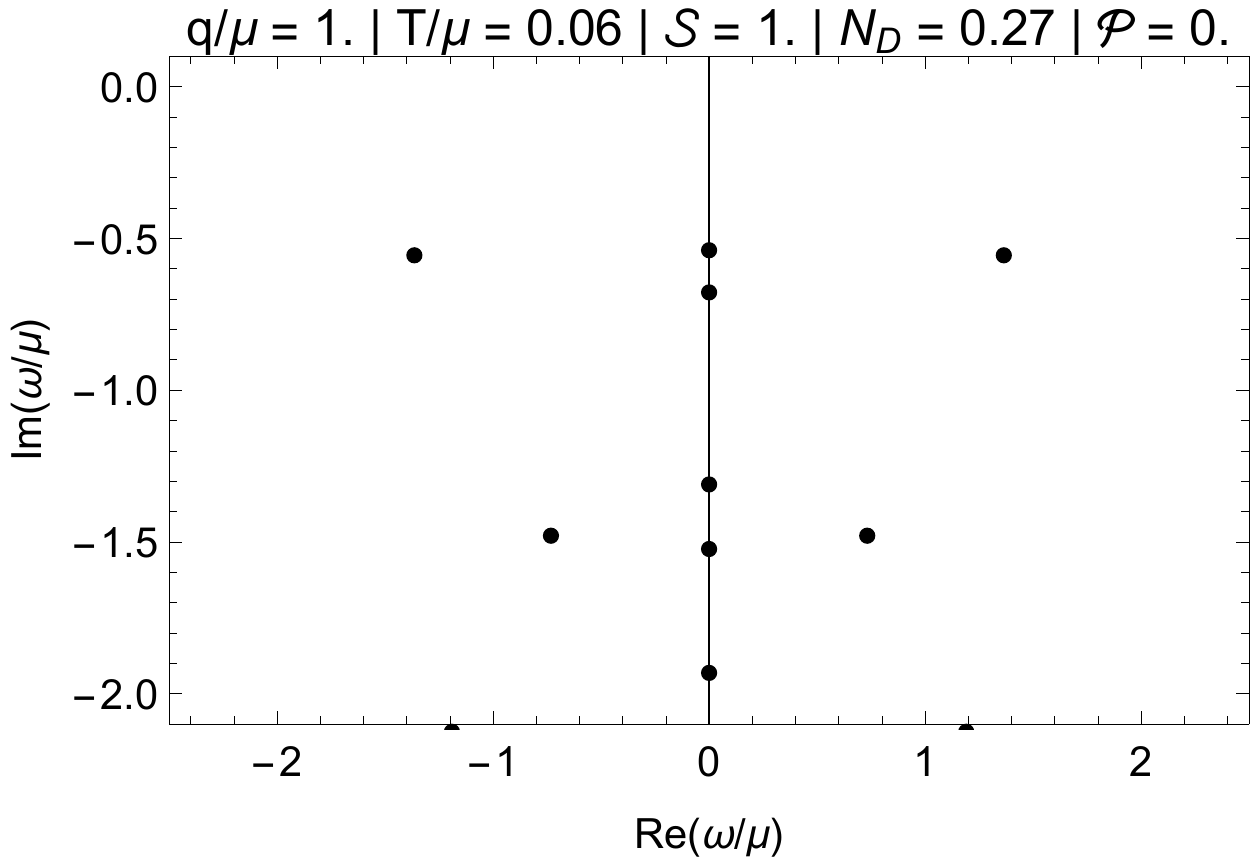}
\includegraphics[width=0.495\textwidth]{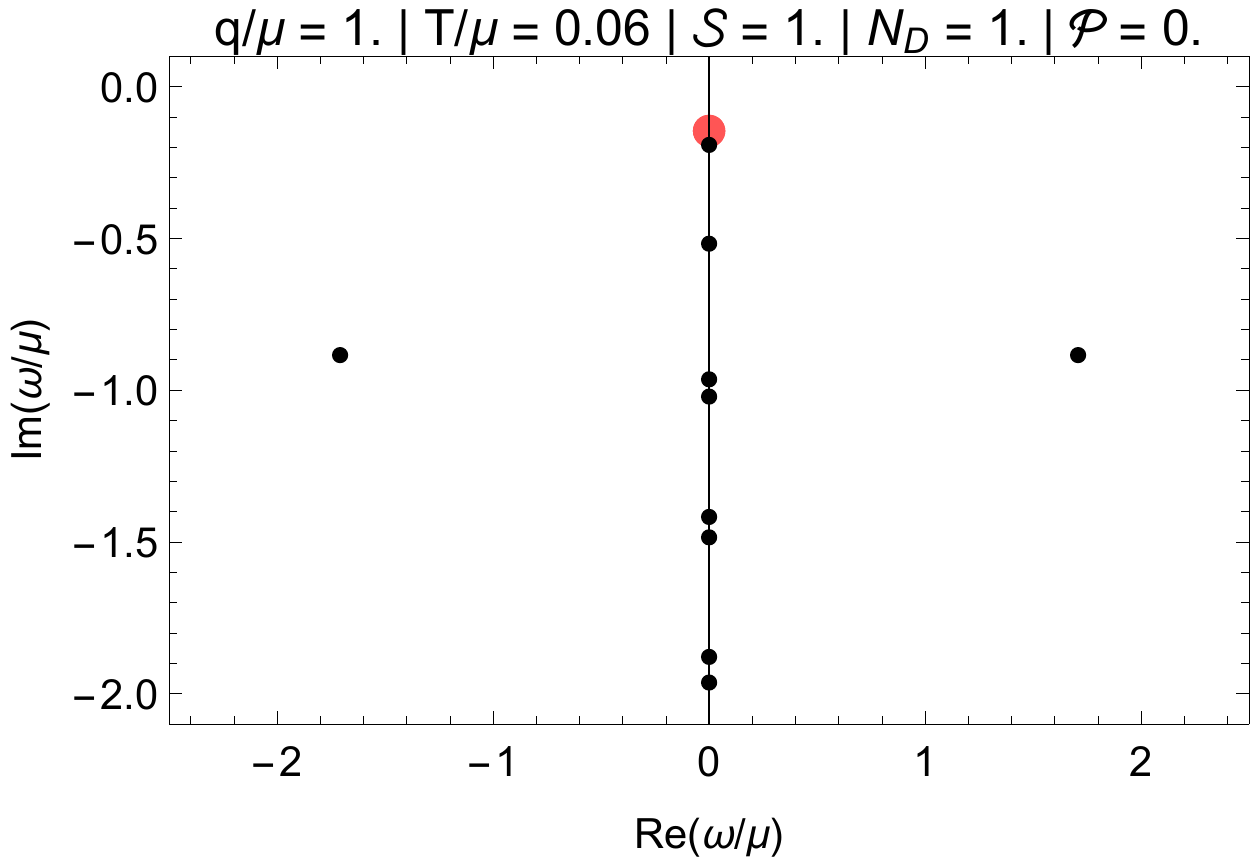}\caption{\label{F11} The QNMs in the complex $\omega/\mu$ plane,
at different values of the back-reaction parameter 
$N_D$, with fixed $T/\mu = 0.06$, $q/\mu=1$, and for $(\CS=1,~\CP=0)$. In comparison to Fig.\ref{F10}, we observe several differences such as the disappearance of the purely imaginary mode at $N_D=0$, and also the absence of the ``repulsion" between the two imaginary modes, which now start off the imaginary axis and merge at $N_D\approx0.27$, to eventually become part of the forming branch cut and the diffusion mode.}
\end{figure}

The presence of the diffusive mode  at $N_D=0$, as well as the ``repulsion" between imaginary modes can be traced back 
to the $\CP$ term,  as described in the previous section. Setting $\CP=0$ and repeating the analysis, we observe that the formation of the branch cut occurs in  a completely different way than in the $\CP=1$ case (see Fig.~\ref{F11}).
In particular, the imaginary mode at $N_D=0$ is no longer present, and the rest of the off-axis QNMs 
look much more similar to the probe linear case. This is yet another example where we see that the term proportional to $\CP$ has a very noticeable effect on the QNMs. This effect is more significant than the one due to the presence of the 
 other non-linearity controlled by $\CS$. 
 
 Finally, the two modes that eventually become the top of the forming branch cut, are already split up and not located on the imaginary axis. They move up off the imaginary axis and at $N_D\approx0.27$ (bottom left panel in Fig.~\ref{F11}) merge on the imaginary axis and become two least  damped modes on the axis, one of them being the diffusion mode. The ``repulsion" effect also seems to make the hydrodynamic description more accurate, since $\Delta(0)\geq\Delta(\CP)$. 

\section{Summary and Discussion}\label{summary}
In this paper, we have explored the effect of non-linearities in the shear channel of a spacetime-filling brane, described by $AdS_4 DBI$, which we deformed with an extra parameter $\CP$, to keep track of the term proportional to $\CP \det (F_{\mu\nu})$ in the Lagrangian.

In section 2, we derived several thermodynamic quantities associated with the system and showed that it is thermodynamically stable, in agreement with ref.~\citep{Tarrio:2013tta}. We showed that the thermodynamics is independent of the parameter $\CP$. 
We also discussed  the probe limit of the entropy density and correctly recovered the results of the probe brane models \citep{Karch:2008fa}. 

In section 3, we derived the equations of motion for the linearized gauge-invariant fluctuations of the dual gravity background, 
which we used to compute the Green's functions of the energy-momentum tensor $T^{ab}$ and the global U(1) current $J^{a}$ in the hydrodynamic approximation. We observed that while the equations for the fluctuations do have a contribution from the parameter $\CP$, it does not affect the quadratic part of the diffusion dispersion relation
\begin{equation}
\omega(q)=-i{\cal D} q^2-i\Delta({\cal P}) q^4+{\cal O}(q^6),
\end{equation}
because it is completely determined by thermodynamics through the universality of $\eta/s$. 

We observed that the quartic correction  $\propto \Delta(\CP)q^4$ decreases when we increase $\CP$,  $\Delta(0)\geq\Delta(\CP)$. In the  limit $\CS\ll 1$ (refs.~\citep{Erdmenger:2008rm, Banerjee:2008th}), 
or in the large temperature limit $T\gg \mu$ (ref.~\citep{Grozdanov:2015kqa}), there is a closed form expression 
for $\Delta$ as a  linear combination of transport coefficients. In our case this combination seems to depend explicitly on the microscopic details of the theory and therefore can not be determined by hydrodynamics or be universal.

This raises the question of whether one could reverse the logic to find possible universal coefficients, apart from the ones already known, in a given holographic system. For example, if one could use a parameter such as $\CP$ to deform the original Lagrangian by a term that vanishes on-shell and sources the E.o.M. for the fluctuations (without being a boundary term and changing the holographic dictionary), then one could argue that the transport coefficients dependent on $\CP$ cannot be universal. However, if    one  could find some higher order transport coefficient (or a combination of them) that are $\CP$-independent, one may argue that those could be universal. As long as one can find a good candidate for the extra term, one could then use this approach to identify - numerically or analytically - possible new universal relations in higher order hydrodynamics, similar to the the ones proposed in \citep{Erdmenger:2008rm,Haack:2008xx,Grozdanov:2016fkt}.

In section 4, we performed numerical studies of the general properties of the QNMs (the poles of the two-point functions involving the transverse components of the energy-momentum tensor and  a $U(1)$ current operators),    as well as the corresponding  spectral functions. In particular, we demonstrated that the hydrodynamic result for the momentum diffusion coefficient $\CD$, derived in the $T \gg \omega$ limit, is still valid even for $T \lesssim \omega$, as long as we are in the large charge density limit $\omega \ll \mu$, $T \ll\mu$,  thus extending the results of ref.~\citep{Davison:2013bxa} to include the non-linearities. There exists a clear maximum of the function  $\mu \CD$  vs $T/\mu$, whose position efectively separates the high-temperature hydrodynamic regime from the low-temperature regime. In the limit $\CS\ll1$, the position and the height of the maximum are 
determined analytically by Eqs.~(\ref{tauM}) and (\ref{tauM-1}), respectively. 
%
%
The behavior of the spectral functions in the regime of large chemical potential is mostly influenced by the back-reaction parameter, whereas the corrections resulting from the field strength non-linearities $\CS$ are highly suppressed, even when $\CS=1$.

We then explored  the effects that the non-linearities $\CS,~\CP$ have on the QNMs at high momenta and low temperature. We found that the $\CP$-dependent  term has a strong impact on the overall structure of the QNMs. We observed, in particular,
that removing this term changes  ``repulsion" between the imaginary axis QNMs into  ``attraction".  The extra ``repulsion"  between the modes seems to  increase the convergence to hydrodynamics predictions at low temperatures. We have shown that the $\CP=1$ results are more accurately described by the hydrodynamic approximation $\omega=-i\CD q^2$, since 
 the correction $\propto\Delta(\CP)$ gets smaller with increasing $\CP$. 

We explored the effect of non-linearities $\CS,\CP$ and back-reaction $N_D$ on the formation of a branch cut along the imaginary axis at very low temperatures and fixed momentum. As the back-reaction increases, new QNMs enter the finite 
region of the complex plane arounfd the origin, concentrating on  the imaginary axis, pairing up, and forming a dense structure reminiscent of the formation of the branch cut \cite{Moore:2018mma}, expected to exist in the limit  $T/\mu \rightarrow 0$. The effect of having  $\CP=1$ 
here is  similar to the previous case, making the convergence to hydrodynamics faster compared to $\CP=0$.

Throughout the paper, we have only considered the behavior of QNMs (i.e. the behavior of the poles of relevant components of correlators involving $T^{ab}$ and $J^{a}$), ignoring the residues. We expect all the residues to remain non-zero even in the very high temperature case, but this requires further study.

Finally, we note that the term $ \det (F_{~\nu}^{\mu})\propto(\tilde{F}F)^2\propto(\vec{B}\cdot\vec{E} )^{2}$  in the Lagrangian density is essential for the electric-magnetic duality in $AdS_4 DBI$ \citep{Gaillard:1997rt, Ketov:2001dq, Gibbons:1995cv}. It would be interesting to  extend the results of ref.~\citep{Hartnoll:2007ip}, where effects of the electric-magnetic duality were studied in the  $AdS_4 RN$ case, to the non-linear $AdS_4 DBI$ case.

\section*{Acknowledgements}
I would like to thank T.~Andrade, R.~Davison, S.~Grozdanov, C.~Herzog, N.~Kaplis, A.~Lucas, G.~Policastro, and A.~Pribitoks for useful conversations and correspondence. I would also like to thank A.~O'Bannon and A.~Starinets for useful comments and advice. This work was supported by the Royal Society research grant ``Strange Metals and String Theory'' (RG130401) and by the European Research Council under the European Union's Seventh Framework Programme (ERC Grant agreement 307955).

\appendix
\section{Matching Procedure and Green's functions}
%
Here, we describe  details of the procedure  to obtain the retarded Green's functions in the hydrodynamic limit, 
following  the recipes in ref.~\citep{Davison:2013bxa}.

We need to distinguish two special regions, where we can solve the E.o.M.  First, we have the near-horizon, or inner, region 
($u-1= \delta\ll1$), where we solve the E.o.M.  imposing the ingoing boundary conditions. Then we have the hydrodynamic, or outer, region  $(\omega/4\pi T\ll1)$, valid everywhere including the boundary. These two regions overlap over a range of $\delta$, called the matching region $\omega/4\pi T \ll\delta\ll1$, which can be used to determine the dependence of the momenta $\Pi_{X,Y}$ on the boundary values $X^{(0)}, Y^{(0)}$, consistent with the ingoing boundary condition near the horizon.

\subsection*{Near-horizon (inner) region}
Near the black brane's horizon $(\delta \ll 1)$, it is useful to introduce $u\rightarrow1+\delta$ and $f\rightarrow f'(1)\delta$, where $f'(1)=\frac{4\pi L^{2}T}{r_{H}}$ and $\partial_{u}\approx1/\delta$. In this limit, the equations for gauge-invariant variables decouple and can be written in a compact form as:
\begin{equation}
\phi_{1,2}''(\delta)+\frac{1}{\delta}\phi_{1,2}'(\delta)+\left(\frac{\omega}{4\pi T}\right)^{2}\frac{1}{\delta^{2}}\phi_{1,2}(\delta)\approx0\,,
\end{equation} 
where we 
used
\begin{equation}
\phi_{1}(u)\equiv\bar q X(u),\quad \phi_{2}(u)\equiv Y(u), \qquad \frac{{ \CF}Q\bar{\omega}}{f'(1)}=\frac{\omega}{4\pi T}.
\end{equation}
The inner solution can be written as
\begin{equation}
\phi_{1,2}^{inner}(\delta)=a_{1,2}^{+}e^{+\frac{i\omega}{4\pi T}\log(\delta)}+a_{1,2}^{-}e^{-\frac{i\omega}{4\pi T}\log(\delta)},
\end{equation}
where $a_{1,2}^{\pm}$ are integration constants. Requiring the ingoing boundary condition to obtain retarded Green's functions \cite{Son:2002sd} implies $a_{1,2}^{+}=0$. Expanding this solution in the hydrodynamic limit ($\omega /4\pi T \ll 1$), we obtain
\begin{equation}
\phi_{1,2}^{inner}(\delta)\approx a_{1,2}^{-}\left(1-\frac{i\omega}{4\pi T}\log(\delta)\right)+...\,.
\end{equation}
%

\subsection*{Hydrodynamic (outer) region}
The outer region is defined by
\begin{equation} \frac{{ \CF^{2}}Q^{2}\bar{\omega}^{2}}{u^{2}f(u)^{2}}\ll1,\qquad\frac{{ \CF^{2}}Q^{2}\bar{q}^{2}}{u^{2}f(u)}\ll1.
\end{equation}
In this limit, the non-derivative terms drop and the equations can be trivially integrated
\begin{align}
\label{aeq1}
u^{4}\phi_{1}'(u)+u^{6}f'(u){ { \CF}\CG}\phi_{2}'(u)&\approx\frac{\bar{\omega}^{2}c_{1}}{f(u)}\,, \\
u^{4}\phi_{1}'(u)+4Q^{2}{ T_{D}}\phi_{2}(u)&\approx\frac{\left[\bar{\omega}^{2}-\bar{q}^{2}f(u)\right]c_{2}}{f(u)}\,,
\label{aeq2}
\end{align}
where $c_{1,2}$ are integration constants. It is easier to solve the system by solving for  $\phi_{1}'$ and getting the following equation for $\phi_{2}$ 
\begin{equation}
\phi_{2}'(u)-\frac{4Q^{2}{ T_{D}}}{u^{6}f'(u){ \CG}}\phi_{2}(u)\approx\frac{\bar{q}^{2}c_{2}}{u^{6}f'(u){ \CF \CG}}+\frac{\bar{\omega}^{2}\left(c_{1}-c_{2}\right)}{u^{6}f'(u)f(u){ \CF \CG}}.
\end{equation}
An equation of the form
\begin{equation}
\phi_{2}'(u)+P(u)\phi_{2}(u)=W(u)
\end{equation}
is solved by 
\begin{equation}
\phi_{2}^{outer}(u)=e^{-\int^{u}P(z)dz}\left[b_{2}+\int^{u}W(z)e^{\int^{z}P(s)ds}dz\right].
\end{equation}
In our case, the exponential reduces to a very simple form:
\begin{equation}
e^{-\int^{u}P(z)dz}=\frac{u^{4}f'(u)}{3M}\qquad \text{using}\qquad \left[u^{4}f'(u)\right]'=\frac{4Q^{2}}{u^{2}}{ \frac{{N_D}}{\CG[u]}}.
\end{equation}
This allow us to find the solution for $\phi_2^{outer}$:
\begin{eqnarray}
\label{phi2}
\phi_{2}^{outer}(u) &=& \frac{u^{4}f'(u)}{3M}\Biggl[\phi_{2}^{(0)}-\frac{\bar{q}^{2}c_{2}}{4Q^{2}{ N_{D}\CF}}\frac{\left[3M-u^{4}f'(u)\right]}{u^{4}f'(u)}\nonumber \\
&+& \int^{u}\frac{3M\bar{\omega}^{2}\left(c_{1}-c_{2}\right)dz}{z^{10}f(z)f'(z)^{2}{ { \CF}\CG[z]}}\Biggr]\,,
\end{eqnarray}
where 
 $\phi_{2}^{(0)}$ is the boundary value of the $\phi_2$ field,
\begin{equation}
\phi_{2}^{(0)}=\phi_{2}^{outer}(u\rightarrow\infty).
\end{equation}
Now that we have the solution for $\phi_2$, we can find $\phi _1$ using Eqs.~(\ref{aeq1})-(\ref{aeq2}):
\begin{equation}
\phi_{1}'(u)\approx\frac{\left[\bar{\omega}^{2}-\bar{q}^{2}f(u)\right]c_{2}}{u^{4}f(u)}-\frac{4Q^{2}{ N_D \CF}}{u^{4}}\phi_{2}(u)
\end{equation}
This allows us to find the solution for $\phi_1^{outer}$ as well
\begin{align}\label{phi1}
\phi_{1}^{outer}(u)	&=\phi_{1}^{(0)}+\left(\frac{1-f(u)}{3M}\right)\left[\bar{q}^{2}c_{2}+4Q^{2}{ \CF}{ N_D}\phi_{2}^{(0)}\right] \notag \\
&+\bar{\omega}^{2}\int^{u}\left(\frac{c_{2}}{z^{4}f(z)}-f'(z)\int^{z}\frac{4Q^{2}{ N_D}\left(c_{1}-c_{2}\right)ds}{s^{10}f(s)f'(s)^{2}{ \CG[s]}}\right)dz\,,
\end{align}
where we defined the boundary value of the $\phi_1$ field as
\begin{equation}
\phi_{1}^{(0)}={\phi_1}^{outer}(u\rightarrow\infty).
\end{equation}
%

\subsection*{Matching}
\label{app-matching}
To completely determine the outer solutions and hence the Green's functions, we must impose the ingoing boundary conditions at the horizon and obtain the integration constants $c_{1,2}$ as functions of the boundary fields. This can be done by requiring consistency in the matching region, where we have an overlapping parameter space to compare the solutions of both regions. As the inner region solutions are valid near the horizon, we expand the outer region solutions near $u\rightarrow1+\delta$ for small $\delta\ll1$. After the expansion, and keeping terms up to order $\bar{\omega}^2$ for consistency, the outer solutions reduce to
\begin{align}
\phi_{2}^{outer}(\delta)&=\frac{f'(1)}{3M}\phi_{2}^{(0)}-\frac{\bar{q}^{2}c_{2}}{3M}\left[\frac{3M-f'(1)}{4Q^{2}{ N_{D}\CF}}\right]+\frac{\bar{\omega}^{2}\left(c_{1}-c_{2}\right)}{f'(1)^{2}{ \CF}{ \CG[1]}}\log(\delta)+...\,, \\
\phi_{1}^{outer}(\delta)&=\phi_{1}^{(0)}+\frac{1}{3M}\left[\bar{q}^{2}c_{2}+4Q^{2}{ \CF}{ N_{D}}\phi_{2}^{(0)}\right]+\frac{\bar{\omega}^{2}c_{2}}{f'(1)}\log(\delta)+...,
\end{align}
where we assumed that  
\begin{equation}
\frac{Q\bar{\omega}{ \CF}}{f'(1)}=\frac{\omega}{4\pi T}\ll\delta\ll1\,.
\end{equation}
We can now compare, term by term, the hydrodynamic expansion of $\phi^{inner}$ with the near-horizon expansion  of $\phi^{outer}$ in this matching region where both should be valid
\begin{eqnarray}
\phi_{1}^{(0)}&+&\frac{1}{3M}\left[\bar{q}^{2}c_{2}+4Q^{2}{ \CF}{ N_D}\phi_{2}^{(0)}\right]+\frac{\bar{\omega}^{2}c_{2}}{f'(1)}\log(\delta)\nonumber \\ &=& a_{1}^{-}\left(1-\frac{i\omega}{4\pi T}\log(\delta)\right)\\
\frac{f'(1)}{3M}\phi_{2}^{(0)} &-&  \frac{\bar{q}^{2}c_{2}}{3M}\left[\frac{3M-f'(1)}{4Q^{2}{ \CF}{ N_D}}\right]+\frac{\bar{\omega}^{2}\left(c_{1}-c_{2}\right)}{f'(1)^{2}{ \CF}{ \CG[1]}}\log(\delta)\nonumber \\  &=& a_{2}^{-}\left(1-\frac{i\omega}{4\pi T}\log(\delta)\right).
\end{eqnarray}
This implies that
\begin{equation}
c_{2}=\frac{Q{ \CF}\left(\phi_{1}^{(0)}+\frac{4Q^{2}{ \CF}{ N_{D}}}{3M}\phi_{2}^{(0)}\right)}{i\bar{\omega}-\frac{Q{ \CF}}{3M}\bar{q}^{2}},
\end{equation}
and we can therefore obtain $c_1$ from
\begin{equation}
\frac{i\bar{\omega}\left(c_{1}-c_{2}\right)}{f'(1)Q{ \CF^{2}}{ \CG[1]}}=\frac{f'(1)}{3M}\phi_{2}^{(0)}-\frac{\bar{q}^{2}c_{2}}{3M}\left[\frac{3M-f'(1)}{4Q^{2}{ \CF}{ N_{D}}}\right]\,.
\end{equation}
Once we determine $c_{1,2}$, we can find $\Pi_X$ and $\Pi_Y$ from (\ref{phi2}) and (\ref{phi1}) as functions of the boundary data $X^{(0)}$ and $Y^{(0)}$ in the hydrodynamic approximation. This allows us to finally determine the Green's functions using (\ref{S2}) and the standard holographic prescription \cite{Son:2002sd}:
\begin{equation}
G_{T^{ty}T^{ty}}^{R}=\frac{r_{H}^{2}q^{2}}{2\kappa^{2}L^{2}\left(i\omega-{\cal D}q^{2}\right)},\qquad G_{J^{y}J^{y}}^{R}=\frac{8L^{2}Q^{2}q^{2}{ N_{D}^{2}}}{9\kappa^{2}M^{2}\left(i\omega-{\cal D}q^{2}\right)}\,,
\end{equation}
\begin{equation}
G_{J^{y}T^{ty}}^{R}	=G_{T^{ty}J^{y}}^{R}=\frac{2r_{H}Qq^{2}{ N_{D}}}{3\kappa^{2}M\left(i\omega-{\cal D}q^{2}\right)}\,.
\end{equation}

\bibliographystyle{JHEP}
\bibliography{refs}

\providecommand{\href}[2]{#2}\begingroup\raggedright\begin{thebibliography}{10}

\bibitem{Maldacena:1997re}
J.~M. Maldacena, {\it {The Large N limit of superconformal field theories and
  supergravity}},  {\em Int. J. Theor. Phys.} {\bf 38} (1999) 1113--1133,
  [\href{http://arxiv.org/abs/hep-th/9711200}{{\tt hep-th/9711200}}]. [Adv.
  Theor. Math. Phys.2,231(1998)].

\bibitem{Gubser:1998bc}
S.~S. Gubser, I.~R. Klebanov, and A.~M. Polyakov, {\it {Gauge theory
  correlators from noncritical string theory}},  {\em Phys. Lett.} {\bf B428}
  (1998) 105--114, [\href{http://arxiv.org/abs/hep-th/9802109}{{\tt
  hep-th/9802109}}].

\bibitem{Witten:1998qj}
E.~Witten, {\it {Anti-de Sitter space and holography}},  {\em Adv. Theor. Math.
  Phys.} {\bf 2} (1998) 253--291,
  [\href{http://arxiv.org/abs/hep-th/9802150}{{\tt hep-th/9802150}}].

\bibitem{Aharony:1999ti}
O.~Aharony, S.~S. Gubser, J.~M. Maldacena, H.~Ooguri, and Y.~Oz, {\it {Large N
  field theories, string theory and gravity}},  {\em Phys. Rept.} {\bf 323}
  (2000) 183--386, [\href{http://arxiv.org/abs/hep-th/9905111}{{\tt
  hep-th/9905111}}].

\bibitem{Hartnoll:2009sz}
S.~A. Hartnoll, {\it {Lectures on holographic methods for condensed matter
  physics}},  {\em Class. Quant. Grav.} {\bf 26} (2009) 224002,
  [\href{http://arxiv.org/abs/0903.3246}{{\tt arXiv:0903.3246}}].

\bibitem{McGreevy:2009xe}
J.~McGreevy, {\it {Holographic duality with a view toward many-body physics}},
  {\em Adv. High Energy Phys.} {\bf 2010} (2010) 723105,
  [\href{http://arxiv.org/abs/0909.0518}{{\tt arXiv:0909.0518}}].

\bibitem{Benini:2012iq}
F.~Benini, {\it {Holography and condensed matter}},  {\em Fortsch. Phys.} {\bf
  60} (2012) 810--821, [\href{http://arxiv.org/abs/1202.6008}{{\tt
  arXiv:1202.6008}}].

\bibitem{Hartnoll:2016apf}
S.~A. Hartnoll, A.~Lucas, and S.~Sachdev, {\it {Holographic quantum matter}},
  \href{http://arxiv.org/abs/1612.07324}{{\tt arXiv:1612.07324}}.

\bibitem{Born:1934gh}
M.~Born and L.~Infeld, {\it {Foundations of the new field theory}},  {\em Proc.
  Roy. Soc. Lond.} {\bf A144} (1934), no.~852 425--451.

\bibitem{Dirac:1962iy}
P.~A.~M. Dirac, {\it {An Extensible model of the electron}},  {\em Proc. Roy.
  Soc. Lond.} {\bf A268} (1962) 57--67.

\bibitem{Cederwall:1996uu}
M.~Cederwall, A.~von Gussich, A.~R. Mikovic, B.~E.~W. Nilsson, and
  A.~Westerberg, {\it {On the Dirac-Born-Infeld action for d-branes}},  {\em
  Phys. Lett.} {\bf B390} (1997) 148--152,
  [\href{http://arxiv.org/abs/hep-th/9606173}{{\tt hep-th/9606173}}].

\bibitem{Ketov:2001dq}
S.~V. Ketov, {\it {Many faces of Born-Infeld theory}},  in {\em {7th
  International Wigner Symposium (Wigsym 7) College Park, Maryland, August
  24-29, 2001}}, 2001.
\newblock \href{http://arxiv.org/abs/hep-th/0108189}{{\tt hep-th/0108189}}.

\bibitem{Pal:2012zn}
S.~S. Pal, {\it {Fermi-like Liquid From Einstein-DBI-Dilaton System}},  {\em
  JHEP} {\bf 04} (2013) 007, [\href{http://arxiv.org/abs/1209.3559}{{\tt
  arXiv:1209.3559}}].

\bibitem{Tarrio:2013tta}
J.~Tarrio, {\it {Transport properties of spacetime-filling branes}},  {\em
  JHEP} {\bf 04} (2014) 042, [\href{http://arxiv.org/abs/1312.2902}{{\tt
  arXiv:1312.2902}}].

\bibitem{Kundu:2017cfj}
A.~Kundu, {\it {Flavours and Infra-red Instability in Holography}},  {\em JHEP}
  {\bf 11} (2017) 101, [\href{http://arxiv.org/abs/1708.01775}{{\tt
  arXiv:1708.01775}}].

\bibitem{Pal:2017hai}
S.~S. Pal, {\it {Thermodynamics of Einstein-DBI System}},
  \href{http://arxiv.org/abs/1712.09249}{{\tt arXiv:1712.09249}}.

\bibitem{Fernando:2003tz}
S.~Fernando and D.~Krug, {\it {Charged black hole solutions in
  Einstein-Born-Infeld gravity with a cosmological constant}},  {\em Gen. Rel.
  Grav.} {\bf 35} (2003) 129--137,
  [\href{http://arxiv.org/abs/hep-th/0306120}{{\tt hep-th/0306120}}].

\bibitem{Dey:2004yt}
T.~K. Dey, {\it {Born-Infeld black holes in the presence of a cosmological
  constant}},  {\em Phys. Lett.} {\bf B595} (2004) 484--490,
  [\href{http://arxiv.org/abs/hep-th/0406169}{{\tt hep-th/0406169}}].

\bibitem{Gaillard:1997rt}
M.~K. Gaillard and B.~Zumino, {\it {Nonlinear electromagnetic selfduality and
  Legendre transformations}},  in {\em {Duality and supersymmetric theories.
  Proceedings, Easter School, Newton Institute, Euroconference, Cambridge, UK,
  April 7-18, 1997}}, pp.~33--48, 1997.
\newblock \href{http://arxiv.org/abs/hep-th/9712103}{{\tt hep-th/9712103}}.

\bibitem{Gibbons:1995cv}
G.~W. Gibbons and D.~A. Rasheed, {\it {Electric - magnetic duality rotations in
  nonlinear electrodynamics}},  {\em Nucl. Phys.} {\bf B454} (1995) 185--206,
  [\href{http://arxiv.org/abs/hep-th/9506035}{{\tt hep-th/9506035}}].

\bibitem{andy}
N.~I. Gushterov, A.~O'Bannon, and R.~Rodgers, {\it {Holographic Zero Sound from
  Spacetime-Filling Branes}},  {\em to appear} (2018).

\bibitem{Davison:2013bxa}
R.~A. Davison and A.~Parnachev, {\it {Hydrodynamics of cold holographic
  matter}},  {\em JHEP} {\bf 06} (2013) 100,
  [\href{http://arxiv.org/abs/1303.6334}{{\tt arXiv:1303.6334}}].

\bibitem{Edalati:2010hk}
M.~Edalati, J.~I. Jottar, and R.~G. Leigh, {\it {Shear Modes, Criticality and
  Extremal Black Holes}},  {\em JHEP} {\bf 04} (2010) 075,
  [\href{http://arxiv.org/abs/1001.0779}{{\tt arXiv:1001.0779}}].

\bibitem{Karch:2002sh}
A.~Karch and E.~Katz, {\it {Adding flavor to AdS / CFT}},  {\em JHEP} {\bf 06}
  (2002) 043, [\href{http://arxiv.org/abs/hep-th/0205236}{{\tt
  hep-th/0205236}}].

\bibitem{deBoer:1999tgo}
J.~de~Boer, E.~P. Verlinde, and H.~L. Verlinde, {\it {On the holographic
  renormalization group}},  {\em JHEP} {\bf 08} (2000) 003,
  [\href{http://arxiv.org/abs/hep-th/9912012}{{\tt hep-th/9912012}}].

\bibitem{Fukuma:2002sb}
M.~Fukuma, S.~Matsuura, and T.~Sakai, {\it {Holographic renormalization
  group}},  {\em Prog. Theor. Phys.} {\bf 109} (2003) 489--562,
  [\href{http://arxiv.org/abs/hep-th/0212314}{{\tt hep-th/0212314}}].

\bibitem{tHooft:1998qmr}
G.~'t~Hooft, {\it {When was asymptotic freedom discovered? or the
  rehabilitation of quantum field theory}},  {\em Nucl. Phys. Proc. Suppl.}
  {\bf 74} (1999) 413--425, [\href{http://arxiv.org/abs/hep-th/9808154}{{\tt
  hep-th/9808154}}].

\bibitem{Politzer:1973fx}
H.~D. Politzer, {\it {Reliable Perturbative Results for Strong Interactions?}},
   {\em Phys. Rev. Lett.} {\bf 30} (1973) 1346--1349. [,274(1973)].

\bibitem{Gross:1973id}
D.~J. Gross and F.~Wilczek, {\it {Ultraviolet Behavior of Nonabelian Gauge
  Theories}},  {\em Phys. Rev. Lett.} {\bf 30} (1973) 1343--1346. [,271(1973)].

\bibitem{Faulkner:2009wj}
T.~Faulkner, H.~Liu, J.~McGreevy, and D.~Vegh, {\it {Emergent quantum
  criticality, Fermi surfaces, and AdS(2)}},  {\em Phys. Rev.} {\bf D83} (2011)
  125002, [\href{http://arxiv.org/abs/0907.2694}{{\tt arXiv:0907.2694}}].

\bibitem{LLfluids}
L.~L.D. and E.~M. Lifshitz, {\em Fluid Mechanics}, vol.~6 of {\em Course of
  Theoretical Physics}.
\newblock Pergamon Press, Oxford, 1987.

\bibitem{Kovtun:2012rj}
P.~Kovtun, {\it {Lectures on hydrodynamic fluctuations in relativistic
  theories}},  {\em J. Phys.} {\bf A45} (2012) 473001,
  [\href{http://arxiv.org/abs/1205.5040}{{\tt arXiv:1205.5040}}].

\bibitem{Policastro:2002se}
G.~Policastro, D.~T. Son, and A.~O. Starinets, {\it {From AdS / CFT
  correspondence to hydrodynamics}},  {\em JHEP} {\bf 09} (2002) 043,
  [\href{http://arxiv.org/abs/hep-th/0205052}{{\tt hep-th/0205052}}].

\bibitem{Son:2006em}
D.~T. Son and A.~O. Starinets, {\it {Hydrodynamics of r-charged black holes}},
  {\em JHEP} {\bf 03} (2006) 052,
  [\href{http://arxiv.org/abs/hep-th/0601157}{{\tt hep-th/0601157}}].

\bibitem{Policastro:2001yc}
G.~Policastro, D.~T. Son, and A.~O. Starinets, {\it {The Shear viscosity of
  strongly coupled N=4 supersymmetric Yang-Mills plasma}},  {\em Phys. Rev.
  Lett.} {\bf 87} (2001) 081601,
  [\href{http://arxiv.org/abs/hep-th/0104066}{{\tt hep-th/0104066}}].

\bibitem{Kovtun:2003wp}
P.~Kovtun, D.~T. Son, and A.~O. Starinets, {\it {Holography and hydrodynamics:
  Diffusion on stretched horizons}},  {\em JHEP} {\bf 10} (2003) 064,
  [\href{http://arxiv.org/abs/hep-th/0309213}{{\tt hep-th/0309213}}].

\bibitem{Kovtun:2005ev}
P.~K. Kovtun and A.~O. Starinets, {\it {Quasinormal modes and holography}},
  {\em Phys. Rev.} {\bf D72} (2005) 086009,
  [\href{http://arxiv.org/abs/hep-th/0506184}{{\tt hep-th/0506184}}].

\bibitem{Buchel:2003tz}
A.~Buchel and J.~T. Liu, {\it {Universality of the shear viscosity in
  supergravity}},  {\em Phys. Rev. Lett.} {\bf 93} (2004) 090602,
  [\href{http://arxiv.org/abs/hep-th/0311175}{{\tt hep-th/0311175}}].

\bibitem{Starinets:2008fb}
A.~O. Starinets, {\it {Quasinormal spectrum and the black hole membrane
  paradigm}},  {\em Phys. Lett.} {\bf B670} (2009) 442--445,
  [\href{http://arxiv.org/abs/0806.3797}{{\tt arXiv:0806.3797}}].

\bibitem{Adams:2012th}
A.~Adams, L.~D. Carr, T.~Schäfer, P.~Steinberg, and J.~E. Thomas, {\it
  {Strongly Correlated Quantum Fluids: Ultracold Quantum Gases, Quantum
  Chromodynamic Plasmas, and Holographic Duality}},  {\em New J. Phys.} {\bf
  14} (2012) 115009, [\href{http://arxiv.org/abs/1205.5180}{{\tt
  arXiv:1205.5180}}].

\bibitem{PhysRevB.56.8714}
K.~Damle and S.~Sachdev, {\it Nonzero-temperature transport near quantum
  critical points},  {\em Phys. Rev. B} {\bf 56} (Oct, 1997) 8714--8733.

\bibitem{Kovtun:2008kx}
P.~Kovtun and A.~Ritz, {\it {Universal conductivity and central charges}},
  {\em Phys. Rev.} {\bf D78} (2008) 066009,
  [\href{http://arxiv.org/abs/0806.0110}{{\tt arXiv:0806.0110}}].

\bibitem{Iqbal:2008by}
N.~Iqbal and H.~Liu, {\it {Universality of the hydrodynamic limit in AdS/CFT
  and the membrane paradigm}},  {\em Phys. Rev.} {\bf D79} (2009) 025023,
  [\href{http://arxiv.org/abs/0809.3808}{{\tt arXiv:0809.3808}}].

\bibitem{Iqbal:2011in}
N.~Iqbal, H.~Liu, and M.~Mezei, {\it {Semi-local quantum liquids}},  {\em JHEP}
  {\bf 04} (2012) 086, [\href{http://arxiv.org/abs/1105.4621}{{\tt
  arXiv:1105.4621}}].

\bibitem{Brattan:2010pq}
D.~K. Brattan and S.~A. Gentle, {\it {Shear channel correlators from hot
  charged black holes}},  {\em JHEP} {\bf 04} (2011) 082,
  [\href{http://arxiv.org/abs/1012.1280}{{\tt arXiv:1012.1280}}].

\bibitem{Denef:2009yy}
F.~Denef, S.~A. Hartnoll, and S.~Sachdev, {\it {Quantum oscillations and black
  hole ringing}},  {\em Phys. Rev.} {\bf D80} (2009) 126016,
  [\href{http://arxiv.org/abs/0908.1788}{{\tt arXiv:0908.1788}}].

\bibitem{Karch:2008fa}
A.~Karch, D.~T. Son, and A.~O. Starinets, {\it {Zero Sound from Holography}},
  \href{http://arxiv.org/abs/0806.3796}{{\tt arXiv:0806.3796}}.

\bibitem{Karch:2009zz}
A.~Karch, D.~T. Son, and A.~O. Starinets, {\it {Holographic Quantum Liquid}},
  {\em Phys. Rev. Lett.} {\bf 102} (2009) 051602.

\bibitem{Davison:2013uha}
R.~A. Davison, M.~Goykhman, and A.~Parnachev, {\it {AdS/CFT and Landau Fermi
  liquids}},  {\em JHEP} {\bf 07} (2014) 109,
  [\href{http://arxiv.org/abs/1312.0463}{{\tt arXiv:1312.0463}}].

\bibitem{Anantua:2012nj}
R.~J. Anantua, S.~A. Hartnoll, V.~L. Martin, and D.~M. Ramirez, {\it {The Pauli
  exclusion principle at strong coupling: Holographic matter and momentum
  space}},  {\em JHEP} {\bf 03} (2013) 104,
  [\href{http://arxiv.org/abs/1210.1590}{{\tt arXiv:1210.1590}}].

\bibitem{Herzog:2002fn}
C.~P. Herzog, {\it {The Hydrodynamics of M theory}},  {\em JHEP} {\bf 12}
  (2002) 026, [\href{http://arxiv.org/abs/hep-th/0210126}{{\tt
  hep-th/0210126}}].

\bibitem{Edalati:2010pn}
M.~Edalati, J.~I. Jottar, and R.~G. Leigh, {\it {Holography and the sound of
  criticality}},  {\em JHEP} {\bf 10} (2010) 058,
  [\href{http://arxiv.org/abs/1005.4075}{{\tt arXiv:1005.4075}}].

\bibitem{Kodama:2003kk}
H.~Kodama and A.~Ishibashi, {\it {Master equations for perturbations of
  generalized static black holes with charge in higher dimensions}},  {\em
  Prog. Theor. Phys.} {\bf 111} (2004) 29--73,
  [\href{http://arxiv.org/abs/hep-th/0308128}{{\tt hep-th/0308128}}].

\bibitem{Kapusta:2006pm}
J.~I. Kapusta and C.~Gale, {\em {Finite-temperature field theory: Principles
  and applications}}.
\newblock Cambridge Monographs on Mathematical Physics. Cambridge University
  Press, 2011.

\bibitem{Son:2002sd}
D.~T. Son and A.~O. Starinets, {\it {Minkowski space correlators in AdS / CFT
  correspondence: Recipe and applications}},  {\em JHEP} {\bf 09} (2002) 042,
  [\href{http://arxiv.org/abs/hep-th/0205051}{{\tt hep-th/0205051}}].

\bibitem{Erdmenger:2008rm}
J.~Erdmenger, M.~Haack, M.~Kaminski, and A.~Yarom, {\it {Fluid dynamics of
  R-charged black holes}},  {\em JHEP} {\bf 01} (2009) 055,
  [\href{http://arxiv.org/abs/0809.2488}{{\tt arXiv:0809.2488}}].

\bibitem{Haack:2008xx}
M.~Haack and A.~Yarom, {\it {Universality of second order transport
  coefficients from the gauge-string duality}},  {\em Nucl. Phys.} {\bf B813}
  (2009) 140--155, [\href{http://arxiv.org/abs/0811.1794}{{\tt
  arXiv:0811.1794}}].

\bibitem{Grozdanov:2016fkt}
S.~Grozdanov and A.~O. Starinets, {\it {Second-order transport, quasinormal
  modes and zero-viscosity limit in the Gauss-Bonnet holographic fluid}},  {\em
  JHEP} {\bf 03} (2017) 166, [\href{http://arxiv.org/abs/1611.07053}{{\tt
  arXiv:1611.07053}}].

\bibitem{Leaver:1990zz}
E.~W. Leaver, {\it {Quasinormal modes of Reissner-Nordstrom black holes}},
  {\em Phys. Rev.} {\bf D41} (1990) 2986--2997.

\bibitem{Moore:2018mma}
G.~D. Moore, {\it {Stress-stress correlator in $\phi^{4}$ theory: poles or a
  cut?}},  {\em JHEP} {\bf 05} (2018) 084,
  [\href{http://arxiv.org/abs/1803.00736}{{\tt arXiv:1803.00736}}].

\bibitem{Banerjee:2008th}
N.~Banerjee, J.~Bhattacharya, S.~Bhattacharyya, S.~Dutta, R.~Loganayagam, and
  P.~Surowka, {\it {Hydrodynamics from charged black branes}},  {\em JHEP} {\bf
  01} (2011) 094, [\href{http://arxiv.org/abs/0809.2596}{{\tt
  arXiv:0809.2596}}].

\bibitem{Grozdanov:2015kqa}
S.~Grozdanov and N.~Kaplis, {\it {Constructing higher-order hydrodynamics: The
  third order}},  {\em Phys. Rev.} {\bf D93} (2016), no.~6 066012,
  [\href{http://arxiv.org/abs/1507.02461}{{\tt arXiv:1507.02461}}].

\bibitem{Hartnoll:2007ip}
S.~A. Hartnoll and C.~P. Herzog, {\it {Ohm's Law at strong coupling: S duality
  and the cyclotron resonance}},  {\em Phys. Rev.} {\bf D76} (2007) 106012,
  [\href{http://arxiv.org/abs/0706.3228}{{\tt arXiv:0706.3228}}].

\end{thebibliography}\endgroup

\end{document}